\documentclass[useAMS,amsmath,amssymb,usegraphicx,usenatbib,usedcolumn]{mn2e}
\usepackage{rotating}
\usepackage{stfloats}
\usepackage{amssymb,amsmath}
\usepackage{longtable,lscape}
\usepackage{acronym}
\usepackage{subfigure}
\usepackage[dvips]{graphicx}
\usepackage{color}
\usepackage{epsf}
\usepackage{multicol}
\usepackage{graphics,latexsym,longtable,lscape}

\newif\ifAMStwofonts
\AMStwofontstrue

\newcommand{\abs}[1]{\lvert#1\rvert}

\newcommand{\SIII}{S\,{\sc iii}}
\newcommand{\HA}{H{$\alpha$}}

\newcommand{\msol}{M$_\odot$}

\newcommand{\miriad}{{\sc miriad}}

\newcommand{\Tb}{T$_b$}

\newcommand{\MlogTb}{log{(\textrm{T}_b)}}

\title[Radio-continuum detections of GPNe: I. MASH PNe ]{Radio-continuum detections of Galactic Planetary Nebulae\\ I. MASH PNe detected in large-scale radio surveys.}
\author[I. S. Boji\v{c}i\'c et al.]{I. S. Boji\v{c}i\'c,$^{1}$
Q. A.~Parker,$^{1,2}$
M. D. Filipovi\'c,$^3$
D. J. Frew$^1$\\
$^1$Department of Physics and Astronomy, Macquarie University, Sydney, NSW 2109, Australia\\
$^2$Australian Astronomical Observatory, Epping, NSW 1710, Australia\\
$^3$University of Western Sydney, Locked Bag 1797, Penrith South DC, NSW 1797, Australia\\
}
\begin{document}

\date{Accepted ... Received ...; in original form ...}

\pagerange{\pageref{firstpage}--\pageref{lastpage}} \pubyear{2010}

\maketitle

\label{firstpage}

\begin{abstract}

We present an updated and newly compiled radio-continuum data-base for MASH PNe detected in the extant large scale ``blind'' radio-continuum surveys (NVSS, SUMSS/MGPS-2 and PMN) and, for a small number of MASH PNe, observed and detected in targeted radio-continuum observations. We found radio counterparts for approximately 250 MASH PNe. In comparison with the percentage of previously known Galactic PNe detected in the NVSS and MGPS-2 radio-continuum surveys and according to their position on the flux density-angular diameter and the radio brightness temperature evolutionary diagrams we conclude, unsurprisingly, that the MASH sample presents the radio-faint end of the known Galactic PNe population. Also, we present radio-continuum spectral properties of a small sub-sample of MASH PNe located in the strip between declinations -30\degr\ and -40\degr, that are detected in both the NVSS and MGPS-2 radio surveys. 

\end{abstract}


\section{Introduction}\label{intro}

Planetary nebulae (PNe) are ionized, gaseous envelopes ejected from intermediate-mass stars (1-8~\msol) in the final stage of their evolution. At radio frequencies the dominant emission mechanism from ionised nebulae is bremsstrahlung (or free-free radiation). Due to the direct dependance of the bremsstrahlung emissivity on the square of the electron density the radio-continuum observations of PNe are an important source of information concerning the overall physical structure and mass of the ionised gas. Also, the radio brightness is especially effective as an evolutionary tracer due to its
intrinsic dependence on the ionised gas density and the degree of ionisation. In the initial stage of the post-AGB evolution, the radio flux density will be proportional to the number of ionising photons from the central star \citep{1990A&A...234..387Z}, while from the moment when the shell becomes fully ionised, the radio-evolution starts to be governed mostly by the expansion of the ionised gas.

Targeted radio-continuum observations of PNe are usually based on optically identified objects. However, the effects of interstellar reddening, especially in directions where most PNe are expected to be found (eg. Galactic plane and Galactic bulge) have strongly biased optically detected PNe toward intrinsically radio brighter objects. The new Macquarie/AAO/Strasbourg \HA\ (MASH) catalogues of Galactic PNe \citep[GPNe;][]{2006MNRAS.373...79P,2008MNRAS.384..525M} have increased by nearly 40\% the known population of Galactic PNe, which now stands at nearly 3000 in total \citep{2010PASA...27..129F} and made a major impact in the domain of PNe with low and extremely low luminosities which were previously poorly represented. These significant discoveries derived from the innovative AAO/UKST SuperCOSMOS H$\alpha$ survey of the Southern Galactic Plane \citep[SHS;][]{2005MNRAS.362..689P} whose depth, arcsecond resolution, uniformity, and 4000 square degree areal coverage opened fresh parameter discovery space.

Key problems in PN research are some of the main aims of current and future studies of the MASH team which will fully exploit this new large sample  \citep[e.g.][]{2006IAUS..234...49F,2007ApJ...669..343C,2008ASPC..391..181M,2009A&A...496..813M,2009A&A...505..249M,2009ASPC..404..337K}. These problems include the distance problem, unravelling the optical and radio-continuum PN luminosity function from significant new samples in the Galactic Bulge \citep{KOVAC2010}, understanding the quantitative differences in the multi-wavelength characteristics of PNe \citep{COH2010} and examination of correlations between evolutionary stage and observable properties of PNe.

The radio-continuum data form an important component in the multi-wavelength evolutionary scheme for this new large sample of Galactic PNe. It enables direct calculation of electron densities, degrees of ionization and interstellar extinction in the direction of PNe. In this paper we present the newly compiled and complete database of radio-continuum detected MASH PNe. 

\section{Radio-continuum identifications of PNe from the MASH catalogue}

Prior to the extensive radio-continuum survey of MASH PNe (Boji\v{c}i\'c et al. in prep.) only a handful of MASH PNe objects have been observed in PNe-targeted radio-continuum surveys. \cite{1991A&AS...91..481R}, using the Westerbork Radio Telescope (WSRT) and the Very Large Array (VLA) radio telescope, observed a large set of PN candidates selected from the IRAS Point Source catalogue and placed in the direction of the Galactic Bulge. The chosen sample is based on the far infrared selection criteria described in \cite{1988A&A...205..248P}. The authors noted that approximately 20\% of observed objects were detectable at 6~cm though not all of detected objects were confirmed PNe. From that sample some eight objects have been recently identified as likely PN in SHS \HA\ images and, after confirmatory optical spectroscopy, made their way into the MASH catalogue. Similarly, based on the [\SIII]$\lambda$9532 survey of a 4$\times$4 degree field centred on the Galactic Centre, \cite{2001A&A...373..536V} reported on $\sim$100 possible identifications of PNe.  For 63 PN candidates from this sample the obtained spectra appear consistent with highly reddened PN \citep{2001A&A...373..536V}. Using the Australia Telescope Compact Array (ATCA), some 64 PN candidates were observed and 57 and 54 detected at 6 and 3~cm, respectively. The MASH catalogue contains five PNe observed in that survey from which only two have been positively detected in the radio-continuum.

On the other hand,  several large-scale radio surveys like the NRAO VLA Sky Survey \citep[NVSS;][]{1998AJ....115.1693C}, the Sydney University Molonglo Sky Survey \citep[SUMSS;][]{1999AJ....117.1578B,2003MNRAS.342.1117M} and its complementary Molonglo Galactic Plane Surveys  \citep[MGPS and MGPS-2;][]{1999ApJS..122..207G,  2007MNRAS.382..382M}, and the Parkes-MIT-NRAO (PMN) survey \citep{1996yCat.8038....0W, 1994ApJS...91..111W, 1994ApJS...90..179G, 1994ApJS...90..173G} have proven to be an excellent source of PNe radio data \citep[e.g.][]{1998ApJS..117..361C, 1999ApJS..123..219C, 2001A&A...373.1032S,2003MNRAS.346..719M, 2005ApJS..159..282L, 2007ApJ...669..343C, 2008A&A...482..529U, 2009A&A...504..291V}. These surveys form the basis for this study. 

\subsection{MASH PNe detected in NVSS}\label{NVSSdetections}

The NVSS is a blind radio survey that covers  $\sim80$\% of the sky north of $\delta$~=~--40$^o$ at 1.4~GHz (20~cm). The detection threshold limit is $\sim$2.5~mJy (for sources with angular size comparable with the FWHM of the produced synthesised beam) with positional uncertainties of the order of 7~arcsec at the survey limit (for sources brighter that 15~mJy the {\it rms} uncertainties are as low as 1~arcsec). The NVSS is $\ge90$\% complete at flux densities above 5~mJy except near the Galactic Bulge, where, for the given flux threshold level, the catalogue completeness is estimated to be $\ge80$\% \citep{1998AJ....115.1693C}. For unresolved sources the incremental completeness is 50\% at 2.5~mJy and it rises rapidly to 99\% at 3.4~mJy.

\cite{1998ApJS..117..361C} (hereafter CK98) reported detections of 680 of the 885 known PNe \citep[listed in the Strasbourg-ESO Catalogue of Galactic Planetary Nebulae;][]{1992secg.book.....A} with $\delta>$--40\degr. An additional 22 known PNe and 122 PNe candidates satisfying the IR colour criteria from \cite{1988A&AS...76..317P} have been presented in \cite{1999ApJS..123..219C}.

Another large data set of PNe radio-continuum identifications in the NVSS catalogue was presented in \cite{2005ApJS..159..282L} (hereafter LCY05). Based on the First Supplement to the Strasbourg-ESO Catalogue of Galactic Planetary Nebulae (Acker et al. 1996), PNe catalogued in \cite{2001A&A...377.1035C, 2001A&A...378..843K,2002AN....323...57K,2003A&A...408.1029K,2003MNRAS.339..735B} and on the set of 1047 positions from the preliminary MASH catalogue they identified 315 correlated radio-continuum detections.

However, from 178 objects detected from the preliminary MASH list, some 33 have been subsequently rejected as non-PN by the MASH team prior to publication of the MASH catalogue\footnote{Based on control evaluation of their multi-wavelenght properties \citep[see ][]{2010PASA...27..129F}.}. Additionally, the MASH-II supplement  \citep{2008MNRAS.384..525M} introduced a large set of PNe which have not been previously correlated with the NVSS catalogue. Thus, an updated cleaned list of MASH-NVSS radio identifications and fluxes is presented here for the first time.

Similarly as in LCY05, we first compared catalogued optical positions with NVSS positions. Cross-identifications between the two catalogues were considered as ``possible'' if the offset between the radio peak and the optical centroid was:

\begin{itemize}
\item less than 25~arcsec for objects with $\theta_{opt}<25$~arcsec,
\item less than 1.2$\times\theta_{opt}$ for objects with $25~\textrm{arcsec}\le\theta_{opt}<45$~arcsec and
\item less than $\theta_{opt}$ for objects with $\theta_{opt}\ge45$~arcsec,
\end{itemize}

\noindent where $\theta_{opt}$ is the optically determined angular diameter in MASH (we will use this notation throughout this paper unless stated otherwise). All possible identifications were visually inspected using the finding charts (approximately 7$\times$7~arcmin) created from radio-continuum and \HA\ images. The 1.4~GHz images were obtained from the NVSS postage stamp server\footnote{www.cv.nrao.edu/nvss/postage.shtml}.

The updated list of positive NVSS radio-continuum identifications now contains 201 MASH PNe. It includes 145 confirmed MASH PNe listed in LCY05 and additional 56 objects mostly from the MASH-II supplement.  Furthermore, 14 radio detections from the updated list have been flagged as suspect due to the larger offset of the radio-peak position from the optical centroid or because their radio counterparts are just below the threshold level and haven't been picked up by the NVSS cataloguing algorithm. In the later case we quote 2~mJy as a rough estimate of the flux density. All detected PNe are presented in Table~\ref{tab:nvssdetection}. The first, second and third row of Table~\ref{tab:nvssdetection} represent the official IAU PNG designation, the unique MASH catalogue identifier as described in \cite{2006MNRAS.373...79P} and designation of the corresponding radio source from the original NVSS catalogue \citep{1998AJ....115.1693C}, respectively. The fourth, fifth and sixth columns contain the equatorial RAJ2000 and DECJ2000 coordinates of the radio source, and the angular offset from the catalogued optical position of the MASH PN (in arcsec), respectively. The integrated flux density, as given in the NVSS, and optically determined angular diameter are presented in columns seven and eight. All suspect detections have been designated with a preceding colon in the flux density column and no uncertainty in the radio flux is reported. The final column gives the identification key of a comment to some specifics of an object, usually found in comparison with optical imagery or, if available, in comparison with independent observational data.  Full comments are given in Appendix~\ref{sec:notesonpne}.

\begin{figure}
\begin{center}
\includegraphics[scale=1]{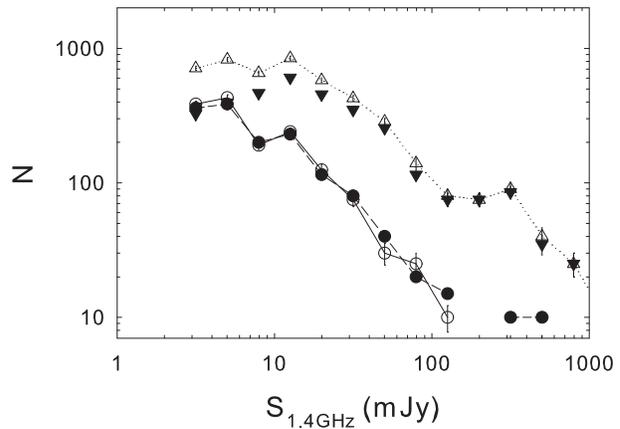}
\caption{Numbers N of detected PNe per decade of 1.4~GHz flux density $S_{1.4\textrm{GHz}}$. Open circles connected with the dashed line and black-filled circles connected with the solid line represent the new MASH + 137 non-MASH PNe from LCY05 and the ``old'' sample from LCY05 which contains non-PNe contaminants. Black-filled triangles are known PNe from CK98. Open triangles, connected with the dotted line, represent the new full sample presented in this paper.}\label{fig:fluxdistsNVSS}
\end{center}
\end{figure}

In total, only about 25\% of MASH PNe have been detected in the NVSS. In this statistic we include all NVSS ``detectable'' MASH objects i.e. north of $\delta$~=~--40\degr\ and with angular diameters smaller than 100~arcsec. Off course, relatively bright objects, larger than 100~arcsec, are detectable in VLA configurations used in the NVSS \citep{1998AJ....115.1693C}. However, due to the generally low radio brightness of MASH PNe, we adapted 100~arcsec as a reasonable upper limit. In comparison with $\sim$75\% NVSS detections of known PNe (CK98) it is clear that the MASH sample contains an intrinsically radio-fainter population at 1.4~GHz than previously observed. Following the analysis given in CK98 and LCY05, we plot in Fig.~\ref{fig:fluxdistsNVSS} the number of detected sources per decade of flux density. We plot known PNe catalogued in CK98 (black-filled triangles), an updated sample of MASH PNe combined with the list of 137 non-MASH PNe from LCY05 (open circles) and the full sample (open triangles). Overplotted with filled circles is the ``old'' sample from LCY05 which contains non-PNe contaminants. As can be seen from Fig.~\ref{fig:fluxdistsNVSS}, the distribution did not change significantly from LCY05 \cite[see Fig.~2 in][]{2005ApJS..159..282L}. However, it is important to note that the number of objects with $S_{1.4\textrm{GHz}}<$10~mJy increased by about 10\% while almost all objects listed in LCY05 with $S_{1.4\textrm{GHz}}>$100~mJy, have now been excluded from the final MASH catalogue as being PNe contaminants.

\subsection{MASH PNe detected in SUMSS/MGPS-2}\label{SUMSSdetections}

\begin{figure}
\begin{center}
\includegraphics[scale=1]{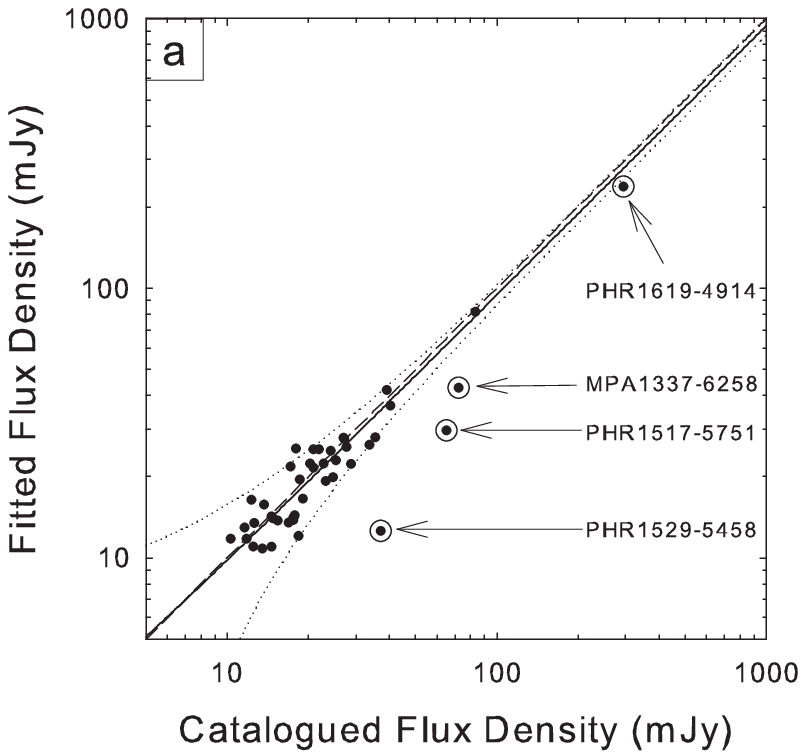}
\includegraphics[scale=1]{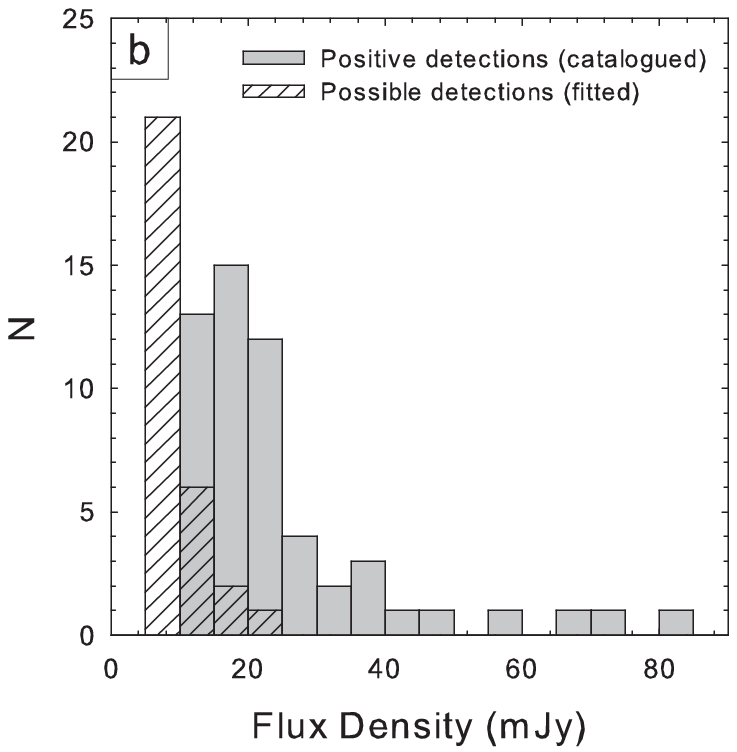}
\caption[Comparison between catalogued and fitted 0.843~GHz flux densities and the distribution histogram of flux
densities for MASH PNe detected and possibly detected  in SUMSS/MGPS-2.]{{\bf a)} Comparison between catalogued (abscissa) and fitted (ordinate) 0.843~GHz flux densities. Four marked PNe (PHR1529-5458, PHR1517-5751, MPA1337-5751, PHR1619-4914) appear to be partially resolved in SUMSS/MGPS-2 radio-continuum images. The solid line represents the 1-1 relation and 90\% prediction bands were plotted with dotted lines. {\bf b)} The distribution histogram of flux densities for detected (grey filled) and possibly detected (open hatched) MASH PNe in MGPS-2.}\label{fig:mgpsfitt}
\end{center}
\end{figure}

The SUMSS and the MGPS-2 are 35.6~cm (0.843~GHz) complementary radio surveys carried out with the Molonglo Observatory Synthesis Telescope (MOST). The main products of both surveys are 4.3\degr$\times$4.3\degr\ mosaic images with 45\arcsec$\times$45\arcsec$\cdot$cosec~$\delta$ resolution making them the highest resolution large scale radio-continuum surveys of the southern Galactic plane \citep{2003MNRAS.342.1117M,2007MNRAS.382..382M}.

The SUMSS catalogue is concentrated on the extragalactic radio population covering approximately 8000 square degrees with $\delta<$~--30\degr\ and $\abs{b}>$10\degr. The MGPS-2 is the Galactic plane counterpart to  SUMSS covering the range $\abs{b}<$~10\degr\ and 245\degr$<l<$365\degr. Positional uncertainties of both surveys are usually smaller than 2~arcsec. Version 2.0 of the SUMSS catalogue contains sources brighter than 6~mJy~Beam$^{-1}$ at $\delta\le$~--50\degr\ and 10~mJy~Beam$^{-1}$ at $\delta>$~--50\degr\ with estimated errors in the  internal flux density scale smaller than 3\%. The sensitivity limits and flux uncertainties in the MGPS-2 catalogue are similar to SUMSS.

\begin{table*}
\begin{footnotesize}
\caption{The nine MASH PNe detected and possibly detected in the PMN survey.}\label{tab:PMNdetections}
\begin{center}
\begin{tabular}{cccccrrrr}

\hline \noalign{\smallskip}

PNG & MASH Name & PMN source & RAJ2000 & DEJ2000 & Offset  & S$_{5\textrm{GHz}}$ & $\theta_{opt}$ & c.\\
\noalign{\smallskip}
& & &  & & [\arcsec] &[mJy] & [\arcsec] &\\
\hline \noalign{\smallskip}

267.4+01.3	&PHR0907$-$4532	&J0907$-$4532&09 07 31.2&$-$45 32 51	&20.0	&79$\pm$9	&199		&-\\
318.9+00.7	&PHR1457$-$5812	&J1457$-$5812&14 57 35.3&$-$58 12 00	&9.5	&	80$\pm$8		&27.8	&-\\
324.3+01.1	&PHR1529$-$5458	&J1529$-$5458&15 29 28.9&$-$54 58 32	&27.1	&50$\pm$8	&92.5	&-\\
329.8$-$03.0	&PHR1617$-$5445	&J1617$-$5445&16 17 26.0&$-$54 45 36	&58.8	&44$\pm$8	&17.4	&-\\
333.9+00.6	&PHR1619$-$4914	&J1619$-$4913&16 19 38.9&$-$49 13 51	&14.8	&244$\pm$15	&33.9	&-\\
337.4+02.6	&PHR1625$-$4522	&J1625$-$4522&16 25 52.5&$-$45 22 06	&37.1	&165$\pm$12	&314.6	&-\\
335.4$-$01.9	&PHR1637$-$4957	&J1637$-$4958&16 37 46.3&$-$49 58 21	&33.8	&42$\pm$8	&19.2	&-\\
003.5$-$01.2	&PPA1758$-$2628	&J1758$-$2628&17 58 38.2&$-$26 28 55	&18.0	&:76$\pm$11	&4.5		&-\\
003.6$-$01.3	&PHR1759$-$2630	&J1759$-$2631&17 59 11.5&$-$26 31 11	&47.7	&:50$\pm$11	&8.5		&-\\

\hline \noalign{\smallskip}
\end{tabular}
\end{center}
\end{footnotesize}
\end{table*}

Selection criteria based on position, angular dimension and angular resolution, following the scheme as for NVSS (see \S~\ref{NVSSdetections}), are applied to these two catalogues. For visual inspection we used available total intensity images from the SUMSS postage stamp server\footnote{http://www.astrop.physics.usyd.edu.au/cgi-bin/postage.pl} (SUMSS/MGPS-2 radio-continuum images hereafter). Finding charts were composed for 527 MASH PNe positions for which SUMSS/MGPS-2 radio-continuum images were available.

None of the 13 MASH PNe with $\delta<$~--30\degr\ and $\abs{b}>$10\degr\ are catalogued in the SUMSS catalogue. However, it should be noted that four of these 13 objects posses much larger angular diameters than the FWHM of the MOST restoring beam (965~arcsec, 401~arcsec, 420~arcsec\ and 250~arcsec, respectively). The other nine MASH PNe are clearly below the sensitivity limit of the SUMSS. In the first iteration, from 672 MASH PNe in the $\abs{b}<$~10\degr\ and 245\degr$<l<$365\degr\ region, we cross-identified 65 radio objects from the MGPS-2 catalogue.  After visual inspection of finding charts, 50 objects have been assigned with a ``positive detection'' flag. From the rest of the preliminary list three objects were flagged as ``suspect'' detections and 12 objects as ``non-detection''.  Also, additional 28 MASH PNe have been assigned with a ``possible detection'' flag. Radio counterparts from the later group were found from the correlation between the optical position and an obvious flux excess over the surrounding noise. None of these objects is catalogued in the MGPS-2 catalogue. Possible detections for three of these 28 (PHR1115$-$6059, PHR1346$-$6116 and PHR1625$-$4522) are discussed in more details in Appendix~\ref{sec:notesonpne}. The radio contour plots for other 25 possible detections are presented in Appendix~\ref{sec:notesonpne}: Fig.~\ref{pysumss1} and Fig.~\ref{pysumss2}. We made an initial estimate of flux densities for these radio objects. We used \miriad's IMFIT task to fit an elliptical (45\arcsec$\times$45\arcsec$\cdot$cosec~$\delta$) Gaussian to all possible detections. For an estimate of cut-off noise we used a 3$\sigma_{rms}$ level where $\sigma_{rms}$ is the local {\it rms} noise level.

In order to examine the quality of the fits we also measured flux densities of all available detections from the catalogued entries. Figure~\ref{fig:mgpsfitt}a shows a comparison between catalogued (abscissa) and fitted (ordinate) flux densities. Three PNe with the largest offset from the catalogued flux (PHR1529-5458, PHR1517-5751 and MPA1337-5751) and the brightest object in this sample (PHR1619-4914) appear to be partially resolved in the SUMSS/MGPS-2 radio-continuum images. The divergence in flux density estimates, for resolved or partially resolved objects, are caused by the simplified fitting method we used. 

As can be seen from Fig.~\ref{fig:mgpsfitt}a most fitted values in the 10-20~mJy region are well contained within the $\sim$20\% deviation from the expected (catalogued) value. However, the 90\% prediction band (dotted line) in the region below 10~mJy imply that errors in our fitted flux densities could be substantial. Thus, we stress that quoted values must be taken only as rough estimates for flux densities.

Fig.~\ref{fig:mgpsfitt}b shows the histogram distribution of flux densities for detected and possibly detected MASH PNe. Approximately 10 objects, flagged as possible detections, have fitted flux densities above the average 10~mJy threshold level for catalogued MGPS-2 sources. Some of these  objects appear to be partially resolved or to be only marginal detections (e.g. PHR1115-6059 and PHR1625-4522) and some were found in particularly noisy regions with the local noise much higher than the usual 1-1.5~mJy~Beam$^{-1}$ (e.g. PHR1619-5131) or on the top of the some larger radio structure (e.g. MPA1523-5710).

The final list of MASH PNe detected (including suspect detections) and possibly detected in SUMSS/MGPS-2 contains 53 and 28 objects, respectively. It is important to remember that for $\sim$25\% of MASH PNe, with positions south of $\delta$=--30\degr, SUMSS/MGPS-2 mosaics were not available. Therefore, another $\sim$10 objects could have 0.843~GHz flux densities larger than 5~mJy. All detected PNe are presented in Table~\ref{tab:mgps2detection}. Possible detections are presented in Table~\ref{tab:sumssdetection}. We list the PNG and MASH designations in columns 1 and 2, and the original MGPS-2 designation, if the object is catalogued, or designation produced by following the usual radio source nomenclature i.e. J{\it HHMMSS-DDMMSS} in the case of possible detection. Next we list RAJ2000 and DECJ2000 of the radio source (columns 3, 4 and 5, respectively), the angular offset from the catalogued optical position in arcsec (column 6) and the total flux density as given in the MGPS-2 catalogue and optically determined angular diameter (columns 7 and 8). The final column gives the identification key of a comment to some specifics of an object, usually found in comparison with optical imagery or, if available, in comparison with independent observational data.  Full comments are given in Appendix~\ref{sec:notesonpne}.

\subsection{MASH PNe detected in PMN}

The PMN survey \citep{1994ApJS...91..111W} fully covers the complete angular distribution of the MASH catalogue. It was made at 4.8~GHz using the NRAO multi-beam receiver mounted at the prime focus of the Parkes 64-m radio telescope. The survey was divided into four zones covering declinations between 10\degr\ to -9\degr.5 (Equatorial), -9\degr.5 to -29\degr\ (Tropical), -29\degr\ to -37\degr\ (Zenith) and -37\degr\ to -87\degr.5 (Southern) and with approximate flux limits of 40~mJy, 42~mJy, 72~mJy and 20-50~mJy for each zone, respectively. The resolution of the survey was 4\arcmin.2.

\begin{center}
\begin{table*}
\caption{List of MASH PNe with 5GHz detections compiled from the literature.}\label{tab:othersdetection}
\begin{tabular}{ccccrrrrr}
\hline \noalign{\smallskip}
PNG & MASH Name & RAJ2000 & DEJ2000 & Offset    & $S_{5\textrm{GHz}}$ & $\theta_{opt}$& Ref.& c.\\
\noalign{\smallskip}
    &           &         &         & [\arcsec]  & [mJy]     &  [\arcsec] &              & \\
\hline \noalign{\smallskip}

293.1$-$00.0&	BMP1128$-$6121&	11 28 12.8&	$-$61 21 02	&5	&34.5	&7		&6	&-\\
322.2$-$00.7&	BMP1524$-$5746&	15 24 24.1&	$-$57 46 21	&1	&39.8	&7		&6	&-\\
350.4+02.0&	PHR1712$-$3543&	17 12 34.1&	$-$35 43 17	&3	&12.6	&10.1	&1	&1\\
350.8+01.7&	MPA1714$-$3535&	17 14 49.8&	$-$35 35 40	&0	&37.2	&5		&1	&-\\
344.8$-$02.6&	MPA1715$-$4303&	17 15 16.0&	$-$43 03 53	&1	&9.6		&7		&6	&-\\
355.0+02.6&	PPA1722$-$3139&	17 22 42.2&	$-$32 30 18	&	&43		&8.5		&1	&-\\
356.0+02.8&	PPA1724$-$3043&	17 24 58.3&	$-$30 43 04	&0	&5.8		&4.9		&2	&-\\
000.4+04.4&	PPA1729$-$2611&	17 29 52.4&	$-$26 11 13	&0	&11.4	&9		&1	&1\\
000.3+04.2&	PPA1730$-$2621&	17 30 12.3&	$-$26 21 01	&0	&2.3		&6		&1	&-\\
353.9+00.0&	PPA1730$-$3400&		-       &	-              		&	&$<$2.5	&4.5		&5;7	&-\\
353.6$-$02.6&	PPA1740$-$3543&	17 40 45.5&	$-$35 43 59	&2	&5.8		&9		&1	&1\\
356.6$-$01.9&	PHR1745$-$3246&	17 45 09.8&	$-$32 46 14	&3	&23.1	&43.2	&1	&1\\
000.1$-$01.7&	PHR1752$-$2941&	17 52 49.1&	$-$29 41 55	&5	&4.4		&13.9	&3	&-\\
000.3$-$01.6&	PHR1752$-$2930&	17 52 52.2&	$-$29 30 00	&1	&8.5		&8		&3	&-\\
004.0$-$00.4&	PHR1756$-$2538&		-       &	-			&	&$<$2.5	&8.4		&5;7	&-\\
010.2+02.7&	PHR1758$-$1841&	17 58 14.3&	$-$18 41 30	&3	&6.1		&8		&4	&1\\
007.2+00.0&	PPA1801$-$2238&		-       &	-			&	&$<$2.5	&4.5		&5;7	&-\\
015.5$-$00.0&	PHR1818$-$1526&		-       &	-			&	&$<$2.5	&24.6	&5;7	&-\\
026.8$-$00.1&	MPA1840$-$0529&	18 40 49.4&	$-$05 29 44	&1	&7.5		&5.9		&5;7	&-\\
028.9+00.2&	PHR1843$-$0325&	18 43 15.3&	$-$03 25 27	& 1	&19		&9.5		&5;7	&-\\
032.5$-$00.0&	PHR1850$-$0021&		-       &	-  			&	&$<$2.5	&12.8	&5;7	&-\\
032.5$-$00.3&	MPA1851$-$0028&	18 51 47.6&	$-$00 28 31	&2	&4.4		&13.4	&7	&-\\

\hline \noalign{\smallskip}
\end{tabular}
\medskip\\
\flushleft
References: {\bf (1)}: \cite{1990A&A...233..181R}, {\bf (2)}: \cite{1988A&A...205..248P}, {\bf (3)}: \cite{2001A&A...373..536V},
{\bf (4)}: \cite{1991A&AS...91..481R}, {\bf (5)}: \cite{1994ApJS...91..347B}, {\bf (6)}: \cite{2007A&A...461...11U},
{\bf (7)}: \cite{2005AJ....130..586W}.\\
\end{table*}
\end{center}

A preliminary list of possible detections was created from comparison between positions from the MASH and PMN catalogues. PMN radio sources within 60~arcsec\ from the MASH optical position have been taken into consideration. All possible detections have been examined using the total intensity maps obtained from the Australia Telescope National Facility's (ATNF's) FTP server\footnote{ftp://ftp.atnf.csiro.au/pub/data/pmn/surveys/} for the PMN survey. Finding charts, for all detected objects, are presented in Appendix~\ref{sec:notesonpne}: Figure~\ref{pmn1}. 

The low resolution and the relatively low sensitivity of the PMN survey allow detection of only nine brighter objects (considering the expected low radio-continuum brightness of MASH PNe). The positive detection flag is applied to 7 MASH PNe and the suspect detection to two MASH PN. Table~\ref{tab:PMNdetections} presents the PNG and MASH designation of the detected object, PMN J2000-based source name, RAJ2000 and DECJ2000 of the radio source, offset from the MASH optical position in arcsec, flux density as published in the original catalogue, optically determined angular diameter and any comment. Suspect detections are marked with a preceding colon in the flux density column.

\subsection{Chance coincidence estimation for NVSS and MGPS-2 radio detections}

In order to estimate the number of matches between MASH PNe positions and catalogued radio sources from NVSS and MGPS-2 that could arise purely by chance, we produced an off-source catalogue.

All MASH PNe positions were shifted by $\pm$10~arcmin in RA and DEC (4 different positions) and matched with radio positions from NVSS and MGPS-2 catalogues. Only cross-correlations within 25~arcsec (which is approximately 1/2 of the synthesised beam FWHM for both surveys) have been catalogued. The average number of chance coincidence for NVSS is 12 from 695 MASH PNe (1.5\%) and 3 from 549 MASH PNe (0.5\%) for the MGPS-2 catalogue. This result implies that all radio-continuum detections for the two analysed catalogues and within 25~arcsec, are highly likely to be real associations.

Due to the extremely small detection rate in the PMN we assumed that all positive correlations between MASH and PMN catalogue (seven out of nine) are likely to be real. On the other hand two PMN detections flagged as suspect (radio counterparts for MASH PNe PPA1758-2628 and PHR1759-2630) are very likely caused by chance coincidence.

\subsection{Other sources of MASH radio-continuum data at 6~cm}

In an attempt to compile all available radio-continuum data for MASH PNe, we examined the VizieR\footnote{http://vizier.u-strasbg.fr/viz-bin/VizieR} database of astronomical catalogues \citep{2000A&AS..143...23O} and also an extensive literature search was undertaken using the SIMBAD\footnote{http://simbad.u-strasbg.fr/simbad/} astronomical database.

\begin{figure*}
\begin{center}
\includegraphics[scale=0.8]{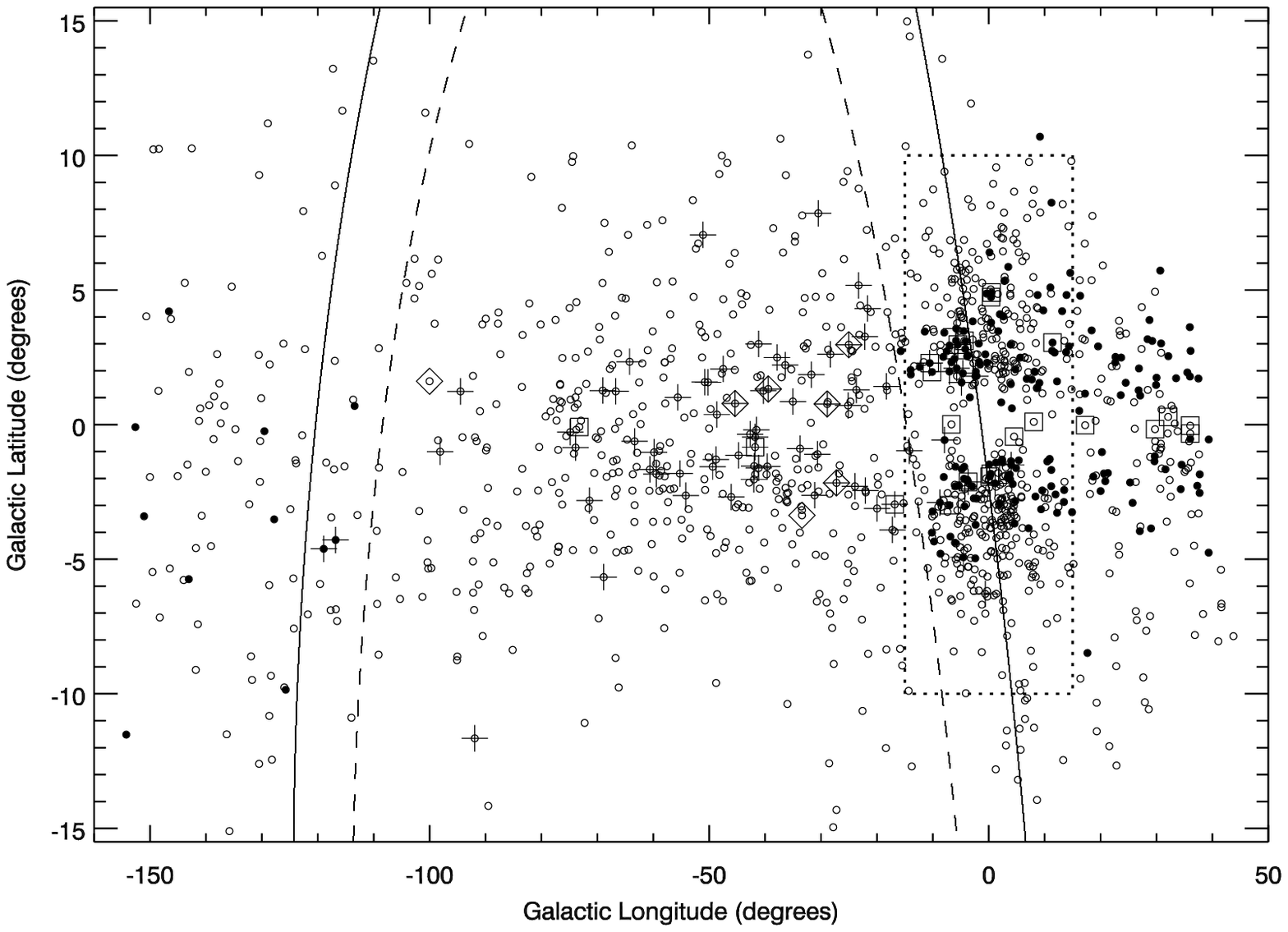}
\includegraphics[scale=0.8]{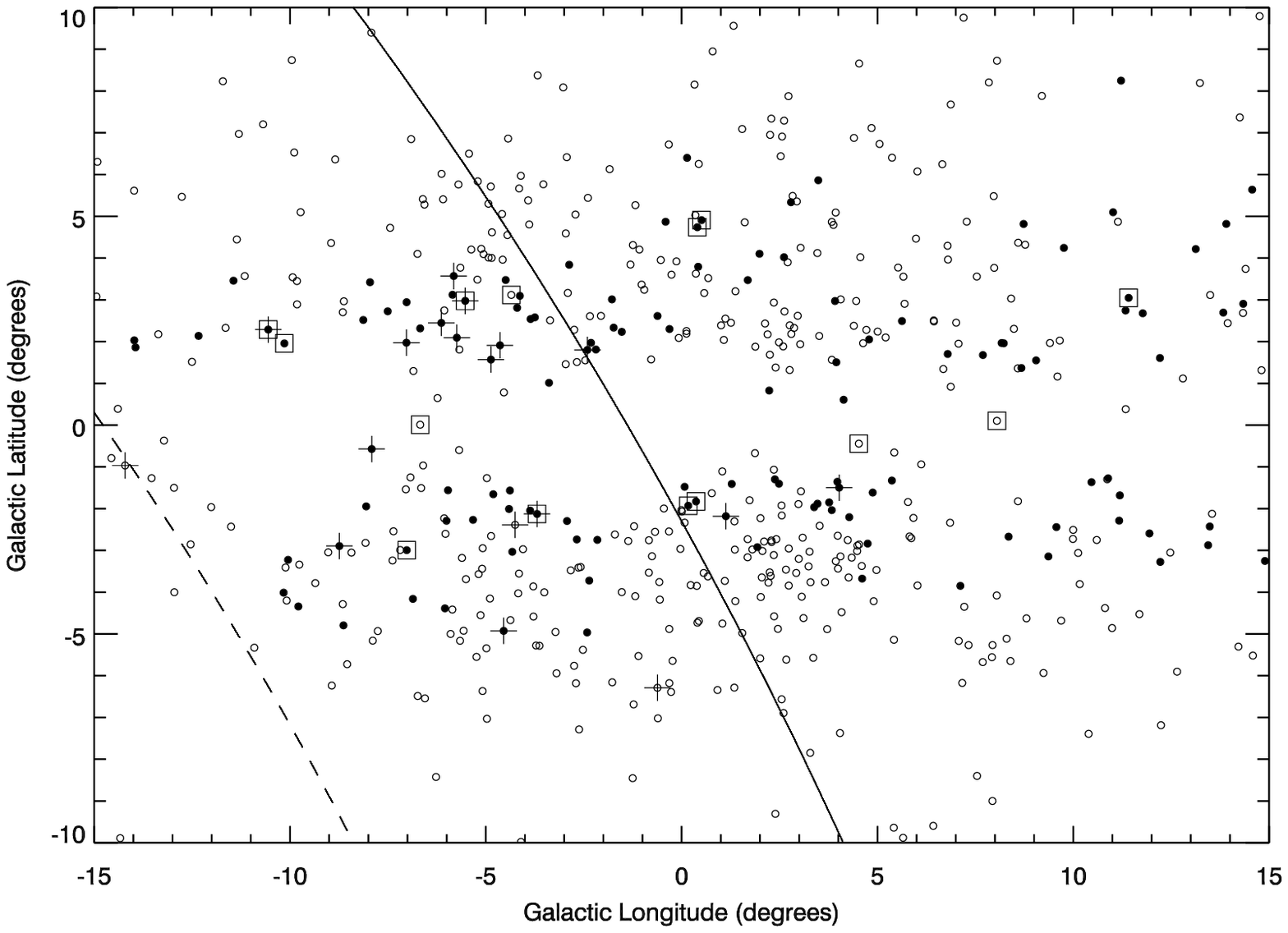}
\end{center}
\caption{Galactic distribution of PNe from the MASH catalogue (empty circles) showing the full MASH sample (top panel) and the portion of the MASH sample located in the direction of the Galactic Bulge (bottom panel). NVSS, MGPS-2 and PMN detections are marked with black filled circles, crosses and diamonds, respectively. Other radio-continuum detections are marked with boxes. Continuos and dashed lines represent DEC=$-$30$^o$ and DEC=$-$40$^o$, respectively. The zoom window, centred on the position of the Galactic centre, is represented with dotted line. Note that points are plotted on top of each other where the original, empty circles, are at the lowest layer.}\label{fig:galdistsDet}
\end{figure*}

For 17 MASH PNe we found reliable radio-continuum data at higher frequencies. In Table.~\ref{tab:othersdetection} we list positions, 6~cm flux densities and optically determined angular diameters of the detected objects.  As stated before, some (11) of these objects have been observed as PNe candidates by \cite{1988A&A...205..248P}, \cite{1990A&A...233..181R}, \cite{1991A&AS...91..481R} and \cite{2001A&A...373..536V}. We also found three objects radio detected by \cite{2007A&A...461...11U} and designated as potential massive young stellar objects (MYSOs). Finally, three objects have been detected in a ``blind'' radio-continuum survey of the Galactic plane (in the -10\degr$<l<$42\degr, $\abs{b}<$~0\degr.4 region) at 6~cm \citep{1994ApJS...91..347B,2005AJ....130..586W}. Another five compact ($\theta_{opt}\le$10~arcsec) MASH PNe are located in this region of the Galactic plane. Regarding the threshold level for detection of $\sim$2.5~mJy achieved in this survey we assigned upper limits of 2.5~mJy for the flux density for these five objects (designated with a preceding $<$ in the flux density column).

Finally, the spatial distribution of PNe from the MASH catalogue, with marked radio-detected objects, is presented in Fig.~\ref{fig:galdistsDet}. NVSS, MGPS-2 and PMN detections are marked with green, yellow and red-filled circles, respectively. Other radio-continuum detections are marked with blue-filled circles. Black filled circles represent MASH PNe not detected in the radio-continuum. The ``southern'' part of the catalogue (the sub-sample below DEC=$-$30$^o$) is under-represented due to the lack of the radio-continuum survey of similar (or better) sensitivity compared to the NVSS. While the brighter portion of MASH PNe is  well covered by the MGPS-2 survey, future observational studies (Boji\v{c}i\'c et al. in prep.) will try to improve the completeness of the radio detected MASH PNe. 

\section{Comparison with radio detected population of Galactic PNe}

The initial comparison of radio-continuum properties between previously known and new MASH PNe detected in the NVSS survey (see Fig.~\ref{fig:fluxdistsNVSS}) clearly suggested that MASH planetaries do not simply present a population of PNe detectable but missed in previous surveys but are an intrinsically faint class of radio objects. The measurable, total radio-continuum emission from PNe, mainly produced in the bremsstrahlung mechanism, principally depends on the distance to the object and the mass, density stratification and chemical abundance of the ionised material. Thus, the low radio-continuum flux density of MASH PNe should be related to their spatial and/or evolutionary properties.

In order to quantitatively examine the radio-continuum properties of MASH PNe in comparison with the previously known part of the Galactic PNe population we compiled a comprehensive database of 5~GHz radio-continuum measurements of GPNe from the literature. Prior to this study two large databases of PNe radio properties were given by \cite{1992secg.book.....A} and \cite{1992A&AS...94..399C}. Also a number of studies used and extended these databases or presented new, refined samples, e.g. \cite{1993ApJS...88..137Z, 1994A&A...289..225S, Z1995, 1995ApJ...446..279B, 2001A&A...374..599B, 2001A&A...373.1032S, PH2002, UVA09}. All these samples were biased toward more accurate measurements which naturally arise from the more radio-bright PNe. Since MASH PNe clearly represents the low end of the PNe radio brightness distribution it was very important to assemble the deepest possible set of flux densities for GPNe for comparison. 

We restricted our search to 5~GHz (6~cm) observations because at this frequency most PNe are optically thin so the observed flux reflects the intrinsic physical properties of the ionised nebula (assuming that a valid distance determination can be achieved). Also, the expected background radiation is weaker than at lower frequencies and so the possible confusion with nearby sources is less (this property is especially important for the data obtained via single-dish observations). These two assets of high frequency radio observations of PNe are in fact the major reason why the majority of PNe targeted surveys were performed at 5~GHz. We based our literature search on the \cite{2003A&A...408.1029K} catalogue of accurate positions of 1312 Galactic PNe originally listed in the Strasbourg ESO Catalogue, its supplement and  version 2000 of the Catalogue of Planetary Nebulae as well as the new 2~kpc volume limited sample of \cite{FREW2008} (hereafter F08).  All radio detections have been traced to the original observational data or to the first citation. If more than one observation was available, and results from different sources appear to be in agreement, we used a simple average between reported values. Interferometric data were preferred over the single-dish measurements, except in the case of objects with quoted large angular sizes (two or more times larger than the FWHM of a synthesised beam). Secondly, results from targeted surveys were preferred over results from ``blind'' surveys. 

The majority of flux densities in the compiled catalogue originate from the VLA surveys of \cite{ZPB89} and \cite{1990A&AS...84..229A} ($\sim50$\%) and the Parkes radio telescope surveys of \cite{1975A&A....38..183M} and \cite{MILNE1979}  ($\sim20$\%). The full catalogue of $\sim$600 Galactic PNe for which we found reliable 5~GHz flux densities will be presented in a future paper. 

\begin{table}
\begin{center}
{\footnotesize
\caption{PN impostors found in current radio catalogues.}
\medskip
\label{table:PN_mimics}
\begin{tabular}{lllll}
\hline \noalign{\smallskip}
Name	     & $\theta_{opt}$ &	S$_{6\textrm{cm}}$	&	Type		&	Ref.\\
	     & [\arcsec]& [mJy]	&	&	\\
\hline \noalign{\smallskip}
Abell 35	 &	772.0	&	255.0	&	Bowshock neb?	&	1	\\
Ns 238	 &	56.0	&	4173.0&	compact HII				&	2	\\
M 1-67	 &	120.4	&	198.0	&	Pop I WR shell		&	3	\\
PHL 932	 &	270.0	&	10.0	&	HII region				&	4	\\
Mz 3		 &	32.9	&	649.0	&	B[e] or Sy*			&	5, 6, 7\\
He 2-146	 &	34.2	&	186.0	&	compact HII			&	6\\
PP 40		 &	30.0	&	213.0	&	compact HII		&	3	\\
M 2-9		 &	20.2	&	36.0	&	B[e] or Sy*			&	5	\\
FP0840-5754	 &	339.9	&	18.4	&	HII region?		&	8\\
PHR1517-5751 &	104.2	&	54.7	&	HII region			&	8\\
\hline									
\end{tabular}}
\end{center}
{\footnotesize
References:  
1: \cite{FREW2008}; 
2: \cite{2007A&A...472..847C};  
3: \cite{2001A&A...378..843K};  
4: \cite{2010PASA...27..203F};  
5: \cite{2010PASA...27..129F};  
6: \cite{COH2010};  
7: \cite{2003ApJ...591L..37K};
8: Parker et al. (2010), in preparation.}
\end{table}

Note that a number of emission nebulae that are often classified as PN in the literature have been excluded from Fig.~\ref{fig:dbcomparison}.  In Table~\ref{table:PN_mimics} we list these objects, giving the correct classification and a reference from the literature justifying why each is a probable PN mimic (Frew \& Parker 2010).

\subsection{The $S_{\nu}-\theta$ evolutionary diagram}

\begin{figure}
\begin{center}
\includegraphics[scale=1]{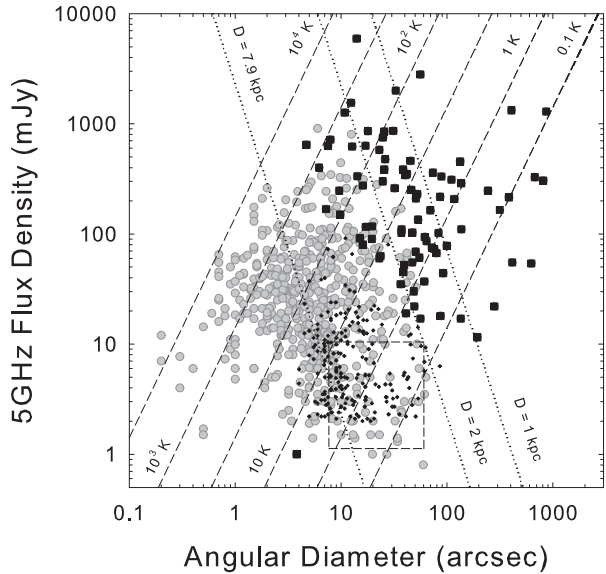}
\caption[Distribution of the catalogued sample of Galactic PNe in the $S_{\nu}-\theta$ diagram.]{Distribution of the catalogued sample of Galactic PNe in the $S_{\nu}-\theta$ diagram. Grey-filled circles and black dots represents the previously known and new MASH samples, respectively (see text for more details). Black-filled boxes represent the 2~kpc volume limited sample from F08. The discrepant point in the F08 sample plotted near the bottom is KjPn~8 (see text for details). Levels of constant brightness temperature are shown as dashed lines. Evolutionary tracks at 1, 2 and 7.9~kpc are plotted with dotted lines. The dashed-line rectangle presents the initial estimate of the region where we expect to improve completeness of the GPNe sample with our new ATCA high-frequency observations (Boji\v{c}i\'c et al. in prep.).}\label{fig:dbcomparison}
\end{center}
\end{figure}

In Fig.~\ref{fig:dbcomparison} we compare the positions of radio detected Galactic PNe in the flux density ($S_{\nu}$) versus angular diameter ($\theta$) plot. This plot essentially represents a radio-evolutionary diagram \citep{1985ApJ...290..568K,1990A&A...234..387Z} of a mixture of PNe at different distances and with a variety of intrinsic physical properties (mostly related to the mass of the progenitor star). 

The 5~GHz flux densities of $\sim$600 previously known Galactic PNe are plotted with grey-filled circles and then overplotted with the 84 radio detected PNe within the 2~kpc volume limited sample (F08; black-filled boxes). Since we found only a limited number of 5~GHz detections for MASH PNe we estimated 5~GHz flux densities from low frequency measurements (i.e. from 1.4~GHz and 0.843~GHz flux densities reported in NVSS and MGPS-2 catalogues, respectively) using equations~\ref{eq:simpspindex} and \ref{eq:alpha_tb} (see \S~\ref{sec:prelspectr}) and adopting the canonical electron temperature of 10$^{4}$~K for all MASH PNe. The adopted method accounts for the radio optical thickness effect only to some extent because it uses the same (optically determined) angular diameter in both the optically thick and optically thin regime. If a MASH PN is detected in both the NVSS and MGPS-2 surveys we used the flux density from the 1.4~GHz observation. The calculated 5~GHz flux densities versus optically determined angular diameters, for some 210 radio-continuum detected MASH PNe are plotted with black dots. 

The difference in distances can be roughly seen from comparison of PNe positions at lines of constant brightness temperature (\Tb). \Tb\ is a distance independent evolutionary property of the evolving ionised nebulosity such that the changing distance to objects can only displace the position of an object in the $S_{\nu}-\theta$ diagram along a line of constant \Tb\ \citep{1990A&A...234..387Z}. Assuming a simple, constant expansion velocity of a spherically symmetric, constant density and fully ionised nebula, the optically thin flux density $S_{\nu}$, at a distance $D$, will be related to its angular diameter $\theta$ as \citep{1982ApJ...260..612D}:

\begin{equation}\label{eq:simpevotrack}
S_{\nu}\propto\theta^{-3}D^{-5}
\end{equation}

In Fig.~\ref{fig:dbcomparison} lines of constant \Tb\ were plotted with dashed lines. The simplified evolutionary tracks, calculated from eq.~\ref{eq:simpevotrack} at 1~kpc, 2~kpc and at the mean distance of the Galactic Bulge \citep[7.9~kpc;][]{2003ApJ...597L.121E} were plotted with dotted lines.

It is important to note that the presented radio evolution of PNe is clearly a strong simplification of the real picture. It doesn't account for the optically thick phase (\Tb$>10^3$K), density and temperature gradients, and assumes an uniform expansion of a spherically symmetric and fully ionised nebular shell around the non-evolving central star (CS).  In the more realistic case the CS and nebular properties are changing mutually.  The gas density stratification in a PN is far from constant and it will continue to change during the subsequent evolution because of the energy-input from the fast wind and due to the progress of the strong ionisation front throughout the neutral gas. Thus, the optical depth at 5~GHz can be important in the initial stage of the PN evolution. Also, the episode in  CS evolution in turning to the ``cooling path'' could result in recombination in denser nebular regions and cause a drastic drop in the radio flux. However, the goal of this paper is not to provide a complete description of the radio-continuum emissivity from the evolving CS-PN system. We believe that the used simplifications do not corrupt our preliminary analysis of the evolutionary position of radio detected MASH PNe in comparison with local and general radio detected Galactic PN populations.  

From the presented radio-evolutionary diagram (Fig.~\ref{fig:dbcomparison}) a clear separation between the local volume sample (F08) and the rest of the Galactic PNe population can be seen supporting the general reliability of the \cite{FREW2008} distances. In the region above \Tb=100~K, evolutionary tracks for PNe at 1~kpc and 2~kpc start to strongly diverge from the empirical distribution mostly because, as stated before, we used approximations of negligible self-absorption and fully ionised nebulae. The highly discrepant F08 object in Figure~\ref{fig:dbcomparison} is KjPn 8 \citep{1995ApJ...455L..63L}. This very unusual object appears not to be a conventional PN \citep[see][for a discussion]{2010PASA...27..129F}. The compact, PN-like core (detected at 6~cm) is surrounded by a very large, shock-excited, bipolar nebula, which has an emission measure too low to have been detected in extant radio surveys. The small core is very underluminous at both optical and radio wavelengths at the accepted distance of 1.6 kpc \citep{1997MNRAS.292L..11M}, falling off the H$\alpha$ surface brightness -- radius relation \citep{2006IAUS..234...49F,2010PASA...27..129F}.

More importantly, we can also see a gradual increase in the number of MASH PNe below \Tb$\approx100$~K and with no MASH objects with brightness temperatures above that value. If MASH PNe follow the same central star mass distribution as the rest of the GPNe population then this implies that the MASH samples do not contain (at least not a significant number) young PNe and that radio ``faintness'' is strictly correlated to evolutionary properties. However, we cannot eliminate the possibility that some of these nebulae are related to low-mass central stars ($M_{CS}<0.6$~\msol). In that case the ejected layers of gas could already be highly dispersed at the point when the central star increases the effective temperature enough to produce a significant number of photons in the Lyman continuum. The starting point in the \Tb\ evolution of these PNe will be way below $10^3-10^4$~K as expected for ``normal'' PNe.

The expected excess of the number of low brightness PNe around 7.9~kpc can also be seen. Obviously, the large density of MASH PNe along the 7.9~kpc evolutionary track is mostly caused by sampling bias. As we mentioned before the NVSS fully covers the bulge region, where the concentration of MASH PNe is the largest, while the southern sample is only covered above 10~mJy with the MGPS-2.

\begin{figure}
\begin{center}
\includegraphics[scale=1]{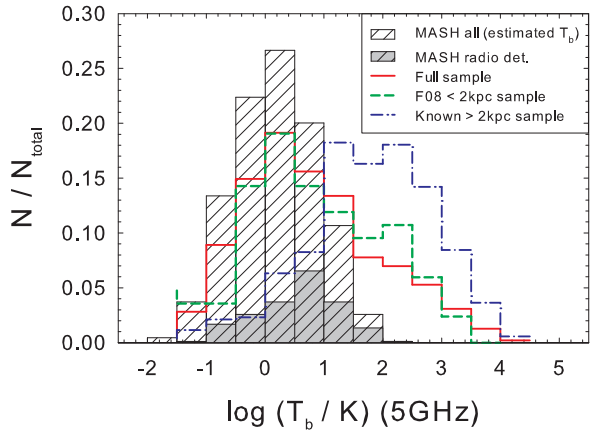}
\end{center}
\caption{The 5~GHz brightness temperature distribution of selected samples of Galactic PNe. Grey hatched and white hatched histograms represent the detected and detected + undetected (see text for more details) MASH PNe, respectively. The green dashed line represents the \Tb\ distribution of 84 radio detected PNe from the 2~kpc volume limited sample (F08), the blue dash-dotted line represents the sample of radio detected Galactic PNe excluding the 2~kpc volume limited sample and the red solid line is the distribution of the full sample (known + MASH detected and undetected). All distributions were normalised to the the total number of elements in the corresponding sample except for the radio detected MASH sample which is normalised to the total number of PNe in the detected + undetected sample. (This figure is in colour in the online version of the Journal.)}\label{fig:Tbcomparison}
\end{figure}

\subsection{The radio surface brightness distribution}

In order to strengthen our claim of intrinsic low radio luminosity of MASH PNe we also examined their position in the radio brightness temperature distribution diagram.  The radio brightness temperature, as a distance independent parameter, is a valuable evolutionary tracer especially after the shocked shell becomes fully ionised, when \Tb\ will start to evolve toward lower values due to the consequent nebular expansion. 

We calculated the radio surface brightness temperatures at 5~GHz for PNe in selected samples: previously known Galactic PNe at distances $>$2~kpc (521 objects; i.e. excluding F08 sample),  F08 sample (84 objects), positively radio detected MASH PNe (175 objects; possible detections were not included) and "radio-undetected" MASH PNe (631 objects). MASH PNe designated as ``true" and with optically determined angular diameters $\theta_{opt}<100$~arcsec were selected for this analysis. With assumption of a negligible optical depth at 1.4~GHz and 0.843~GHz we used the detection limits from NVSS and MGPS-2 to estimate 5~GHz flux densities for ``radio-undetected" MASH PNe. For objects with declinations $\delta\geq-40$\degr\ and $\delta<-40$\degr\ we used 2.5$\times$(4.8/1.4)$^{-0.1}$ mJy Beam$^{-1}$ and 10$\times$(4.8/0.843)$^{-0.1}$ mJy Beam$^{-1}$, respectively. The resulting histogram for comparison between selected samples is presented in Fig.~\ref{fig:Tbcomparison}.

The distinction between known and MASH PNe is even more evident from this diagram. No more than 12\% of PNe from the MASH catalogue were placed within the brightest three magnitudes of the $\MlogTb$ distribution. The radio detected MASH PNe sub-sample shows a peak at $\MlogTb<1$ which is the position of the significant drop in the distribution of known PNe ($>$2~kpc sample). This strongly implies that MASH PNe are the evolutionary complement of the previously known Galactic PNe. The concentration of radio-undetected MASH PNe, which is propagated to the distribution of the full sample (red solid line), around $\MlogTb\approx0$ is artificial and caused by placing of flux densities of this large set of PNe at the detection limits. Thus, it is more likely that the real distribution will have a much milder gradient or even settling to a constant level at  $\MlogTb<1.5$. 



\section{A preliminary radio-continuum spectral analysis of MASH PNe}\label{sec:prelspectr}

\begin{figure*}
\begin{center}
\includegraphics[scale=1.1]{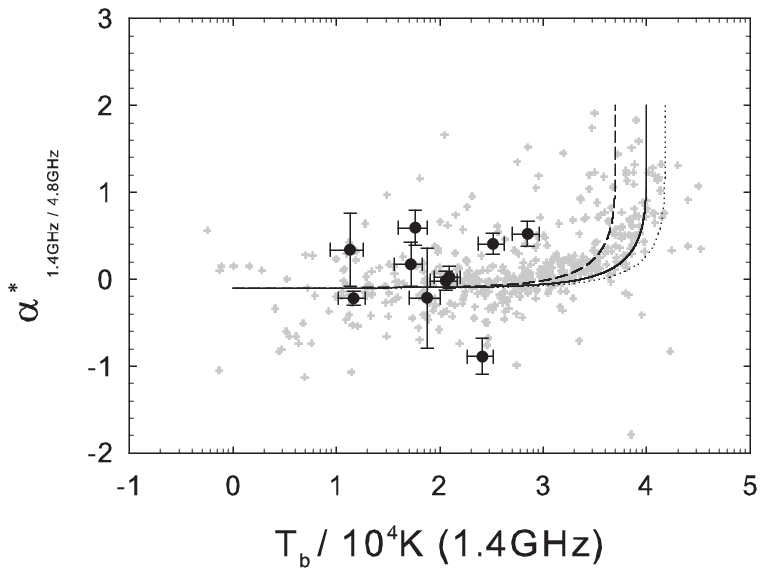}
\includegraphics[scale=1.1]{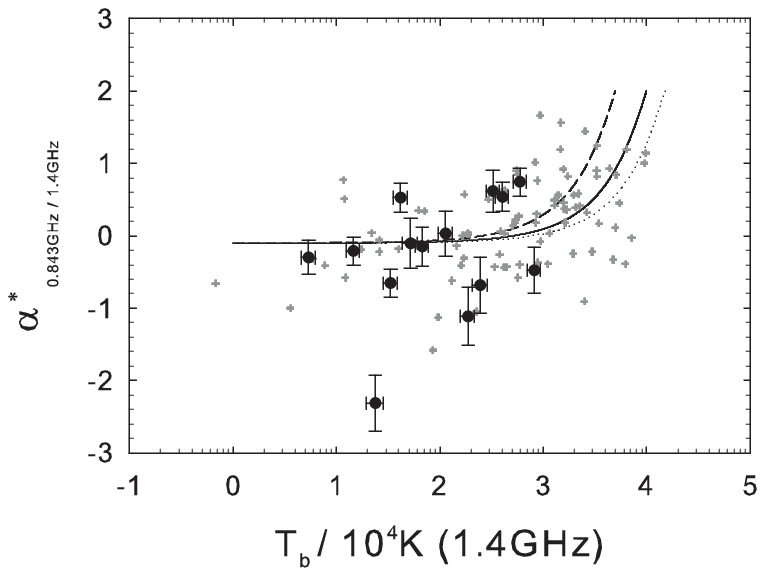}
\end{center}
\caption{A plot of $\alpha^*$ vs. \Tb\ at 1.4~GHz (black dots with error bars) values for MASH PNe with both available 1.4~GHz and 5~GHz flux densities ({\it left}) and with available 0.843~GHz and 1.4~GHz flux densities ({\it right}). Overploted are data-points from the catalogue of previously known PNe. Predictions from the model are shown with a dashed line ($T_e=5\times10^3$~K), solid line ($T_e=10\times10^3$~K) and dotted line ($T_e=15\times10^3$~K.)}\label{fig:spindex_dist_NVSSMGPS2}
\end{figure*}

Finally, using the newly collected data-set, we examined the radio-continuum spectral properties of a small sub-set of multi-wavelength radio-continuum detected MASH PNe. 

\begin{table}
\begin{footnotesize}
\begin{center}
\caption{Spectral indices of multi-wavelength radio-continuum detected MASH PNe. The averaged spectral index between frequencies $\nu_{1}$ and $\nu_{2}$ ($\alpha^*_{\nu_1/\nu_2}$) is calculated using eq.~\ref{eq:simpspindex}. Values of $\alpha^*_{\nu_1/\nu_2}$ calculated from an unreliable flux density (upper/lower limit or possible detection) are designated with a preceding colon.}\label{tab:spindices}
\begin{tabular}{crrr}

\hline \noalign{\smallskip}

Name & $\alpha^*_{0.843/1.4}$ &$\alpha^*_{0.843/5}$ &$\alpha^*_{1.4/5}$\\

\hline \noalign{\medskip}

  PHR0748$-$3258    &  0.5  $\pm$  0.2  &  -  &  -  \\  
  PHR0754$-$3444    &  -0.3  $\pm$  0.2  &  -  &  -  \\  
  PHR1457$-$5812    &  -  &  0.0  $\pm$  0.1  &  -  \\  
  BMP1524$-$5746    &  -  &  0.2  $\pm$  0.1  &  -  \\  
  PHR1529$-$5458    &  -  &  :0.2  $\pm$  0.1  &  -  \\  
  PHR1619$-$4914    &  -  &  -0.11  $\pm$  0.05  &  -  \\  
  PHR1637$-$4957    &  -  &  0.1  $\pm$  0.1  &  -  \\  
  PHR1712$-$3543    &  0.0  $\pm$  0.3  &  0.0  $\pm$  0.1  &  0.0  $\pm$  0.1  \\  
  MPA1714$-$3535    &  -  &  -  &  0.5  $\pm$  0.1  \\  
  MPA1715$-$4303    &  -  &  -0.3  $\pm$  0.1  &  -  \\  
  PHR1719$-$3134    &  :-2.6  $\pm$  0.5  &  -  &  -  \\  
  PPA1722$-$3317    &  -0.5  $\pm$  0.3  &  -  &  -  \\  
  PPA1722$-$3139    &  0.6  $\pm$  0.3  &  0.5  $\pm$  0.1  &  0.4  $\pm$  0.1  \\  
  PPA1723$-$3223    &  :-0.8  $\pm$  0.6  &  -  &  -  \\  
  PPA1725$-$3216    &  -1.1  $\pm$  0.4  &  -  &  -  \\  
  MPA1728$-$3132    &  :0.7  $\pm$  0.6  &  -  &  -  \\  
  PPA1729$-$3152    &  -0.7  $\pm$  0.4  &  -  &  -  \\  
  MPA1729$-$3513    &  0.5  $\pm$  0.2  &  -  &  -  \\  
  PPA1729$-$2611    &  -  &  -  &  0.0  $\pm$  0.1  \\  
  PPA1730$-$2621    &  -  &  -  &  -0.2  $\pm$  0.4  \\  
  PPA1734$-$2954    &  -0.1  $\pm$  0.3  &  -  &  -  \\  
  PHR1736$-$3659    &  -0.1  $\pm$  0.3  &  -  &  -  \\  
  PPA1740$-$3543    &  -  &  -  &  0.2  $\pm$  0.2  \\  
  PHR1745$-$3246    &  -0.2  $\pm$  0.2  &  -0.2  $\pm$  0.1  &  -0.2  $\pm$  0.1  \\  
  PHR1752$-$2941    &  -  &  -  &  0.3  $\pm$  0.3  \\  
  PHR1752$-$2930    &  -  &  -  &  0.6  $\pm$  0.1  \\  
  PHR1753$-$3443    &  -0.6  $\pm$  0.2  &  -  &  -  \\  
  PHR1755$-$2904    &  -2.3  $\pm$  0.4  &  -  &  -  \\  
  PHR1758$-$1841    &  -  &  -  &  -0.9  $\pm$  0.2  \\  
  PPA1758$-$2628    &  -  &  -  &  1.4  $\pm$  0.1  \\  
  PHR1759$-$2630    &  0.7  $\pm$  0.2  &  0.2  $\pm$  0.1  &  0.0  $\pm$  0.2  \\  

\hline \noalign{\smallskip}

\end{tabular}
\end{center}
\end{footnotesize}
\end{table}

The averaged spectral index
$\alpha^*_{\nu_1/\nu_2}$ between measured flux densities $S_{\nu_1}$ and $S_{\nu_2}$ at frequencies $\nu_1$ and
$\nu_2$ can be found from:

\begin{equation}\label{eq:simpspindex}
\alpha^*_{\nu_1/\nu_2}=\frac{\ln{(S_{\nu_1}/S_{\nu_2})}}{\ln{(\nu_1/\nu_2})}
\end{equation}

\noindent and it will vary between -0.1 and 2 i.e. between cases when radio-continuum emission at both frequencies is optically thin or optically thick, respectively. Table~\ref{tab:spindices} lists spectral indices for 33 MASH PNe for which we found flux densities at more than one frequency. Values of $\alpha^*_{\nu_1/\nu_2}$ calculated from an unreliable flux density (upper/lower limit or possible detection) are designated with a preceding colon. 

It is important to note that comparison of flux densities obtained with different instruments must take in to account a range of spatial frequencies which could be present in the observed source and which could be successfully measured by the used radio telescope. This is especially important  when comparing single dish and interferometric measurements of the faint and extended source \citep[comparing to the FWHM of the interferometer's synthesised beam;][]{1994A&A...289..261P}. While the spectral indices between 1.4~GHz and 0.843~GHz (NVSS and MGPS-2) were obtained from flux measurements with the matching synthesised beam size, the 5~GHz flux densities originate from single-dish and interferometric radio observations ranging in resolution from 4\arcmin.2 (PMN) to 2\arcsec (ATCA in the 6A configuration). Except for PHR1529-5358 ($\theta_{opt}=114$~arcsec), for which we already marked the 0.843~GHz flux density as suspect, the rest of the PMN detected sub-sample is comparable or much smaller than the NVSS and MGPS-2 synthesised beams. Thus, we believe that our comparison of NVSS and MGPS-2 flux densities with those from PMN is valid. However, we stress that the accuracy of spectral indices obtained from comparison of NVSS and MGPS-2 flux densities with those from interferometric observations could suffer from filtering out of flux from larger structures and should be taken with caution. 

Assuming that the observed radiation at $\nu_{1}$ and $\nu_{2}$ is coming from the same solid angle, the $S_{\nu_1}/S_{\nu_2}$ ratio can be calculated from \citep{2001A&A...373.1032S}:

\begin{equation}
\frac{S_{\nu_1}}{S_{\nu_2}}=\biggl(\frac{\nu_1}{\nu_2}\biggr)^2\frac{1-e^{-\tau_{\nu_1}}}
{1-e^{-\tau_ { \nu_2 }}}
\end{equation}

The optical depth through the ionised envelope, at frequency $\nu_1$, can be approximated with
\citep{1984ASSL..107.....P}:

\begin{equation}\label{eq:optdepth1}
\tau_{\nu_1}=8.235\times10^{-2}\biggl(\frac{T_e}{K}\biggr)^{-1.35}\biggl(\frac{\nu_1}{\textrm{GHz}}
\biggr)^{
-2.1 } \biggl(\frac{E}{cm^{-6}\cdot pc}\biggr),
\end{equation}

\noindent where $E$ is the emission measure in $cm^{-6}\cdot pc$. Thus, at frequency $\nu_2$, the
optical depth will be:

\begin{equation}
\tau_{\nu_2}=\tau_{\nu_1}(\nu_2/\nu_1)^{-2.1}
\end{equation}

With the usual adopted approximations it can be proved that:

\begin{equation}\label{eq:alpha_tb}
\alpha^*_{\nu_1/\nu_2}=2-\frac{2.1}{\ln{\gamma}}\ln{\frac{1-\xi^{\gamma}}{1-\xi}},
\end{equation}

\noindent where $\xi=1-T_e/T_b$ and $\gamma=(\nu_1/\nu_2)^{-2.1}$.
 
Figure~\ref{fig:spindex_dist_NVSSMGPS2} shows positions of these objects in  $\alpha^*_{\nu_1/\nu_2}$=$f(T_b)$  diagrams. Only data-points with positive detections and catalogued flux densities were plotted (i.e. we did not use our estimates for 0.843~GHz flux densities). Overplotted (gray crosses) are positions of previously known PNe. The expected values for $T_e=0.5\times10^4$~K, $T_e=1.0\times10^4$~K and $T_e=1.5\times10^4$~K are plotted with dashed, full
and dotted lines, respectively. 

On a first glance, comparing to the sample of previously known PNe, it appears that MASH PNe display larger scatter around the theoretical model. However, it is just as likely an effect of a small and generally fainter sample. We do not see a strong systematic effect in divergence from the theoretical curve and if plotted with the same symbols as the sample of previously known PNe the sub-sample of MASH PNe are practically indistinguishable. We examined several objects with apparent steep negative radio-continuum spectra, uncommon for PNe,  in more detail in Appendix~\ref{app:negspectra}.

\section{Summary}

In this paper we present freshly compiled and re-examined radio-continuum data for MASH PNe. In searching for radio-detections we examined three large ``blind'' radio-continuum surveys: NVSS, SUMSS/MGPS-2 and PMN. In the most sensitive survey of these three (NVSS with detection threshold level of about 2.5~mJy) we found radio counterparts for 201 MASH PNe (25\%). This number fell significantly, to 81 positive and possible detections (10\%), for the southern part of the MASH catalogue covered with MGPS-2 (with catalogue detection threshold level of about 10~mJy). The radio detection rates of MASH PNe are considerably smaller than what we see for the previously known population (of about 75\% and 50\% for NVSS and MGPS-2, respectively). 

Also, as we can see from the $S_{\nu}-\theta$ plot (Fig.~\ref{fig:dbcomparison}) and radio brightness temperature distribution (Fig.~\ref{fig:Tbcomparison}), radio detected MASH PNe are concentrated at the faint end of the current PN radio-continuum brightness distribution and appear to be the evolutionary complement of the previously known Galactic PNe. This finding posses an important question: where, in these two evolutionary diagrams, should we expect to see the rest of the MASH PNe? We believe that our deep ATCA observations (Boji\v{c}i\'c et al. in prep.) at least partially answers this question.

Finally, we examine, in some detail, radio continuum spectral properties of several MASH PNe with available multi-wavelength radio data.  Except for five objects from this sub-sample, for which we found a divergence from the expected radio-continuum spectral distribution (a steep negative radio-continuum spectra), we didn't find any strong evidence that radio-detected MASH PNe differ in radio-continuum spectral properties from their previously known ``cousins''.

\section*{ACKNOWLEDGEMENTS}

We used the {\sc miriad} software package developed by the ATNF and the Karma software package developed by Richard Gooch  \citep{1996ASPC..101...80G}. This research has made use of the SIMBAD database and VizieR catalogue access tool, operated at CDS, Strasbourg, France. This research has been supported by the International Macquarie University Research Scholarship (iMURS) and University of Western Sydney research grant (project number {\bf 20721.80758}).  The authors thank the referee for helpful comments that significantly improved this paper.


\begin{thebibliography}{}

\bibitem[\protect\citeauthoryear{Aaquist \& Kwok}{Aaquist \&
  Kwok}{1990}]{1990A&AS...84..229A}
Aaquist O.~B.,  Kwok S.,  1990, A\&AS, 84, 229

\bibitem[\protect\citeauthoryear{Acker, Marcout, Ochsenbein, Stenholm \& Tylenda}{Acker et~al.}{1992}]{1992secg.book.....A}  Acker, A., Marcout, J.,  Ochsenbein, F., Stenholm, B.,  \& Tylenda, R.\ 1992, Strasbourg-ESO Catalogue of Galactic Planetary Nebulae, Garching: European Southern Observatory

\bibitem[\protect\citeauthoryear{Becker, White \& Helfand D.~J.~and Zoonematkermani}{Becker et~al.}{1994}]{1994ApJS...91..347B} 
Becker R.~H., White R.~L., Helfand D.~J.~and Zoonematkermani S., 1994, ApJS, 91, 347

\bibitem[\protect\citeauthoryear{Bensby \& Lundstr{\"o}m}{Bensby \&
  Lundstr{\"o}m}{2001}]{2001A&A...374..599B}
Bensby T.,  Lundstr{\"o}m I.,  2001, A\&A, 374, 599

\bibitem[\protect\citeauthoryear{Bock, Large \& Sadler}{Bock et~al.}{1999}]{1999AJ....117.1578B} 
Bock D.~C.-J., Large M.~I., Sadler E.~M.,  1999, AJ, 117, 1578

\bibitem[\protect\citeauthoryear{Boumis, Paleologou, Mavromatakis \& Papamastorakis}{Boumis et~al.}{2003}]{2003MNRAS.339..735B} 
Boumis P.,  Paleologou E.~V.,  Mavromatakis F.,    Papamastorakis J.,  2003, MNRAS, 339, 735

\bibitem[\protect\citeauthoryear{Buckley \& Schneider}{Buckley \&
  Schneider}{1995}]{1995ApJ...446..279B}
Buckley D.,  Schneider S.~E.,  1995, ApJ, 446, 279

\bibitem[\protect\citeauthoryear{Cahn, Kaler \& Stanghellini}{Cahn
  et~al.}{1992}]{1992A&AS...94..399C}
Cahn J.~H.,  Kaler J.~B.,   Stanghellini L.,  1992, A\&AS, 94, 399

\bibitem[\protect\citeauthoryear{Cappellaro, Sabbadin, Benetti \& Turatto}{Cappellaro et~al.}{2001}]{2001A&A...377.1035C}
Cappellaro E.,  Sabbadin F.,  Benetti S.,    Turatto M.,  2001, A\&A, 377, 1035

\bibitem[\protect\citeauthoryear{Cohen, Parker, Green, Murphy, Miszalski, Frew, Meade, Babler, Indebetouw, Whitney, Watson, Churchwell \& Watson}{Cohen et~al.}{2007}]{2007ApJ...669..343C} 
Cohen M.,  Parker Q.~A.,  Green A.~J. et al.,  2007, ApJ, 669, 343

\bibitem[\protect\citeauthoryear{Cohen et al.}{Cohen et al.}{2010}]{COH2010} 
Cohen M., Parker Q., Green A.,  Miszalski B., Frew D.~J., Murphy T., 2010, {\bf submitted to} MNRAS.

\bibitem[\protect\citeauthoryear{Condon, Cotton, Greisen, Yin, Perley, Taylor \& Broderick}{Condon et~al.}{1998}]{1998AJ....115.1693C} 
Condon J.~J.,  Cotton W.~D.,  Greisen E.~W.,  Yin Q.~F.,  Perley R.~A.,  Taylor G.~B.,    Broderick J.~J.,  1998, AJ, 115, 1693

\bibitem[\protect\citeauthoryear{Condon \& Kaplan}{Condon \& Kaplan}{1998}]{1998ApJS..117..361C} 
Condon J.~J.,  Kaplan D.~L.,  1998, ApJS, 117, 361

\bibitem[\protect\citeauthoryear{Condon, Kaplan \& Terzian}{Condon et~al.}{1999}]{1999ApJS..123..219C}
Condon J.~J.,  Kaplan D.~L.,    Terzian Y.,  1999, ApJS, 123, 219

\bibitem[\protect\citeauthoryear{Copetti, Oliveira, Riffel, Casta{\~n}eda \& Sanmartim}{Copetti et~al.}{2007}]{2007A&A...472..847C} 
Copetti M.~V.~F.,  Oliveira V.~A.,  Riffel R.,  Casta{\~n}eda H.~O., Sanmartim D.,  2007, A\&A, 472, 847

\bibitem[\protect\citeauthoryear{Daub}{Daub}{1982}]{1982ApJ...260..612D} 
Daub C.~T.,  1982, ApJ, 260, 612

\bibitem[\protect\citeauthoryear{DePew et al.}{DePew et al.}{2010}]{DEPEW2010} 
DePew K., Parker Q., Miszalski B., De Marco O., Frew D.~J., Acker A., Kovacevic A., Sharp R., 2010, {\bf submitted to} MNRAS.


\bibitem[\protect\citeauthoryear{Eisenhauer, Sch{\"o}del, Genzel, Ott, Tecza, Abuter, Eckart \& Alexander}{Eisenhauer et~al.}{2003}]{2003ApJ...597L.121E} 
Eisenhauer F.,  Sch{\"o}del R.,  Genzel R.,  Ott T.,  Tecza M.,  Abuter R., Eckart A.,    Alexander T.,  2003, ApJ, 597, L121

\bibitem[\protect\citeauthoryear{Frew}{Frew}{2008}]{FREW2008} 
Frew D.,  2008, PhD thesis, Macquarie University

\bibitem[\protect\citeauthoryear{Frew, Madsen, O'Toole \& Parker}{Frew
  et~al.}{2010}]{2010PASA...27..203F} Frew, D.~J., Madsen, 
G.~J., O'Toole, S.~J., \& Parker, Q.~A.\ 2010, PASA, 27, 203 


\bibitem[\protect\citeauthoryear{Frew \& Parker}{Frew \&
  Parker}{2006}]{2006IAUS..234...49F}
Frew, D.~J., \& Parker, Q.~A.\ 2006, Planetary Nebulae in our Galaxy and Beyond, IAUS, 234, 49 


\bibitem[\protect\citeauthoryear{Frew \& Parker}{Frew \&
  Parker}{2010}]{2010PASA...27..129F}
Frew D.~J.,  Parker Q.~A.,  2010, PASA, 27, 129

\bibitem[\protect\citeauthoryear{Gooch}{Gooch}{1996}]{1996ASPC..101...80G}
Gooch, R.\ 1996, Astronomical 
Data Analysis Software and Systems V, 101, 80 

\bibitem[\protect\citeauthoryear{Green, Cram, Large \& Ye}{Green et~al.}{1999}]{1999ApJS..122..207G} 
Green A.~J.,  Cram L.~E.,  Large M.~I.,    Ye T.,  1999, ApJS, 122, 207

\bibitem[\protect\citeauthoryear{Gregory, Vavasour, Scott \& Condon}{Gregory et~al.}{1994}]{1994ApJS...90..173G} 
Gregory P.~C.,  Vavasour J.~D.,  Scott W.~K.,    Condon J.~J.,  1994, ApJS, 90, 173

\bibitem[\protect\citeauthoryear{Griffith, Wright, Burke \& Ekers}{Griffith et~al.}{1994}]{1994ApJS...90..179G} 
Griffith M.~R.,  Wright A.~E.,  Burke B.~F.,    Ekers R.~D.,  1994, ApJS, 90, 179

\bibitem[\protect\citeauthoryear{Kastner, Balick, Blackman, Frank, Soker,
  Vrt{\'{\i}}lek \& Li}{Kastner et~al.}{2003}]{2003ApJ...591L..37K}
Kastner J.~H.,  Balick B.,  Blackman E.~G.,  Frank A.,  Soker N.,
  Vrt{\'{\i}}lek S.~D.,    Li J.,  2003, ApJ, 591, L37

\bibitem[\protect\citeauthoryear{Kerber, Mignani, Guglielmetti \& Wicenec}{Kerber et~al.}{2003}]{2003A&A...408.1029K} 
Kerber F.,  Mignani R.~P.,  Guglielmetti F.,    Wicenec A.,  2003, A\&A, 408, 1029

\bibitem[\protect\citeauthoryear{Kohoutek}{Kohoutek}{2001}]{2001A&A...378..843K} 
Kohoutek L.,  2001, A\&A, 378, 843

\bibitem[\protect\citeauthoryear{Kohoutek}{Kohoutek}{2002}]{2002AN....323...57K} 
Kohoutek L.,  2002, Astronomische Nachrichten, 323, 57

\bibitem[\protect\citeauthoryear{Kovacevic \& Parker}{Kovacevic \& Parker}{2009}]{2009ASPC..404..337K} 
Kovacevic, A., \& Parker, Q.\ 2009, Astronomical Society of the Pacific Conference Series, 404, 337 

\bibitem[\protect\citeauthoryear{Kovacevic et al.}{Kovacevic et al.}{2010}]{KOVAC2010} 
Kovacevic A., Parker Q., Jacoby G., Sharp R., Miszalski B.,  Frew D.~J., 2010, {\bf submitted to} MNRAS.

\bibitem[\protect\citeauthoryear{Kwok}{Kwok}{1985}]{1985ApJ...290..568K} 
Kwok S.,  1985, ApJ, 290, 568

\bibitem[\protect\citeauthoryear{Lopez, Vazquez \& Rodriguez}{Lopez et~al.}{1995}]{1995ApJ...455L..63L} 
Lopez J.~A.,  Vazquez R.,    Rodriguez L.~F.,  1995, ApJ, 455, L63

\bibitem[\protect\citeauthoryear{Luo, Condon \& Yin}{Luo et~al.}{2005}]{2005ApJS..159..282L} 
Luo S.~G.,  Condon J.~J.,    Yin Q.~F.,  2005, ApJS, 159, 282

\bibitem[\protect\citeauthoryear{Mauch, Murphy, Buttery, Curran, Hunstead, Piestrzynski, Robertson \& Sadler}{Mauch et~al.}{2003}]{2003MNRAS.342.1117M} 
Mauch T.,  Murphy T.,  Buttery H.~J.,  Curran J.,  Hunstead R.~W., Piestrzynski B.,  Robertson J.~G.,    Sadler E.~M.,  2003, MNRAS, 342, 1117

\bibitem[\protect\citeauthoryear{Meaburn}{Meaburn}{1997}]{1997MNRAS.292L..11M} 
Meaburn J.,  1997, MNRAS, 292, L11

\bibitem[\protect\citeauthoryear{Milne}{Milne}{1979}]{MILNE1979}
Milne D.~K.,  1979, A\&AS, 36, 227

\bibitem[\protect\citeauthoryear{Milne \& Aller}{Milne \&
  Aller}{1975}]{1975A&A....38..183M}
Milne D.~K.,  Aller L.~H.,  1975, A\&A, 38, 183

\bibitem[\protect\citeauthoryear{Miszalski, Acker, Moffat, Parker \& Udalski}{Miszalski et~al.}{2009a}]{2009A&A...496..813M} Miszalski B.,  Acker A.,  Moffat A.~F.~J.,  Parker Q.~A., Udalski A., 2009a, A\&A, 496, 813

\bibitem[\protect\citeauthoryear{Miszalski, Acker \& Parker}{Miszalski et~al.}{2008}]{2008ASPC..391..181M} 
Miszalski, B., Acker,  A., \& Parker, Q.~A.\ 2008, Hydrogen-Deficient Stars , 391, 181 



\bibitem[\protect\citeauthoryear{Miszalski, Acker, Parker \& Moffat}{Miszalski et~al.}{2009b}]{2009A&A...505..249M} 
Miszalski B.,  Acker A.,  Parker Q.~A.,    Moffat A.~F.~J.,  2009b, A\&A, 505, 249

\bibitem[\protect\citeauthoryear{Miszalski, Parker, Acker, Birkby, Frew \& Kovacevic}{Miszalski et~al.}{2008}]{2008MNRAS.384..525M} 
Miszalski B.,  Parker Q.~A.,  Acker A.,  Birkby J.~L.,  Frew D.~J., Kovacevic A.,  2008, MNRAS, 384, 525

\bibitem[\protect\citeauthoryear{Morgan, Parker \& Cohen}{Morgan et~al.}{2003}]{2003MNRAS.346..719M} 
Morgan D.~H.,  Parker Q.~A.,    Cohen M.,  2003, MNRAS, 346, 719

\bibitem[\protect\citeauthoryear{Murphy, Mauch, Green, Hunstead R.~W.~andPiestrzynska, Kels \& Sztajer}{Murphy et~al.}{2007}]{2007MNRAS.382..382M} 
Murphy T.,  Mauch T.,  Green A.,  Hunstead R.~W.~and Piestrzynska B.,  Kels A.~P.,    Sztajer P.,  2007, MNRAS, 382, 382

\bibitem[\protect\citeauthoryear{Ochsenbein, Bauer \& Marcout}{Ochsenbein et~al.}{2000}]{2000A&AS..143...23O} 
Ochsenbein F.,  Bauer P.,    Marcout J.,  2000, A\&AS, 143, 23

\bibitem[\protect\citeauthoryear{Parker, Acker, Frew, Hartley, Peyaud, Ochsenbein, Phillipps, Russeil, Beaulieu, Cohen, K{\"o}ppen, Miszalski, Morgan, Morris, Pierce \& Vaughan}{Parker et~al.}{2006}]{2006MNRAS.373...79P} 
Parker Q.~A.,  Acker A.,  Frew D.~J.  et al.,  2006, MNRAS, 373, 79

\bibitem[\protect\citeauthoryear{Parker, Phillipps, Pierce, Hartley, Hambly, Read, MacGillivray, Tritton, Cass, Cannon, Cohen, Drew, Frew, Hopewell, Mader, Malin, Masheder, Morgan, Morris, Russeil, Russell \& Walker}{Parker et~al.}{2005}]{2005MNRAS.362..689P} 
Parker Q.~A.,  Phillipps S.,  Pierce M.~J. et al.,  2005, MNRAS, 362, 689

\bibitem[\protect\citeauthoryear{Phillips}{Phillips}{2002}]{PH2002}
Phillips J.~P.,  2002, ApJS, 139, 199

\bibitem[\protect\citeauthoryear{Pottasch}{Pottasch}{1984}]{1984ASSL..107.....P} 
Pottasch, S.~R.\ 1984,  Astrophysics and Space Science Library, 107

\bibitem[\protect\citeauthoryear{Pottasch, Olling, Bignell \& Zijlstra}{Pottasch et~al.}{1988}]{1988A&A...205..248P}
Pottasch S.~R.,  Olling R.,  Bignell C.,    Zijlstra A.~A.,  1988, A\&A, 205, 248

\bibitem[\protect\citeauthoryear{Pottasch \& Zijlstra}{Pottasch \& Zijlstra}{1994}]{1994A&A...289..261P} 
Pottasch S.~R.,  Zijlstra A.~A.,  1994, A\&A, 289, 261

\bibitem[\protect\citeauthoryear{Preite-Martinez}{Preite-Martinez}{1988}]{1988A&AS...76..317P} 
Preite-Martinez A.,  1988, A\&AS, 76, 317

\bibitem[\protect\citeauthoryear{Ratag \& Pottasch}{Ratag \& Pottasch}{1991}]{1991A&AS...91..481R} 
Ratag M.~A.,  Pottasch S.~R.,  1991, A\&AS, 91, 481

\bibitem[\protect\citeauthoryear{Ratag, Pottasch, Zijlstra \& Menzies}{Ratag et~al.}{1990}]{1990A&A...233..181R} 
Ratag M.~A.,  Pottasch S.~R.,  Zijlstra A.~A.,    Menzies J.,  1990, A\&A, 233, 181

\bibitem[\protect\citeauthoryear{Si{\'o}dmiak \& Tylenda}{Si{\'o}dmiak \& Tylenda}{2001}]{2001A&A...373.1032S} 
Si{\'o}dmiak N.,  Tylenda R.,  2001, A\&A, 373, 1032

\bibitem[\protect\citeauthoryear{Stasinka \& Tylenda}{Stasinka \&
  Tylenda}{1994}]{1994A&A...289..225S}
Stasinka G.,  Tylenda R.,  1994, A\&A, 289, 225

\bibitem[\protect\citeauthoryear{Umana, Leto, Trigilio, Buemi, Manzitto, Toscano, Dolei \& Cerrigone}{Umana et~al.}{2008}]{2008A&A...482..529U} 
Umana G.,  Leto P.,  Trigilio C.,  Buemi C.~S.,  Manzitto P.,  Toscano S., Dolei S.,    Cerrigone L.,  2008, A\&A, 482, 529

\bibitem[\protect\citeauthoryear{Uro{\v s}evi{\'c}, Vukoti{\'c}, Arbutina, Ili{\'c}, Filipovi{\'c}, Boji{\v c}i{\'c}, Segan \& Vidojevi{\'c}}{Uro{\v s}evi{\'c} et~al.}{2009}]{UVA09}
Uro{\v s}evi{\'c}, D., Vukoti{\'c}, B., Arbutina, B., Ili{\'c}, D., Filipovi{\'c}, M., Boji{\v c}i{\'c}, I., Segan, S., \& Vidojevi{\'c}, S.\ 2009, A\&A, 495, 537 

\bibitem[\protect\citeauthoryear{Urquhart, Busfield, Hoare, Lumsden, Clarke, Moore, Mottram \& Oudmaijer}{Urquhart et~al.}{2007}]{2007A&A...461...11U} 
Urquhart J.~S.,  Busfield A.~L.,  Hoare M.~G.,  Lumsden S.~L.,  Clarke A.~J., Moore T.~J.~T.,  Mottram J.~C.,    Oudmaijer R.~D.,  2007, A\&A, 461, 11

\bibitem[\protect\citeauthoryear{Van~de Steene \& Jacoby}{Van~de Steene \& Jacoby}{2001}]{2001A&A...373..536V} 
Van~de Steene G.~C.,  Jacoby G.~H.,  2001, A\&A, 373, 536

\bibitem[\protect\citeauthoryear{Viironen, Greimel, Corradi, Mampaso, Rodriguez, Sabin, Delgado-Inglada, Drew, Giammanco, Gonzalez-Solares, Irwin, Miszalski, Parker, Rodriguez-Flores \& Zijlstra}{Viironen et~al.}{2009}]{2009A&A...504..291V} 
Viironen K.,  Greimel R.,  Corradi R.~L.~M. et al.,  2009, A\&A, 504, 291

\bibitem[\protect\citeauthoryear{White, Becker \& Helfand}{White et~al.}{2005}]{2005AJ....130..586W} 
White R.~L.,  Becker R.~H.,    Helfand D.~J.,  2005, AJ, 130, 586

\bibitem[\protect\citeauthoryear{Whiteoak}{Whiteoak}{1992}]{1992MNRAS.256..121W} 
Whiteoak J.~B.~Z.,  1992, MNRAS, 256, 121

\bibitem[\protect\citeauthoryear{Wright, Griffith, Burke \& Ekers}{Wright et~al.}{1994}]{1994ApJS...91..111W} 
Wright A.~E.,  Griffith M.~R.,  Burke B.~F.,    Ekers R.~D.,  1994, ApJS, 91, 111

\bibitem[\protect\citeauthoryear{Wright, Griffith, Burke \& Ekers}{Wright et~al.}{1996}]{1996yCat.8038....0W} 
Wright A.~E.,  Griffith M.~R.,  Burke B.~F.,    Ekers R.~D.,  1996, VizieR Online Data Catalog, 8038, 0

\bibitem[\protect\citeauthoryear{Zhang}{Zhang}{1995}]{Z1995}
Zhang C.~Y.,  1995, ApJS, 98, 659

\bibitem[\protect\citeauthoryear{Zhang \& Kwok}{Zhang \&
  Kwok}{1993}]{1993ApJS...88..137Z}
Zhang C.~Y.,  Kwok S.,  1993, ApJS, 88, 137

\bibitem[\protect\citeauthoryear{Zijlstra}{Zijlstra}{1990}]{1990A&A...234..387Z} 
Zijlstra A.~A.,  1990, A\&A, 234, 387

\bibitem[\protect\citeauthoryear{Zijlstra, Pottasch \& Bignell}{Zijlstra
  et~al.}{1989}]{ZPB89}
Zijlstra A.~A.,  Pottasch S.~R.,    Bignell C.,  1989, A\&AS, 79, 329

\end{thebibliography}

\onecolumn

\newpage

\begin{center}
{\scriptsize
\begin{longtable}{cccccrrrr}
\caption{MASH PNe detected in the NVSS.}\label{tab:nvssdetection}\\

\hline \noalign{\smallskip}

PNG & MASH Name & NVSS source & RAJ2000 & DEJ2000 & Offset  & S$_{1.4\textrm{GHz}}$ & $\theta_{opt}$& c.\\
\noalign{\smallskip}
& & &  & & [\arcsec] &[mJy]&[\arcsec]&\\
\hline \noalign{\smallskip}

\endfirsthead

\multicolumn{8}{c}{{\tablename} \thetable{} -- Continued} \\[0.5ex]

\hline \noalign{\smallskip}

PNG & MASH Name & NVSS source & RAJ2000 & DEJ2000 & Offset  & S$_{1.4\textrm{GHz}}$ & $\theta_{opt}$& c.\\
\noalign{\smallskip}
& & &  & & [\arcsec] &[mJy]&[\arcsec]&\\
\hline \noalign{\smallskip}
\endhead

\hline \noalign{\smallskip}
\endfoot

\hline \noalign{\smallskip}
\endlastfoot
209.1$-$08.2&PHR0615$-$0025&061521$-$002544&06 15 21.25&$-$00 25 44.9&13.4&21.1$\pm$0.8	&100		&1\\
214.2$-$02.4&PHR0645$-$0217&064504$-$021739&06 45 04.09&$-$02 17 39.8&15.1&2.5$\pm$0.5	&50.5	&1\\
212.6$-$00.0&PHR0650+0013&065040+001341&06 50 40.45&+00 13 41.6&1.8&5.4$\pm$0.5			&42		&-\\
222.8$-$04.2&PHR0654$-$1045&065413$-$104533&06 54 13.26&$-$10 45 33.8&4.7&2.7$\pm$0.5	&20.8	&-\\
219.1+03.0&MPA0713$-$0405&071347$-$040521&07 13 47.93&$-$04 05 21.3&13.3&10.5$\pm$0.5	&62.8	&-\\
\noalign{\smallskip}

240.6$-$07.7&BMP0715$-$2805&071502$-$280541&07 15 02.35&$-$28 05 41.5&2.6&2.3$\pm$0.5	&19		&-\\
239.3$-$02.7&PHR0731$-$2439&073200$-$243916&07 32 00.35&$-$24 39 16.4&16.1&4.0$\pm$0.7	&19		&1\\
222.5+07.6&BMP0736$-$0500&073623$-$050001&07 36 23.02&$-$05 00 01.9&18.0&:2.7			&69.6	&20\\
237.6$-$00.1&BMP0738$-$2155&073806$-$215529&07 38 06.36&$-$21 55 29.4&2.5&3.1$\pm$0.5	&10		&-\\
247.5$-$04.7&PHR0742$-$3247&074219$-$324713&07 42 19.84&$-$32 47 13.9&56.7&:3.8		&166.2	&21\\
\noalign{\smallskip}

248.3$-$03.6&PHR0748$-$3258&074832$-$325821&07 48 32.08&$-$32 58 21.3&5.3&23.5$\pm$0.9	&22.5	&-\\
250.5$-$03.4&PHR0754$-$3444&075455$-$344408&07 54 55.83&$-$34 44 08.5&1.7&21.2$\pm$1.6	&59.5	&-\\
254.5$-$02.7&PHR0808$-$3745&&&&-&:2.0												&173.9	&22\\
254.1+00.5&BMP0820$-$3536&082050$-$353554&08 20 50.76&$-$35 35 54.0&8.3&3.3$\pm$0.6		&11		&-\\
345.8+02.4&MPA1656$-$3912&165640$-$391226&16 56 40.80&$-$39 12 26.3&13.4&4.9$\pm$0.8		&4		&-\\
\noalign{\smallskip}

347.4+01.8&PHR1704$-$3819&170416$-$381957&17 04 16.89&$-$38 19 57.8&3.0&4.6$\pm$0.6		&22.7	&-\\
347.4+01.6&PPA1704$-$3824&170459$-$382412&17 04 59.32&$-$38 24 12.6&9.0&5.6$\pm$0.6		&7.7		&-\\
349.6+03.1&PHR1706$-$3544&170601$-$354437&17 06 01.23&$-$35 44 37.8&8.2&3.6$\pm$0.5		&53.5	&-\\
348.8+01.9&PPA1708$-$3705&170823$-$370544&17 08 23.38&$-$37 05 44.4&8.4&3.4$\pm$0.5		&5.7		&-\\
350.4+02.0&PHR1712$-$3543&171234$-$354317&17 12 34.02&$-$35 43 17.6&2.4&12.9$\pm$0.6		&10.1	&1\\
\noalign{\smallskip}

350.8+01.7&MPA1714$-$3535&171449$-$353541&17 14 49.58&$-$35 35 41.4&3.0&19.6$\pm$0.8		&5		&-\\
352.8+03.0&PPA1715$-$3313&171510$-$331351&17 15 10.65&$-$33 13 51.6&4.8&2.5$\pm$0.5		&6.5		&-\\
352.6+02.2&PPA1717$-$3349&171756$-$334926&17 17 56.28&$-$33 49 26.0&1.4&3.7$\pm$0.5		&7.5		&10\\
353.2+02.4&PPA1718$-$3315&171844$-$331526&17 18 44.54&$-$33 15 26.0&4.9&7.5$\pm$0.6		&8		&-\\
353.6+02.6&MPA1719$-$3247&171910$-$324718&17 19 10.83&$-$32 47 18.4&9.2&3.3$\pm$0.5		&7.5		&-\\
\noalign{\smallskip}

354.7+03.2&PHR1719$-$3134&171954$-$313443&17 19 54.53&$-$31 34 43.8&7.1&3.9$\pm$0.5		&6.7		&-\\
354.7+02.8&PPA1721$-$3149&172123$-$314952&17 21 23.89&$-$31 49 52.3&2.4&5.9$\pm$0.6		&7.9		&-\\
353.9+02.0&PHR1722$-$3251&172213$-$325116&17 22 13.09&$-$32 51 16.7&9.6&6.1$\pm$0.6		&28.8	&-\\
353.6+01.7&PPA1722$-$3317&172235$-$331714&17 22 35.43&$-$33 17 14.6&0.6&14.6$\pm$0.7		&4		&-\\
355.0+02.6&PPA1722$-$3139&172240$-$313956&17 22 40.85&$-$31 39 56.9&2.0&26.1$\pm$0.9		&8.5		&-\\
\noalign{\smallskip}

354.4+02.2&PPA1723$-$3223&172304$-$322309&17 23 04.22&$-$32 23 09.4&0.7&7.1$\pm$0.6		&9.4		&-\\
355.9+03.1&PPA1723$-$3038&172323$-$303840&17 23 23.47&$-$30 38 40.3&0.5&10.1$\pm$0.6		&4		&-\\
000.1+05.7&PHR1724$-$2543&172404$-$254328&17 24 04.47&$-$25 43 28.4&14.6&3.9$\pm$0.6		&10.5	&-\\
354.8+01.8&PPA1725$-$3216&172515$-$321608&17 25 15.80&$-$32 16 08.9&2.5&10.0$\pm$0.6		&6.9		&-\\
356.2+02.7&MPA1725$-$3033&172532$-$303410&17 25 32.33&$-$30 34 10.8&19.5&2.8$\pm$0.5		&5		&-\\
\noalign{\smallskip}

357.4+03.4&PPA1725$-$2915&172554$-$291510&17 25 54.45&$-$29 15 10.3&2.4&2.4$\pm$0.5		&9		&-\\
356.2+02.5&PPA1726$-$3045&172623$-$304536&17 26 23.61&$-$30 45 36.8&2.2&12.7$\pm$1.2		&4.5		&-\\
359.6+04.3&PPA1727$-$2653&172758$-$265351&17 27 58.14&$-$26 53 51.1&7.0&2.6$\pm$0.6		&7.5		&-\\
356.5+02.2&PHR1728$-$3038&172807$-$303821&17 28 07.73&$-$30 38 21.3&3.3&4.6$\pm$0.5		&16.9	&-\\
356.6+02.3&PHR1728$-$3032&172814$-$303213&17 28 14.61&$-$30 32 13.2&5.4&2.3$\pm$0.5		&9.8		&-\\
\noalign{\smallskip}

355.8+01.7&MPA1728$-$3132&172831$-$313214&17 28 31.30&$-$31 32 14.6&6.2&14.4$\pm$1.3		&4		&12\\
355.6+01.4&PPA1729$-$3152&172911$-$315242&17 29 11.16&$-$31 52 42.6&3.2&8.2$\pm$0.6		&5.5		&-\\
008.3+09.6&PHR1729$-$1647&172912$-$164745&17 29 12.73&$-$16 47 45.7&9.0&2.4$\pm$0.5		&20.5	&-\\
352.8$-$00.5&MPA1729$-$3513&172937$-$351344&17 29 37.64&$-$35 13 44.1&2.6&72.8$\pm$2.3	&12.7	&-\\
000.4+04.4&PPA1729$-$2611&172952$-$261113&17 29 52.37&$-$26 11 13.6&0.7&11.1$\pm$0.6		&9		&1\\
\noalign{\smallskip}

000.3+04.2&PPA1730$-$2621&173012$-$262054&17 30 12.80&$-$26 20 54.5&9.3&3.0$\pm$0.5		&6		&1\\
358.4+02.7&PHR1731$-$2850&173114$-$285042&17 31 14.16&$-$28 50 42.6&5.9&3.5$\pm$0.6		&17		&-\\
003.1+05.2&PHR1733$-$2327&173307$-$232803&17 33 07.39&$-$23 28 03.2&4.6&3.4$\pm$0.5		&13.4	&-\\
002.5+04.8&PPA1733$-$2415&173322$-$241440&17 33 22.77&$-$24 14 40.8&24.2&4.6$\pm$0.5		&16.9	&-\\
000.3+03.4&PHR1733$-$2647&173327$-$264758&17 33 27.18&$-$26 47 58.6&1.1&3.8$\pm$0.6		&14.5	&-\\
\noalign{\smallskip}

357.9+01.7&PPA1733$-$2945&173338$-$294531&17 33 38.28&$-$29 45 31.1&1.9&4.2$\pm$0.5		&7		&-\\
358.4+02.1&PPA1733$-$2908&173340$-$290830&17 33 40.22&$-$29 08 30.4&5.1&2.7$\pm$0.5		&9.5		&-\\
357.8+01.6&PPA1734$-$2954&173401$-$295433&17 34 01.99&$-$29 54 33.6&4.0&12.8$\pm$0.6		&13		&-\\
006.2+06.9&PHR1734$-$2000&&&&0.0&:2.0												&19.5	&-\\
350.9$-$02.9&PHR1734$-$3809&173420$-$380859&17 34 20.24&$-$38 08 59.4&7.2&7.8$\pm$1.3	&61.1	&-\\
\noalign{\smallskip}

352.7$-$01.7&PPA1734$-$3600&173422$-$360045&17 34 22.20&$-$36 00 45.0&3.2&8.0$\pm$0.6		&9		&-\\
358.0+01.6&PHR1734$-$2944&173429$-$294432&17 34 29.13&$-$29 44 32.3&3.7&2.4$\pm$0.5		&13.5	&-\\
358.6+02.0&PHR1734$-$2902&173429$-$290227&17 34 29.32&$-$29 02 27.1&23.2&16.1$\pm$2.8	&53.1	&1\\
356.9+00.9&PPA1734$-$3102&173434$-$310209&17 34 34.56&$-$31 02 09.6&3.7&12.5$\pm$0.6		&5		&-\\
357.7+01.4&PPA1734$-$3004&&&&0.0&:2.0												&7.7		&-\\
\noalign{\smallskip}

359.4+02.3a&PPA1735$-$2809&173512$-$280933&17 35 12.30&$-$28 09 33.2&4.5&2.9$\pm$0.5		&7.9		&-\\
001.7+03.6&PHR1735$-$2527&173547$-$252739&17 35 47.34&$-$25 27 39.1&4.0&3.9$\pm$0.6		&7	&-\\
352.1$-$02.6&PHR1736$-$3659&173618$-$365950&17 36 18.89&$-$36 59 50.6&19.4&14.6$\pm$1.2	&15.9	&1\\
003.5+04.5&PHR1736$-$2330&&&&0.0&:2.0												&11.4	&-\\
359.7+02.0&PPA1736$-$2804&173657$-$280441&17 36 57.09&$-$28 04 41.0&4.0&10.9$\pm$0.6		&5.7		&-\\
\noalign{\smallskip}

350.8$-$03.6&MPA1737$-$3837&173706$-$383721&17 37 06.13&$-$38 37 21.8&11.9&4.4$\pm$0.7	&8		&-\\
001.5+03.1&PHR1737$-$2559&173716$-$255933&17 37 16.35&$-$25 59 33.7&4.8&3.5$\pm$0.5		&16.9	&-\\
002.3+03.6&PPA1737$-$2501&173724$-$250137&17 37 24.12&$-$25 01 37.7&5.9&5.6$\pm$0.6		&7		&-\\
354.6$-$01.4&PPA1737$-$3414&173753$-$341424&17 37 53.82&$-$34 14 24.3&2.9&7.2$\pm$0.6		&6		&11\\
351.1$-$03.9&PHR1739$-$3829&173917$-$382929&17 39 17.20&$-$38 29 29.8&14.2&:11.2			&45.3	&1;27\\
\noalign{\smallskip}

354.5$-$02.0a&PPA1740$-$3437&174030$-$343713&17 40 30.17&$-$34 37 13.3&5.5&3.4$\pm$0.6	&7		&-\\
353.6$-$02.6&PPA1740$-$3543&174045$-$354359&17 40 45.94&$-$35 43 59.5&4.4&4.7$\pm$0.5		&9		&1\\
355.6$-$01.4&PHR1740$-$3324&174054$-$332418&17 40 54.75&$-$33 24 18.4&1.9&5.5$\pm$0.6	&9		&-\\
010.1+07.4&PHR1741$-$1624&174104$-$162450&17 41 04.87&$-$16 24 50.2&12.9&3.5$\pm$0.5		&12.7	&-\\
356.0$-$01.4&PPA1741$-$3302&174133$-$330208&17 41 33.52&$-$33 02 08.6&6.6&4.8$\pm$0.5		&7.5		&-\\
\noalign{\smallskip}

355.2$-$02.0&PPA1741$-$3405&174159$-$340549&17 41 59.09&$-$34 05 49.7&15.7&3.5$\pm$0.6	&6.5		&1\\
356.0$-$01.8&PPA1743$-$3315&174307$-$331554&17 43 07.44&$-$33 15 54.9&3.1&5.8$\pm$0.5		&5		&-\\
003.5+02.6&PHR1743$-$2431&174338$-$243158&17 43 38.88&$-$24 31 58.2&7.7&4.3$\pm$0.6		&30.4	&-\\
352.2$-$04.3&PHR1743$-$3749&174348$-$375000&17 43 48.57&$-$37 50 00.0&12.1&2.6$\pm$0.6	&35.5	&-\\
356.5$-$01.8&PPA1744$-$3252&174427$-$325211&17 44 27.77&$-$32 52 11.8&3.0&6.1$\pm$0.6		&6		&13\\
\noalign{\smallskip}

356.6$-$01.9&PHR1745$-$3246&174509$-$324616&17 45 09.55&$-$32 46 16.5&3.2&30.3$\pm$1.6	&43.2	&1;18\\
353.8$-$03.7&PHR1745$-$3609&174531$-$361011&17 45 31.56&$-$36 10 11.3&16.3&3.0$\pm$0.6	&30.4	&1\\
356.1$-$02.7&PPA1747$-$3341&174705$-$334112&17 47 05.46&$-$33 41 12.0&12.2&4.7$\pm$0.8	&5.2		&1\\
007.8+04.3&PHR1747$-$1957&174715$-$195723&17 47 15.74&$-$19 57 23.3&5.8&2.3$\pm$0.5		&15.9	&-\\
002.0+00.7&MPA1747$-$2649&174728$-$264948&17 47 28.15&$-$26 49 48.1&2.0&4.1$\pm$0.5		&4		&-\\
\noalign{\smallskip}

357.3$-$02.0&PPA1747$-$3215&174728$-$321546&17 47 28.46&$-$32 15 46.0&0.5&4.2$\pm$0.5		&4.5		&-\\
354.5$-$03.9&PHR1748$-$3538&174815$-$353846&17 48 15.79&$-$35 38 46.4&16.6&:3.2			&46.7	&1;28\\
004.3+01.8a&PHR1748$-$2417&174832$-$241738&17 48 32.87&$-$24 17 38.7&3.2&3.8$\pm$0.6		&13.5	&-\\
003.5+01.3&MPA1748$-$2511&174841$-$251135&17 48 41.74&$-$25 11 35.8&1.9&24.9$\pm$0.9		&4.5		&-\\
005.0+02.2&PHR1748$-$2326&174845$-$232627&17 48 45.10&$-$23 26 27.2&12.4&3.7$\pm$0.6		&10.2	&-\\
\noalign{\smallskip}

357.5$-$02.4&PPA1749$-$3216&174938$-$321630&17 49 38.05&$-$32 16 30.4&3.1&7.3$\pm$0.5		&8.4		&-\\
009.9+04.5&PHR1750$-$1803&175047$-$180334&17 50 47.62&$-$18 03 34.0&4.0&4.9$\pm$0.6		&26.2	&-\\
358.0$-$02.4&PPA1750$-$3152&175048$-$315224&17 50 48.27&$-$31 52 24.7&3.7&4.6$\pm$0.6		&6		&-\\
000.0$-$01.3&PPA1751$-$2933&175059$-$293347&17 50 59.46&$-$29 33 47.7&12.7&4.5$\pm$0.8	&11		&2\\
008.8+03.8&PHR1751$-$1925&175107$-$192540&17 51 07.74&$-$19 25 40.8&14.9&3.0$\pm$0.5		&10		&-\\
\noalign{\smallskip}

010.0+04.3&PHR1751$-$1804&175144$-$180425&17 51 44.84&$-$18 04 25.5&28.9&:3.8			&27.5	&1\\
003.7+00.5&PHR1752$-$2527&175208$-$252746&17 52 08.55&$-$25 27 46.8&1.4&23.1$\pm$1.2		&11.7	&1\\
000.1$-$01.7&PHR1752$-$2941&175248$-$294205&17 52 48.74&$-$29 42 05.7&7.0&2.9$\pm$0.6	&13.9	&-\\
000.3$-$01.6&PHR1752$-$2930&175252$-$293000&17 52 52.19&$-$29 30 00.6&1.2&4.1$\pm$0.5	&8		&14\\
001.1$-$01.2&PPA1753$-$2836&175317$-$283602&17 53 17.42&$-$28 36 02.9&3.1&8.8$\pm$0.6		&10.7	&-\\
\noalign{\smallskip}

355.9$-$04.4&PHR1753$-$3443&175339$-$344340&17 53 39.83&$-$34 43 40.2&5.9&15.0$\pm$0.7	&20.1	&-\\
006.1+01.5&PHR1753$-$2254&175345$-$225402&17 53 45.35&$-$22 54 02.5&1.6&25.2$\pm$1.2		&20.6	&-\\
357.8$-$03.3&PHR1753$-$3228&175354$-$322847&17 53 54.92&$-$32 28 47.6&13.9&3.1$\pm$0.5	&32		&-\\
002.1$-$01.1&MPA1755$-$2741&175510$-$274131&17 55 10.19&$-$27 41 31.9&8.6&2.7$\pm$0.5	&8		&-\\
007.3+01.7&PHR1755$-$2142&175535$-$214244&17 55 35.07&$-$21 42 44.8&8.5&5.0$\pm$0.6		&16		&-\\
\noalign{\smallskip}

006.9+01.5&MPA1755$-$2212&175536$-$221256&17 55 36.72&$-$22 12 56.8&9.8&4.0$\pm$0.5		&4		&-\\
007.4+01.7&PHR1755$-$2140&175542$-$214017&17 55 42.82&$-$21 40 17.8&3.1&3.1$\pm$0.5		&16.7	&-\\
001.0$-$01.9&PHR1755$-$2904&175543$-$290408&17 55 43.07&$-$29 04 08.0&3.0&4.8$\pm$0.6	&13.5	&-\\
002.2$-$01.2&PPA1755$-$2739&175545$-$273942&17 55 45.36&$-$27 39 42.9&3.7&8.0$\pm$0.5		&12.6	&-\\
013.1+05.0&PHR1755$-$1502&175546$-$150247&17 55 46.48&$-$15 02 47.8&4.0&3.7$\pm$0.5		&17.1	&-\\
\noalign{\smallskip}

012.5+04.3&PHR1757$-$1556&175710$-$155620&17 57 10.61&$-$15 56 20.5&3.0&3.7$\pm$0.7		&26.9	&-\\
011.8+03.7&PHR1757$-$1649&175740$-$164915&17 57 40.72&$-$16 49 15.9&16.4&14.5$\pm$1.8	&137.7	&1\\
010.2+02.7&PHR1758$-$1841&175814$-$184125&17 58 14.40&$-$18 41 25.2&0.8&18.3$\pm$0.7		&8		&1\\
357.8$-$04.4&PHR1758$-$3304&175826$-$330501&17 58 26.50&$-$33 05 01.6&8.0&2.4$\pm$0.5	&17.7	&-\\
007.8+01.2&MPA1758$-$2135&175833$-$213519&17 58 33.95&$-$21 35 19.6&4.6&8.8$\pm$0.6		&5		&-\\
\noalign{\smallskip}

003.5$-$01.2&PPA1758$-$2628&175837$-$262848&17 58 37.05&$-$26 28 48.9&2.0&13.7$\pm$0.6	&4.5		&-\\
008.1+01.3&PHR1758$-$2112&175840$-$211240&17 58 40.66&$-$21 12 40.4&5.7&6.1$\pm$0.6		&9.9		&-\\
010.2+02.4&PHR1759$-$1853&175905$-$185337&17 59 05.49&$-$18 53 37.3&16.3&2.9$\pm$0.7		&10.8	&-\\
003.6$-$01.3&PHR1759$-$2630&175912$-$263026&17 59 12.11&$-$26 30 26.8&2.8&47.7$\pm$1.5	&8.5		&1\\
003.1$-$01.6&PHR1759$-$2706&175926$-$270623&17 59 26.08&$-$27 06 23.7&10.3&3.7$\pm$0.6	&33.8	&-\\
\noalign{\smallskip}

003.0$-$01.7&PHR1759$-$2712&175932$-$271245&17 59 32.85&$-$27 12 45.1&6.8&5.0$\pm$0.7	&65.2	&-\\
003.3$-$01.6&PHR1759$-$2651&175955$-$265131&17 59 55.19&$-$26 51 31.8&17.3&4.2$\pm$0.6	&8.8		&1\\
001.7$-$02.6&PPA1800$-$2846&180001$-$284628&18 00 01.27&$-$28 46 28.8&9.0&2.5$\pm$0.5		&14.6	&-\\
010.6+02.4&MPA1800$-$1834&180008$-$183436&18 00 08.49&$-$18 34 36.3&4.7&6.3$\pm$0.6		&9		&-\\
003.4$-$01.8&PHR1800$-$2653&180042$-$265334&18 00 42.26&$-$26 53 34.6&2.5&4.7$\pm$0.5	&11.5	&-\\
\noalign{\smallskip}

004.8$-$01.1&PHR1801$-$2522&180116$-$252239&18 01 16.90&$-$25 22 39.6&1.6&67.3$\pm$2.5	&8.4		&3\\
004.3$-$01.4&PPA1801$-$2553&180118$-$255323&18 01 18.92&$-$25 53 23.2&2.2&12.7$\pm$0.7	&5.5		&-\\
014.7+04.3&MPA1801$-$1402&180146$-$140205&18 01 46.34&$-$14 02 05.2&14.0&2.5$\pm$0.6		&5		&-\\
003.8$-$01.9&PHR1802$-$2637&180210$-$263700&18 02 10.68&$-$26 37 00.6&10.2&2.7$\pm$0.6	&8.2		&-\\
012.4+02.4&MPA1803$-$1657&180355$-$165724&18 03 55.07&$-$16 57 24.2&2.2&21.5$\pm$0.8		&11		&-\\
\noalign{\smallskip}

012.9+02.6&MPA1804$-$1627&180410$-$162743&18 04 10.45&$-$16 27 43.7&0.8&5.6$\pm$0.5		&10		&-\\
011.0+01.4&PHR1804$-$1842&180428$-$184241&18 04 28.62&$-$18 42 41.5&8.3&14.1$\pm$0.7		&17.5	&-\\
004.2$-$02.5&PHR1805$-$2631&180519$-$263146&18 05 19.34&$-$26 31 46.9&10.4&4.6$\pm$0.6	&9.7		&-\\
004.1$-$03.3&PPA1808$-$2700&180800$-$270017&18 08 00.78&$-$27 00 17.4&8.4&3.6$\pm$0.7		&11.4	&-\\
015.5+02.8&BMP1808$-$1406&180825$-$140912&18 08 25.91&$-$14 09 12.9&200.0&:3.8			&470		&29\\
\noalign{\smallskip}

016.6+03.1&PHR1809$-$1300&180935$-$130007&18 09 35.34&$-$13 00 07.6&15.8&6.0$\pm$0.7		&14.4	&-\\
013.3+01.1&PHR1810$-$1647&181011$-$164724&18 10 11.64&$-$16 47 24.1&37.6&$>$5.3			&121		&1;19\\
009.4$-$01.2&PHR1811$-$2123&181110$-$212312&18 11 10.78&$-$21 23 12.7&2.8&18.6$\pm$0.8	&16.2	&-\\
009.8$-$01.1&PHR1811$-$2100&181139$-$210043&18 11 39.20&$-$21 00 43.6&2.8&3.3$\pm$0.6		&19.6	&-\\
007.5$-$02.4&PPA1811$-$2337&181140$-$233717&18 11 40.65&$-$23 37 17.9&2.0&9.2$\pm$0.6		&9.9		&-\\
\noalign{\smallskip}

008.6$-$02.2&PPA1813$-$2233&181311$-$223322&18 13 11.19&$-$22 33 22.7&9.2&3.2$\pm$0.7		&7.2		&-\\
006.4$-$03.4&PHR1813$-$2505&181326$-$250538&18 13 26.82&$-$25 05 38.8&13.4&3.7$\pm$0.8	&26.5	&-\\
014.6+01.0&PHR1813$-$1543&181330$-$154358&18 13 30.06&$-$15 43 58.7&42.5&:8.5			&23.8	&1\\
017.6+02.6&PHR1813$-$1220&181333$-$122047&18 13 33.69&$-$12 20 47.8&0.2&26.6$\pm$0.9		&8		&-\\
010.0$-$01.5&PHR1813$-$2057&181335$-$205705&18 13 35.35&$-$20 57 05.3&0.8&33.4$\pm$1.1	&13.4	&4\\
\noalign{\smallskip}

015.5+01.0&PHR1815$-$1457&181506$-$145723&18 15 06.61&$-$14 57 23.1&2.6&9.2$\pm$0.6		&8.5		&-\\
008.4$-$02.8&PHR1815$-$2300&181513$-$230105&18 15 13.07&$-$23 01 05.2&1.5&3.2$\pm$0.6	&13		&-\\
014.5+00.4&MPA1815$-$1602&181521$-$160259&18 15 21.19&$-$16 02 59.7&3.9&10.5$\pm$0.6		&5.5		&-\\
010.0$-$02.0&MPA1815$-$2113&181537$-$211318&18 15 37.26&$-$21 13 18.0&4.7&19.0$\pm$0.8	&8		&-\\
010.7$-$02.3&MPA1818$-$2044&181804$-$204411&18 18 04.44&$-$20 44 11.8&2.3&5.7$\pm$0.5		&10.8	&-\\
\noalign{\smallskip}

017.5+01.0&MPA1819$-$1307&181903$-$130717&18 19 03.37&$-$13 07 17.2&20.5&:5.2			&5.5		&-\\
020.4+02.2&PHR1820$-$1002&182015$-$100228&18 20 15.28&$-$10 02 28.9&1.8&4.9$\pm$0.7		&21.1	&-\\
012.1$-$02.1&PHR1820$-$1926&182017$-$192637&18 20 17.95&$-$19 26 37.1&6.4&4.4$\pm$0.5	&10.2	&-\\
020.4+02.0&MPA1820$-$1009&182050$-$100945&18 20 50.75&$-$10 09 45.9&3.6&3.3$\pm$0.7		&6		&-\\
011.0$-$02.9&PHR1820$-$2048&182053$-$204813&18 20 53.71&$-$20 48 13.6&2.0&15.0$\pm$1.1	&14.9	&1\\
\noalign{\smallskip}

012.1$-$02.5&PHR1821$-$1939&182143$-$193944&18 21 43.32&$-$19 39 44.5&7.6&2.5$\pm$0.5	&12.2	&-\\
021.3+02.2&MPA1822$-$0914&182206$-$091401&18 22 06.14&$-$09 14 01.0&10.3&4.3$\pm$0.8		&8		&-\\
027.7+05.1&PHR1823$-$0214&182347$-$021433&18 23 47.85&$-$02 14 33.3&5.3&2.6$\pm$0.6		&28.1	&-\\
013.4$-$02.9&PHR1825$-$1839&182535$-$183942&18 25 35.63&$-$18 39 42.8&2.2&4.9$\pm$0.6	&29.2	&-\\
025.9+03.4&PHR1826$-$0435&182617$-$043525&18 26 17.94&$-$04 35 25.6&5.7&18.5$\pm$1.3		&17.5	&5\\
\noalign{\smallskip}

021.2+00.9&PHR1826$-$0953&182626$-$095327&18 26 26.55&$-$09 53 27.5&6.8&8.4$\pm$0.6		&47.6	&-\\
022.0+01.3&PHR1826$-$0859&182627$-$085915&18 26 27.90&$-$08 59 15.1&7.7&4.5$\pm$0.6		&44		&1\\
018.2$-$00.9&MPA1827$-$1328&182729$-$132819&18 27 29.74&$-$13 28 19.1&1.2&34.5$\pm$1.2	&12		&-\\
016.8$-$01.7&BMP1827$-$1504&182750$-$150425&18 27 50.72&$-$15 04 25.4&1.8&11.6$\pm$0.7	&8		&-\\
025.6+02.8&MPA1827$-$0510&182757$-$051019&18 27 57.24&$-$05 10 19.3&4.4&4.7$\pm$0.5		&10		&-\\
\noalign{\smallskip}

024.2+01.8&MPA1828$-$0652&182847$-$065159&18 28 47.85&$-$06 51 59.9&2.2&3.5$\pm$0.5		&7		&-\\
026.4+02.7&PHR1829$-$0431&182934$-$043135&18 29 34.15&$-$04 31 35.5&2.6&10.2$\pm$0.5		&18.2	&6\\
018.5$-$01.6&PHR1830$-$1331&183042$-$133111&18 30 42.31&$-$13 31 11.8&3.1&5.9$\pm$0.6		&18.4	&-\\
024.1+01.1&PHR1831$-$0715&183116$-$071525&18 31 16.87&$-$07 15 25.1&4.0&5.9$\pm$0.5		&16.9	&-\\
018.0$-$02.2&PHR1831$-$1415&183151$-$141519&18 31 51.05&$-$14 15 19.6&11.5&10.9$\pm$1.5	&56.9	&-\\
\noalign{\smallskip}

019.2$-$01.6&MPA1831$-$1256&183152$-$125613&18 31 52.98&$-$12 56 13.7&0.4&15.0$\pm$0.7	&11		&-\\
018.7$-$01.8&PHR1832$-$1326&183204$-$132616&18 32 04.70&$-$13 26 16.3&8.8&5.9$\pm$0.6	&38.9	&-\\
024.4+00.9&MPA1832$-$0706&183222$-$070648&18 32 22.93&$-$07 06 48.0&9.0&5.5$\pm$0.5		&14		&-\\
027.8+02.7&PHR1832$-$0317&183231$-$031743&18 32 31.50&$-$03 17 43.1&3.5&11.3$\pm$1.4		&19.7	&-\\
027.0+01.5&PHR1835$-$0429&183511$-$042903&18 35 11.68&$-$04 29 03.1&3.1&62.0$\pm$2.7		&39.5	&-\\
\noalign{\smallskip}

027.0+01.3&MPA1835$-$0440&183553$-$044009&18 35 53.71&$-$04 40 09.0&1.7&9.5$\pm$0.6		&7		&-\\
029.0+02.2&MPA1836$-$0227&183610$-$022710&18 36 10.77&$-$02 27 10.8&3.4&12.2$\pm$0.6		&10		&-\\
032.5+03.2&MPA1839+0106&183911+010617&18 39 11.80&+01 06 17.9&7.1&6.1$\pm$0.6			&10		&-\\
022.8$-$01.9&MPA1839$-$0953&183944$-$095313&18 39 44.48&$-$09 53 13.7&3.5&5.1$\pm$0.5	&9.4		&-\\
030.2+01.5&MPA1841$-$0140&184104$-$014044&18 41 04.41&$-$01 40 44.1&5.1&4.4$\pm$0.7		&8.9		&-\\
\noalign{\smallskip}

032.6+02.4&MPA1842+0050&184211+005032&18 42 11.43&+00 50 32.8&0.9&15.1$\pm$0.7			&6		&-\\
023.2$-$02.6&MPA1842$-$0949&184257$-$094956&18 42 57.28&$-$09 49 56.5&9.0&11.1$\pm$0.7	&9		&-\\
032.0+01.7&PHR1843+0002&184345+000222&18 43 45.06&+00 02 22.2&1.3&10.7$\pm$0.6			&13		&-\\
029.8+00.5&PHR1843$-$0232&184358$-$023155&18 43 58.05&$-$02 31 55.4&21.3&:33		&57.4	&1\\
026.8$-$01.0&MPA1843$-$0556&184358$-$055626&18 43 58.76&$-$05 56 26.2&14.2&5.5$\pm$0.9	&11.4	&-\\
\noalign{\smallskip}

026.7$-$01.2&PHR1844$-$0603&184428$-$060326&18 44 28.82&$-$06 03 26.8&3.2&2.6$\pm$0.5	&14		&-\\
027.6$-$00.8&PHR1844$-$0503&184445$-$050421&18 44 45.91&$-$05 04 21.7&27.9&:3.2	&20.5	&1\\
032.5+01.5&PHR1845+0021&184500+002116&18 45 00.37&+00 21 16.4&8.2&3.6$\pm$0.5			&30.9	&-\\
033.8+01.5&PHR1847+0132&184745+013248&18 47 45.45&+01 32 48.1&1.3&13.4$\pm$0.6			&43.4	&-\\
016.0$-$07.6&PHR1848$-$1829&184811$-$182927&18 48 11.46&$-$18 29 27.4&15.8&3.0$\pm$0.6	&18.8	&-\\
\noalign{\smallskip}

024.4$-$03.5&PHR1848$-$0912&184832$-$091156&18 48 32.56&$-$09 11 56.6&5.8&6.0$\pm$0.7		&16.9	&-\\
028.5$-$01.4&PHR1848$-$0435&184842$-$043605&18 48 42.02&$-$04 36 05.6&21.2&5.6$\pm$0.7	&43		&1\\
026.2$-$03.4&PHR1851$-$0732&185131$-$073238&18 51 31.17&$-$07 32 38.1&9.3&6.2$\pm$0.5	&39.7	&-\\
032.5$-$00.4&MPA1852$-$0033&185225$-$003323&18 52 25.45&$-$00 33 23.0&3.1&20.5$\pm$0.8	&9		&-\\
031.1$-$01.3&MPA1852$-$0210&185255$-$021057&18 52 55.87&$-$02 10 57.5&1.8&11.7$\pm$0.6	&7.5		&-\\
\noalign{\smallskip}

031.0$-$02.1&MPA1855$-$0240&185534$-$024022&18 55 34.56&$-$02 40 22.6&4.4&8.7$\pm$0.5	&7.5		&-\\
035.5$-$00.4&PHR1857+0207&185759+020708&18 57 59.75&+02 07 08.8&4.2&100$\pm$4			&11		&1\\
034.1$-$01.6&MPA1859+0017&185925+001739&18 59 25.02&+00 17 39.0&12.4&3.7$\pm$0.6		&8.9		&7\\
033.7$-$02.0a&PHR1900$-$0014&190003$-$001356&19 00 03.94&$-$00 13 56.4&14.7&2.5$\pm$0.5	&41		&-\\
034.0$-$02.2&MPA1901+0000&190136+000012&19 01 36.01&+00 00 12.8&1.8&2.5$\pm$0.5			&7		&-\\
035.6$-$04.2&MPA1911+0027&191125+002756&19 11 25.57&+00 27 56.1&16.0&3.4$\pm$0.6		&26.9	&-\\
\end{longtable}
}
\end{center}

\newpage

\begin{center}
\begin{table*}
\caption{MASH PNe positively detected in the MPGS-2.}\label{tab:mgps2detection}
{\scriptsize
\begin{tabular}{cccccrrrr}

\hline \noalign{\smallskip}

PNG & MASH Name & MGPS2 source & RAJ2000 & DEJ2000 & Offset  & S$_{0.843\textrm{GHz}}$ & $\theta_{opt}$& c.\\
\noalign{\smallskip}
& & &  & & [\arcsec] &[mJy] & [\arcsec] &\\
\hline \noalign{\smallskip}

248.3$-$03.6&PHR0748$-$3258&J074831$-$325823&07 48 31.96&$-$32 58 23.6&7.8&18.0$\pm$1.7	&22.5	&-\\
250.5$-$03.4&PHR0754$-$3444&J075455$-$344407&07 54 55.60&$-$34 44 07.3&2.1&24.7$\pm$2.3	&59.5	&-\\
274.2$-$09.7&FP0840$-$5754&J084023$-$575501&08 40 23.81&$-$57 55 01.4&14.0&21.9$\pm$1.1	&339.9	&-\\
269.3$-$00.8&PHR0905$-$4822&J090528$-$482251&09 05 28.79&$-$48 22 51.3&6.1&16.2$\pm$2.2	&23.1	&-\\
292.4$-$00.7&MPA1121$-$6146&J112121$-$614608&11 21 21.73&$-$61 46 08.1&2.1&40.3$\pm$2.6	&6.5		&-\\
\noalign{\smallskip}

294.7$-$02.4&MPA1135$-$6406&J113546$-$640618&11 35 46.43&$-$64 06 18.6&0.4&20.0$\pm$1.3	&11		&-\\
297.0$-$04.9&PHR1150$-$6704&J115028$-$670519&11 50 28.49&$-$67 05 19.6&23.2&:14.6$\pm$1.8	&45.4	&23\\
302.3$-$00.5&PHR1246$-$6324&J124627$-$632431&12 46 27.54&$-$63 24 31.3&7.7&17.8$\pm$2.1	&24.3	&15\\
309.8$-$01.6&MPA1354$-$6337&J135421$-$633714&13 54 21.99&$-$63 37 14.5&4.1&13.7$\pm$1.2	&7.5		&-\\
310.8$-$02.3&MPA1404$-$6403&J140424$-$640402&14 04 24.31&$-$64 04 02.2&8.8&13.5$\pm$1.7	&6		&-\\
\noalign{\smallskip}

314.4+01.3&BMP1423$-$5923&J142359$-$592337&14 23 59.57&$-$59 23 37.3&1.2&14.8$\pm$1.4	&8.5		&-\\
315.9+00.3&PHR1437$-$5949&J143753$-$594917&14 37 53.09&$-$59 49 17.7&7.3&18.4$\pm$2.8	&80.6	&-\\
316.9+01.8&MPA1440$-$5802&J144027$-$580219&14 40 27.00&$-$58 02 19.7&3.3&11.8$\pm$1.0	&11.5	&-\\
315.7$-$01.1&MPA1441$-$6114&J144132$-$611417&14 41 32.71&$-$61 14 17.6&3.7&20.3$\pm$1.7	&6.5		&-\\
318.9+00.7&PHR1457$-$5812&J145735$-$581205&14 57 35.66&$-$58 12 05.4&3.6&83.3$\pm$2.9	&27.8	&-\\
\noalign{\smallskip}

319.5$-$01.0&PHR1507$-$5925&J150750$-$592514&15 07 50.63&$-$59 25 14.0&3.3&24.2$\pm$1.5	&19.3	&-\\
321.3$-$00.3&PHR1517$-$5751&J151731$-$575110&15 17 31.35&$-$57 51 10.7&1.4&65.1$\pm$7.7	&104.2	&-\\
322.2$-$00.4&BMP1522$-$5729&J152258$-$573000&15 22 58.68&$-$57 30 00.7&2.5&17.2$\pm$1.5	&12		&-\\
322.2$-$00.7&BMP1524$-$5746&J152423$-$574621&15 24 23.85&$-$57 46 21.9&1.5&27.0$\pm$1.8	&7		&-\\
323.5+01.1&MPA1525$-$5528&J152506$-$552827&15 25 06.25&$-$55 28 27.7&5.8&22.8$\pm$1.4	&13.8	&-\\
\noalign{\smallskip}

322.0$-$01.3&BMP1525$-$5823&J152559$-$582303&15 25 59.37&$-$58 23 03.7&1.7&27.2$\pm$2.2	&8.9		&-\\
322.0$-$01.8&PHR1527$-$5846&J152735$-$584616&15 27 35.88&$-$58 46 16.8&6.0&15.4$\pm$1.5	&16.3	&-\\
324.3+01.1&PHR1529$-$5458&J152927$-$545914&15 29 27.96&$-$54 59 14.3&22.7&:37.1$\pm$4.6	&92.5	&25\\
322.7$-$01.4&MPA1530$-$5801&J153050$-$580125&15 30 50.50&$-$58 01 25.5&4.7&16.9$\pm$1.7	&6		&-\\
327.1+01.9&MPA1541$-$5243&J154129$-$524350&15 41 29.41&$-$52 43 50.9&2.1&39.1$\pm$1.9	&11		&-\\
\noalign{\smallskip}

332.3+07.0&PHR1547$-$4533&J154743$-$453243&15 47 43.94&$-$45 32 43.4&17.8&:27.7			&118.9	&26\\
328.3+00.7&PHR1552$-$5254&J155257$-$525408&15 52 57.28&$-$52 54 08.5&5.6&20.9$\pm$2.3	&28.4	&-\\
334.4+02.3&BMP1614$-$4742&J161451$-$474211&16 14 51.82&$-$47 42 11.1&4.8&12.3$\pm$1.3	&11.5	&-\\
333.9+00.6&PHR1619$-$4914&J161940$-$491358&16 19 40.18&$-$49 13 58.4&1.8&294.8$\pm$9.7	&33.9	&-\\
338.9+04.6&PHR1624$-$4252&J162400$-$425239&16 24 00.27&$-$42 52 39.8&6.5&10.3$\pm$1.6	&40		&-\\
\noalign{\smallskip}

340.4+03.8&MPA1632$-$4220&J163231$-$422053&16 32 31.40&$-$42 20 53.9&3.3&20.6$\pm$1.2	&14.4	&-\\
337.3+00.6&PHR1633$-$4650&J163358$-$465009&16 33 58.18&$-$46 50 09.0&2.7&47.2$\pm$5.5	&13.9	&-\\
340.0+02.9&PHR1634$-$4318&J163442$-$431759&16 34 42.81&$-$43 17 59.5&3.2&14.6$\pm$1.1	&16.4	&-\\
338.6+01.1&BMP1636$-$4529&J163658$-$452935&16 36 58.20&$-$45 29 35.9&8.7&22.0$\pm$4.2	&9.9		&-\\
335.4$-$01.9&PHR1637$-$4957&J163744$-$495747&16 37 44.86&$-$49 57 47.3&2.7&37.0$\pm$1.9	&19.2	&-\\
\noalign{\smallskip}

338.4$-$02.0&MPA1650$-$4747&J165009$-$474705&16 50 09.87&$-$47 47 05.1&0.7&14.6$\pm$1.6	&5		&-\\
343.5+01.2&PHR1653$-$4143&J165355$-$414357&16 53 55.50&$-$41 43 57.9&3.1&18.7$\pm$1.9	&12.8	&-\\
340.1$-$02.2a&MPA1657$-$4633&J165706$-$463357&16 57 06.27&$-$46 33 57.2&2.8&12.6$\pm$1.6	&7		&-\\
341.9$-$02.8&MPA1706$-$4527&J170624$-$452659&17 06 24.71&$-$45 26 59.0&7.2&22.2$\pm$3.5	&7.5		&-\\
350.4+02.0&PHR1712$-$3543&J171233$-$354321&17 12 33.85&$-$35 43 21.5&2.4&12.7$\pm$1.9	&10.1	&-\\
\noalign{\smallskip}

347.2$-$00.8&PHR1714$-$4006&J171447$-$400600&17 14 47.77&$-$40 06 00.0&19.7&25.6$\pm$6.8	&14.8	&-\\
344.8$-$02.6&MPA1715$-$4303&J171515$-$430357&17 15 15.59&$-$43 03 57.6&5.0&17.6$\pm$2.2	&7		&-\\
353.6+01.7&PPA1722$-$3317&J172235$-$331704&17 22 35.43&$-$33 17 04.4&10.6&18.6$\pm$2.9	&4		&-\\
355.0+02.6&PPA1722$-$3139&J172240$-$313954&17 22 40.68&$-$31 39 54.9&1.5&19.1$\pm$2.7	&8.5		&-\\
354.8+01.8&PPA1725$-$3216&J172515$-$321607&17 25 15.81&$-$32 16 07.6&3.6&17.6$\pm$3.4	&6.9		&-\\
\noalign{\smallskip}

355.6+01.4&PPA1729$-$3152&J172911$-$315243&17 29 11.04&$-$31 52 43.0&2.1&11.6$\pm$2.1		&5.5		&-\\
352.8$-$00.5&MPA1729$-$3513&J172937$-$351339&17 29 37.20&$-$35 13 39.8&7.2&56$\pm$5		&12.7	&-\\
357.8+01.6&PPA1734$-$2954&J173401$-$295442&17 34 01.96&$-$29 54 42.9&8.6&13.8$\pm$1.8	&13		&-\\
352.1$-$02.6&PHR1736$-$3659&J173617$-$365937&17 36 17.86&$-$36 59 37.2&11.9&15.4$\pm$2.4	&15.9	&-\\
356.6$-$01.9&PHR1745$-$3246&J174509$-$324629&17 45 09.46&$-$32 46 29.2&12.9&33.7$\pm$2.8	&43.2	&18\\
\noalign{\smallskip}

355.9$-$04.4&PHR1753$-$3443&J175339$-$344336&17 53 39.87&$-$34 43 36.2&7.2&20.9$\pm$1.8	&20.1	&-\\
001.0$-$01.9&PHR1755$-$2904&J175543$-$290406&17 55 43.23&$-$29 04 06.8&2.5&15.5$\pm$2.3	&13.5	&-\\
003.6$-$01.3&PHR1759$-$2630&J175912$-$263033&17 59 12.00&$-$26 30 33.4&9.5&32.7$\pm$3.0	&8.5		&-\\
\hline \noalign{\smallskip}
\end{tabular}
}
\end{table*}
\end{center}

\newpage

\begin{center}
\begin{table*}
\caption{MASH PNe possibly detected in the MPGS-2.}\label{tab:sumssdetection}
{\scriptsize
\begin{tabular}{cccccrrrr}

\hline \noalign{\smallskip}

PNG & MASH Name & J2000 des. & RAJ2000 & DEJ2000 & Offset  & S$_{0.843\textrm{GHz}}$ & $\theta_{opt}$ & c.\\
\noalign{\smallskip}
& & &  & & [\arcsec] &[mJy] & [\arcsec] &\\
\hline \noalign{\smallskip}

272.8+01.0&PHR0928$-$4936&J0928$-$4936&09 28 41.79&$-$49 36 54.4&20.7&7.5		&85.5	&-\\
291.6$-$00.2&PHR1115$-$6059&J1115$-$6059&11 15 48.16&$-$60 59 13.4&25.0&$>$18.7	&102.5	&16\\
297.0+01.1&PHR1202$-$6112&J1202$-$6112&12 02 17.02&$-$61 12 45.6&14.8&7.3		&13.5	&-\\
299.2+01.0&PHR1220$-$6134&J1220$-$6134&12 20 09.21&$-$61 34 20.3&7.5&7.4		&9.7		&-\\
301.5+02.0&MPA1239$-$6048&J1239$-$6048&12 39 46.20&$-$60 48 15.9&1.9&7.5		&6.9		&-\\
\noalign{\smallskip}
304.8$-$01.4&MPA1309$-$6415&J1309$-$6415&13 09 30.16&$-$64 15 31.9&11.1&5.7		&8.9		&-\\
305.6$-$00.9&MPA1315$-$6338&J1315$-$6338&13 15 30.89&$-$63 38 37.7&9.1&9.8		&6		&-\\
305.9$-$01.6&MPA1319$-$6418&J1319$-$6419&13 19 24.47&$-$64 19 05.7&14.5&6.7		&4		&-\\
309.5+00.8&PHR1346$-$6116&J1346$-$6116&13 46 39.37&$-$61 16 14.2&21.7&:6.8		&61		&24\\
313.4+06.2&MPA1405$-$5507&J1405$-$5507&14 05 31.49&$-$55 07 42.3&7.8&6.0		&8		&-\\
\noalign{\smallskip}
313.9+01.4&PHR1420$-$5933&J1420$-$5933&14 20 07.09&$-$59 33 48.1&10.6&7.7		&28.9	&-\\
315.2$-$01.3&PHR1438$-$6140&J1438$-$6140&14 38 18.03&$-$61 40 07.2&5.3&5.6		&27		&-\\
318.2$-$02.3&PHR1504$-$6113&J1504$-$6113&15 04 39.97&$-$61 13 06.8&7.0&9.0		&15.9	&-\\
322.7+02.6&PHR1514$-$5436&J1514$-$5436&15 14 34.42&$-$54 36 08.6&6.2&6.6		&21.8	&-\\
322.4$-$00.1a&MPA1523$-$5710&J1523$-$5710&15 23 22.09&$-$57 10 54.0&6.1&18.3		&14.5	&8\\
\noalign{\smallskip}
325.7+02.2&BMP1533$-$5319&J1533$-$5319&15 33 20.29&$-$53 19 03.7&5.4&8.7		&17.2	&-\\
324.1$-$01.3&MPA1539$-$5709&J1539$-$5709&15 39 10.56&$-$57 09 56.6&5.6&9.8		&9.4		&-\\
331.3+01.6&PHR1603$-$5016&J1603$-$5017&16 03 54.01&$-$50 17 01.4&10.5&7.6		&14.9	&-\\
329.5$-$00.8&MPA1605$-$5319&J1605$-$5320&16 05 36.95&$-$53 20 03.7&11.8&11.7		&6.9		&-\\
332.3$-$00.9&PHR1619$-$5131&J1619$-$5131&16 19 58.85&$-$51 31 48.2&20.3&13.7		&11		&9\\
\noalign{\smallskip}
331.8$-$02.3&MPA1624$-$5250&J1624$-$5250&16 24 03.72&$-$52 50 17.6&16.9&$>$7.2	&9		&-\\
337.4+02.6&PHR1625$-$4522&J1625$-$4521&16 25 51.21&$-$45 21 22.3&76.8&$>$11.1	&314.6	&17\\
344.4$-$03.5&MPA1717$-$4351&J1717$-$4351&17 17 52.90&$-$43 51 52.9&6.8&8.3		&7		&-\\
354.7+03.2&PHR1719$-$3134&J1719$-$3134&17 19 54.55&$-$31 34 32.5&15.8&14.8		&6.7		&-\\
354.4+02.2&PPA1723$-$3223&J1723$-$3222&17 23 04.00&$-$32 22 52.8&17.5&10.7		&9.4		&-\\
\noalign{\smallskip}
355.8+01.7&MPA1728$-$3132&J1728$-$3132&17 28 30.87&$-$31 32 13.2&5.4&10.2		&4		&-\\
356.1$-$02.1&PHR1744$-$3319&J1744$-$3319&17 44 51.04&$-$33 19 32.6&11.7&9.6		&8.5		&-\\
359.4$-$05.6&BMP1807$-$3215&J1807$-$3215&18 07 05.77&$-$32 15 12.2&20.9&7.9		&20.4	&-\\
\hline \noalign{\smallskip}
\end{tabular}
}
\end{table*}
\end{center}

\appendix

\section{Finding charts and selected notes on individual MASH PNe}\label{sec:notesonpne}

\begin{figure*}
\begin{center}
\includegraphics[scale=0.45]{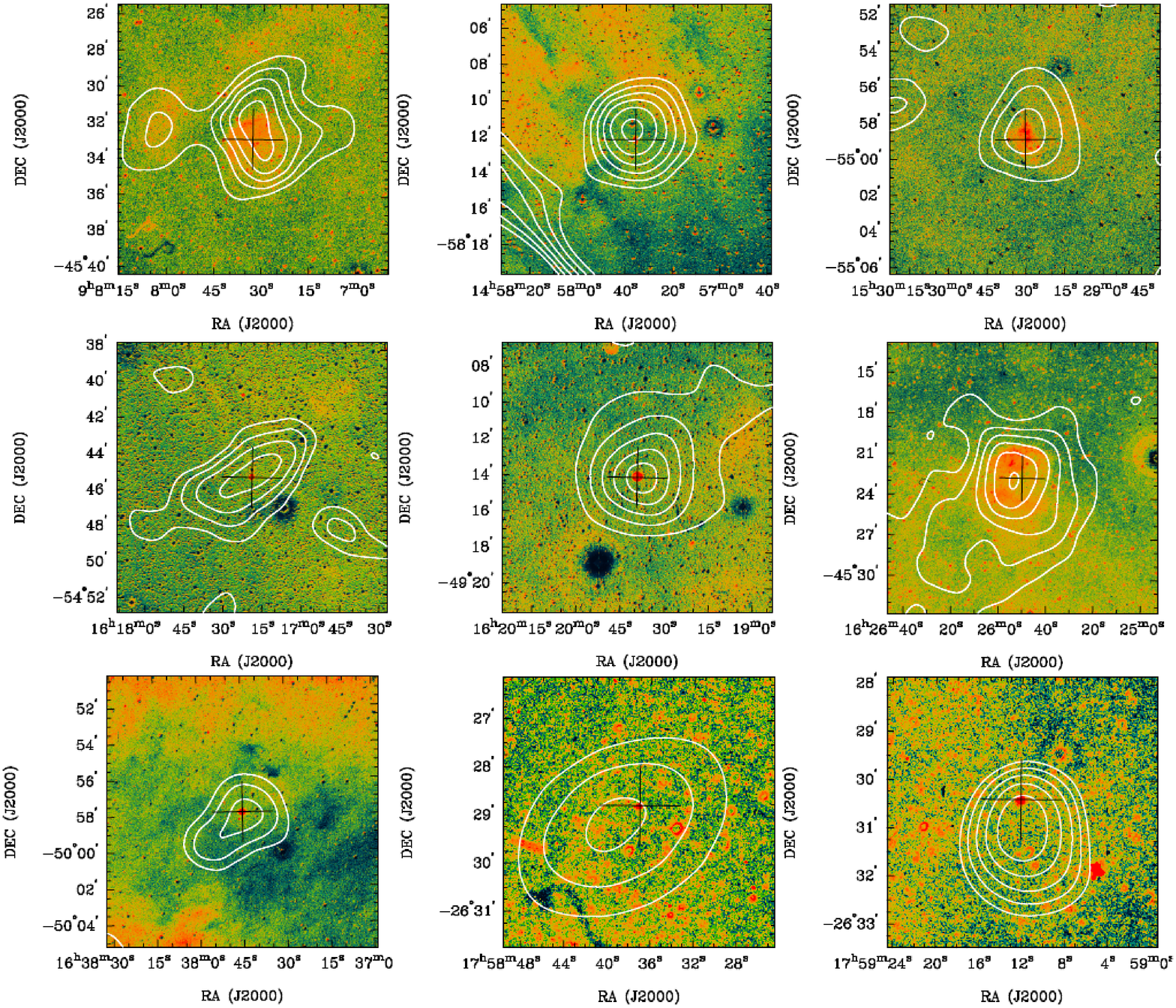}
\caption[Finding charts of MASH PNe with PMN radio counterparts.]
{Finding charts of MASH PNe with PMN radio counterparts. Finding charts are produced as radio-continuum contour maps from the PMN superposed on the SHS {\it quotient} images (see text for more details). 
{\bf First row from left to right}: 
PHR0907-4532, contours are at 20, 30, 40, 60 and 50~mJy;
PHR1457-5812, contours are at 20, 30, 40, 50, 60 and 70~mJy and
PHR1529-5458, contours are at 20, 30 and 40~mJy.
{\bf Second row from left to right}: 
PHR1617-5445, contours are at 20, 25, 30 and 35~mJy;
PHR1619-4914, contours are at 20, 60, 100, 140 and 180~mJy and
PHR1625-4522, contours are at 20, 40, 60, 80, 100 and 120~mJy.
{\bf Third row from left to right}:  
PHR1637-4957, contours are at 20, 30 and 40~mJy;
PPA1758-2628, contours are at 40, 50 and 60~mJy and
PHR1759-2630, contours are at 40, 42, 44, 46 and 48~mJy.}\label{pmn1}
\end{center}
\end{figure*}

\begin{figure*}
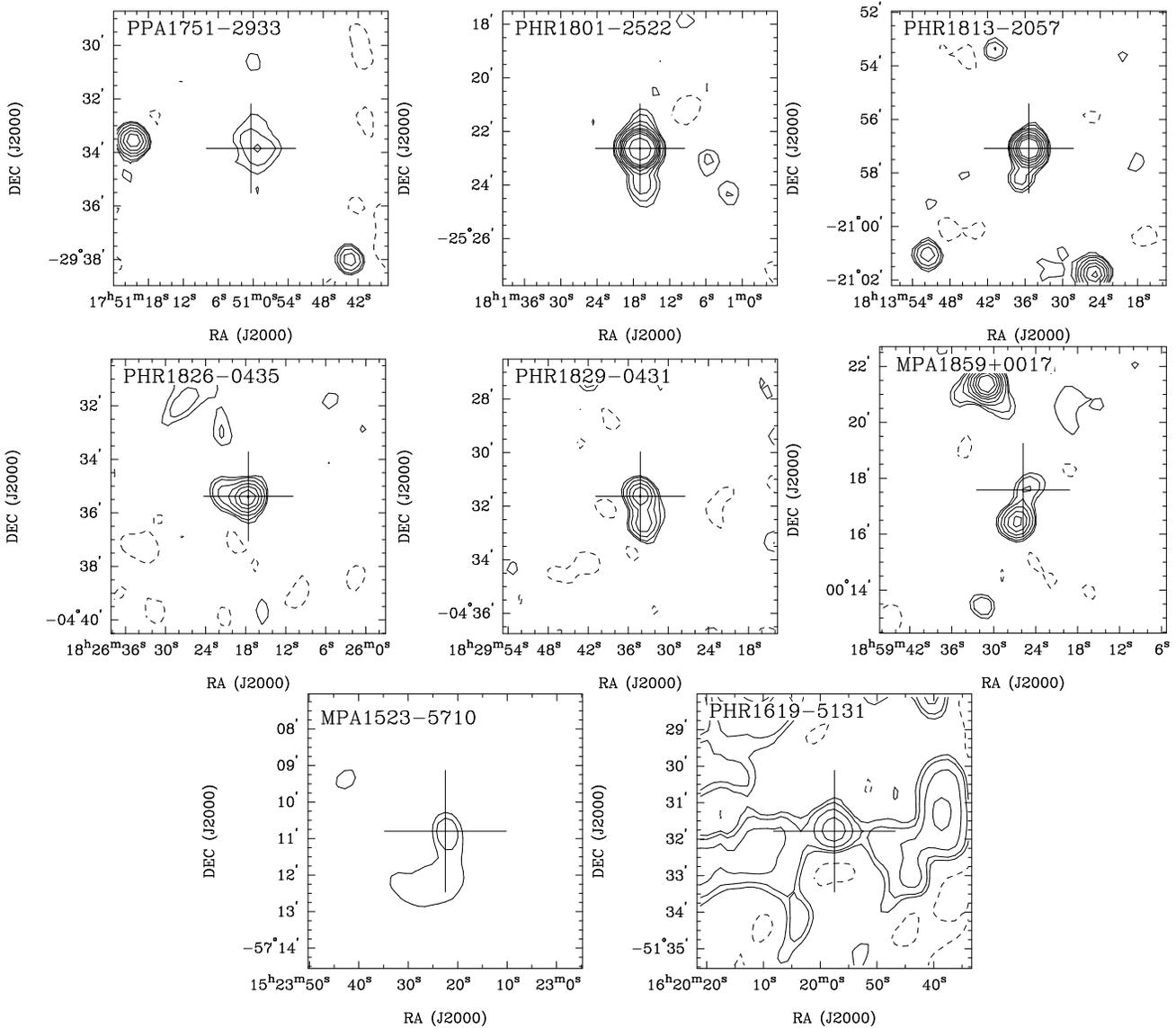

\begin{center}
\includegraphics[scale=0.3]{nvss_PPA1751-2933.eps}
\includegraphics[scale=0.3]{nvss_PHR1801-2522.eps}
\includegraphics[scale=0.3]{nvss_PHR1813-2057.eps}
\includegraphics[scale=0.3]{nvss_PHR1826-0435.eps}
\includegraphics[scale=0.3]{nvss_PHR1829-0431.eps}
\includegraphics[scale=0.3]{nvss_MPA1859+0017.eps}
\includegraphics[scale=0.3]{sumss_MPA1523-5710.eps}
\includegraphics[scale=0.3]{sumss_PHR1619-5131.eps}
\caption[Radio-continuum contour plots of MASH PNe detected or possibly detected in the NVSS/MGPS-2. ]{Radio-continuum contour plots of MASH PNe detected or possibly detected in the NVSS/MGPS-2. 
First row (from left to right): PPA1751$-$2933 (NVSS), PHR1801$-$2522 (NVSS) and PHR1813$-$2057 (NVSS). 
Second row (from left to right): PHR1826$-$0435 (NVSS), PHR1829$-$0431 (NVSS) and MPA1859+0017 (NVSS).  
Third row (from left to right): MPA1523$-$5710 (MGPS-2) and PHR1619$-$5131 (MGPS-2). 
Contours are at -2, 2, 3, 5, 8, 12, 17, 23, 30, 60 and 120 $\times\sigma_{rms}$ (where $\sigma_{rms}$ is a local {\it rms} noise). Negative contour (at -2$\times\sigma_{rms}$) is presented with a dashed line. Radio-continuum contour images are overlaid with a cross centred in the MASH PN optical position. MASH PN designation is in the upper left corner.}\label{fig:exam1}
\end{center}
\end{figure*}

\begin{figure*}
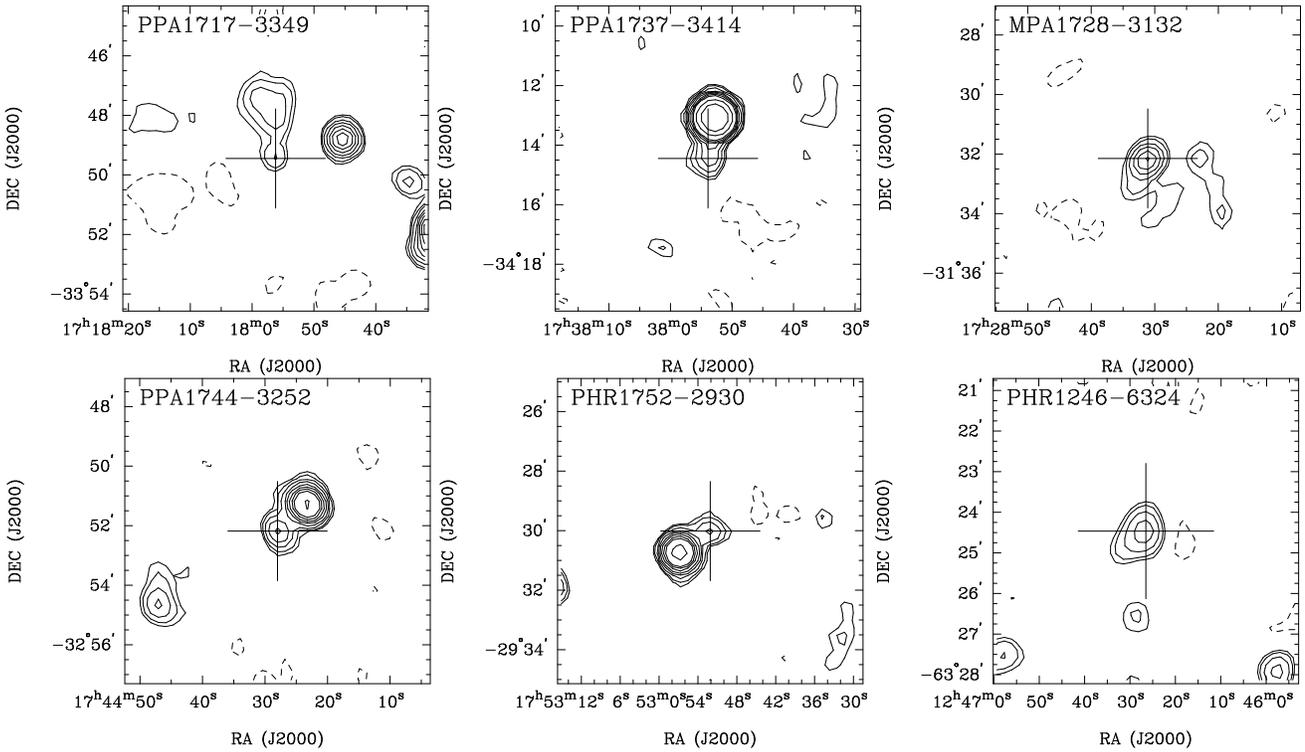

\begin{center}
\includegraphics[scale=0.3]{nvss_PPA1717-3349.eps}
\includegraphics[scale=0.3]{nvss_PPA1737-3414.eps}
\includegraphics[scale=0.3]{nvss_MPA1728-3132.eps}
\includegraphics[scale=0.3]{nvss_PPA1744-3252.eps}
\includegraphics[scale=0.3]{nvss_PHR1752-2930.eps}
\includegraphics[scale=0.3]{sumss_PHR1246-6324.eps}
\caption[Radio-continuum contour plots of MASH PNe detected or possibly detected in the NVSS/MGPS-2.]{Same as in Fig.~\ref{fig:exam1}. First row (from left to right): PPA1717$-$3349 (NVSS), PPA1737$-$3414 (NVSS) and MPA1728$-$3132 (NVSS). Second row (from left to right): PPA1744$-$3252 (NVSS), PHR1752$-$2930 (NVSS) and PHR1246$-$6324 (MGPS-2).}\label{fig:exam2}
\end{center}
\end{figure*}

\begin{figure*}
\begin{center}
\includegraphics[scale=0.45]{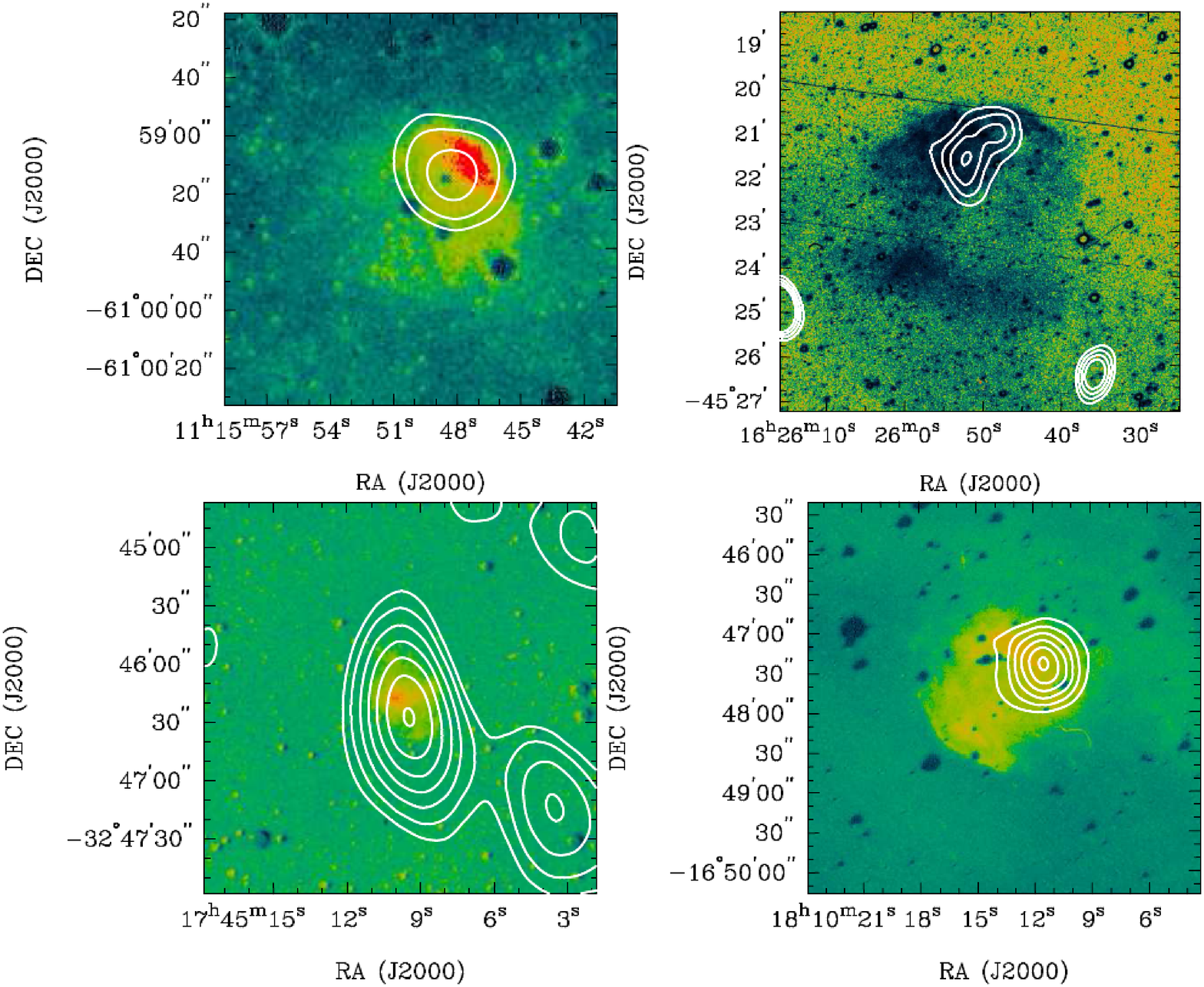} 
\caption[MASH PNe detected in the MGPS2/NVSS.]{Finding charts of MASH PNe with MGPS-2/NVSS radio counterparts. Finding charts are produced as radio-continuum contour maps from the MGPS-2/NVSS superposed on the SHS {\it quotient} images (see text for more details). 
{\bf Top left}: PHR1115-6059, the MGPS2 radio contours are at 6, 10 and 15 mJy; 
{\bf Top right}: PHR1625-4522, the MGPS2 radio contours are at 6, 8, 10 and 12~mJy;
{\bf Bottom left}: PHR1745-3246, the MGPS2 radio contours are at 6, 10, 15, 20, 25, 30, 35~mJy;
{\bf Bottom right}: PHR1810-1647, the NVSS radio contours are at 1.5, 2, 2.5, 3, 3.5, and 4~mJy; 
Images for PHR1115-6059, PHR1745-3246 and PHR1810-1647 were produced as log scaled SHS {\it quotient} images and for PHR1625-4522,as histogram equalized SHS {\it quotient} image.} \label{fig:detextend}
\end{center}
\end{figure*}

\begin{figure*}
\begin{center}
\includegraphics[scale=0.45]{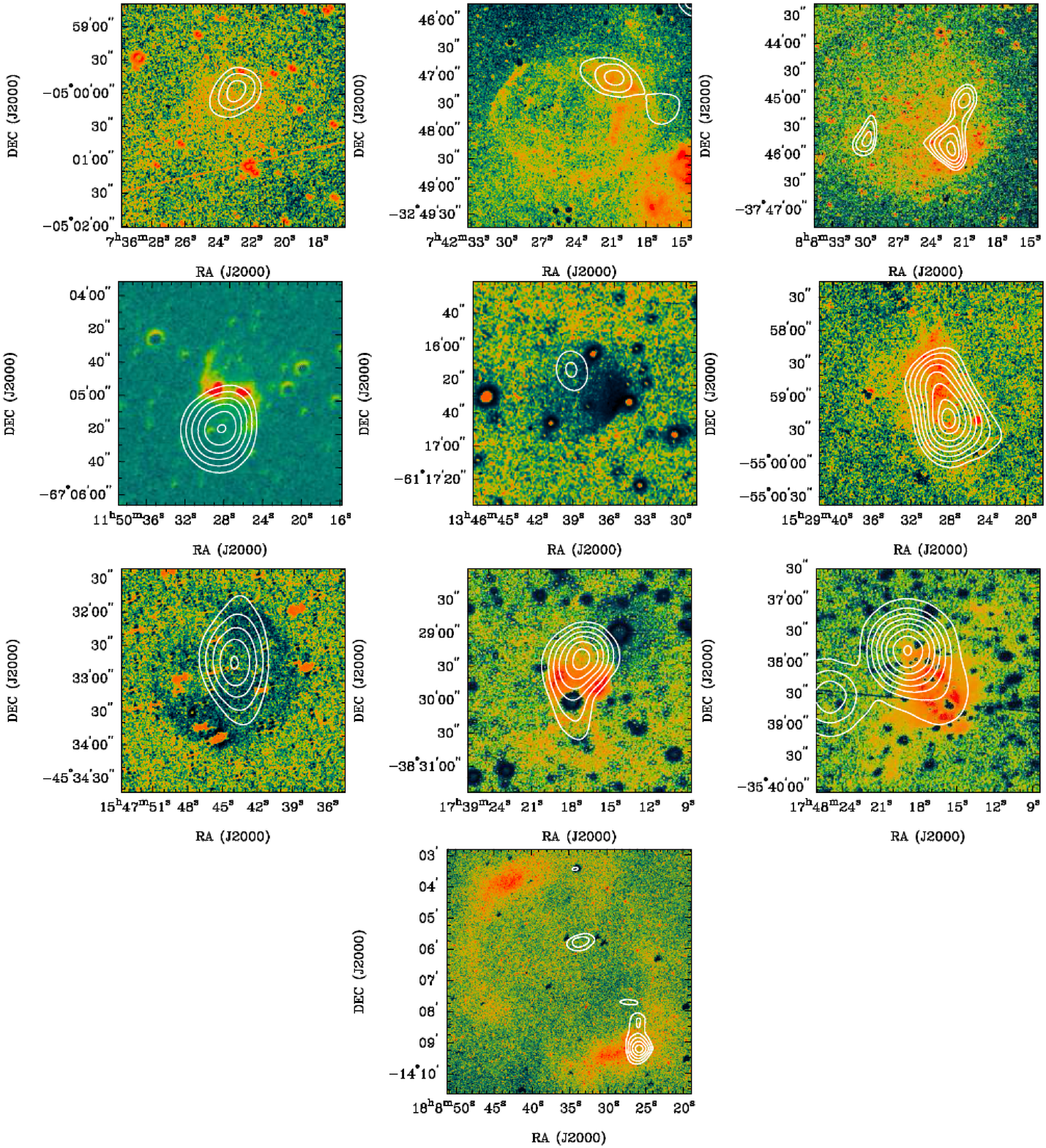} 
\caption[Suspected radio detections of MASH PNe.]{Finding charts of MASH PNe with MGPS-2/NVSS suspected radio counterparts. Finding charts are produced as radio-continuum contour maps from the MGPS-2/NVSS superposed on the SHS {\it quotient} images (see text for more details). 
{\bf First row from left to right}: 
BMP0736-0500, the NVSS radio contours are at 1.5, 2 and 2.5~mJy; 
PHR0742-3247, the NVSS radio contours are at 1.5, 1.8, 2.1~mJy; 
PHR0808-3745, the NVSS radio contours are at 1.5, 1.6, 1.7, 1.8, 1.9 mJy; 
{\bf Second row from left to right}: 
PHR1150-6704, the MGPS2 radio contours are at 6, 7, 8, 9, 10 and 11 mJy.
PHR1346-6116, the MGPS2 radio contours are at 5 and 6~mJy. 
PHR1529-5458, the MGPS2 radio contours are at  6, 7, 8, 9, 10, 11, 12 and 12.5~mJy 
{\bf Third row from left to right}: 
PHR1547-4533, the MGPS2 radio contours are at 6, 10, 15, 20 and 25 mJy; 
PHR1739-3829, the NVSS radio contours are at 2, 2.5, 3, 3.5, 4, 4.5 and 5 mJy; 
PHR1748-3538, the NVSS radio contours are at 2, 5, 8, 12, 17, 23, 30 and 38 mJy;
{\bf Fourth row}: 
BMP1808-1406, the NVSS radio contours are at 1.5, 2, 2.5, 3, 3.5, and 4~mJy; 
Images were produced as histogram equalized SHS {\it quotient} images except for PHR1150-6704 which is produced as log scaled SHS {\it quotient} image.}
\label{fig:detsuspect1}
\end{center}
\end{figure*}

\begin{figure}
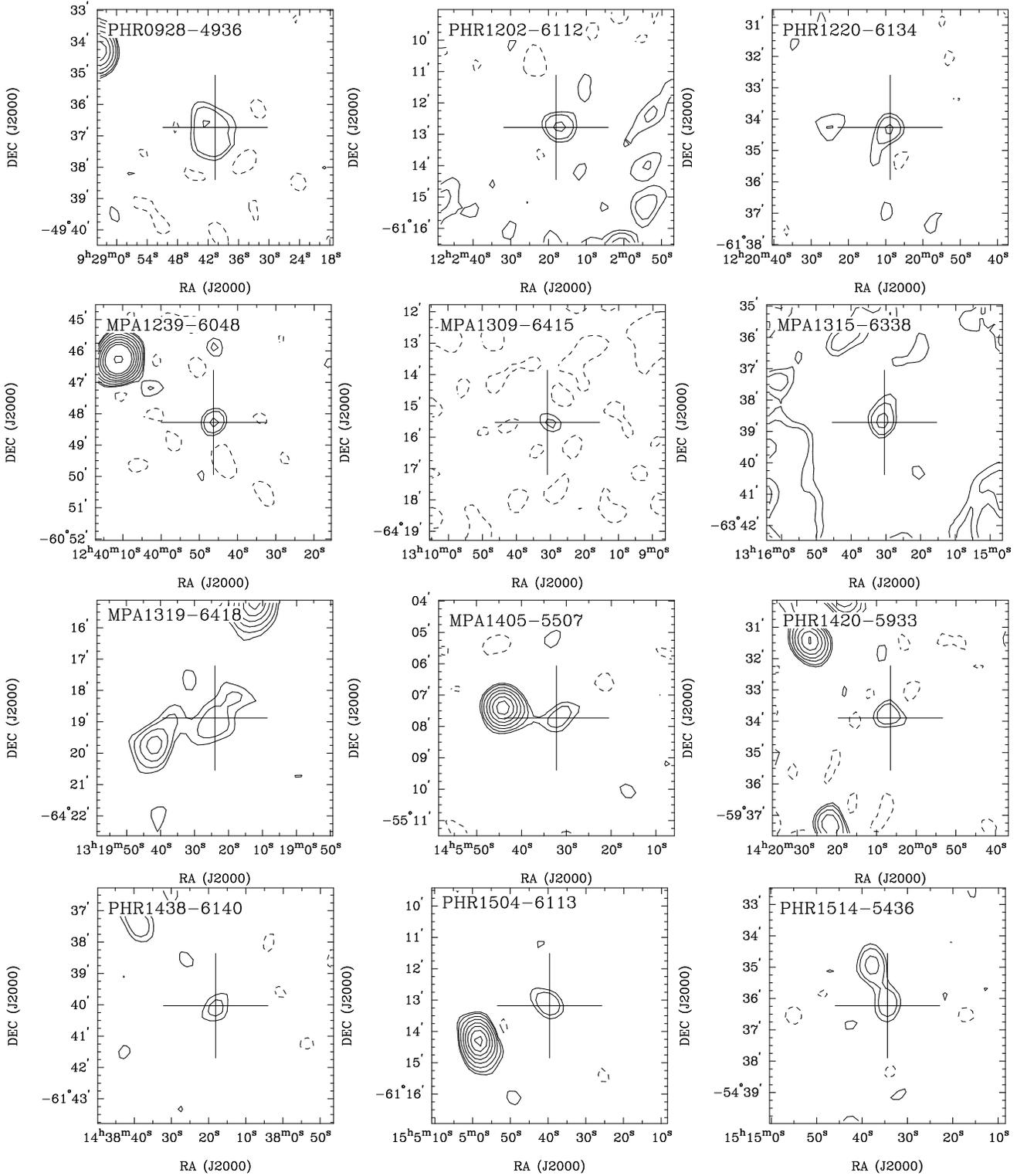

\begin{center}
\includegraphics[scale=0.3]{sumss_PHR0928-4936.eps}\label{sumss_PHR0928-4936}
\includegraphics[scale=0.3]{sumss_PHR1202-6112.eps}\label{sumss_PHR1202-6112}
\includegraphics[scale=0.3]{sumss_PHR1220-6134.eps}\label{sumss_PHR1220-6134}
\includegraphics[scale=0.3]{sumss_MPA1239-6048.eps}\label{sumss_MPA1239-6048}
\includegraphics[scale=0.3]{sumss_MPA1309-6415.eps}\label{sumss_MPA1309-6415}
\includegraphics[scale=0.3]{sumss_MPA1315-6338.eps}\label{sumss_MPA1315-6338}
\includegraphics[scale=0.3]{sumss_MPA1319-6418.eps}\label{sumss_MPA1319-6418}
\includegraphics[scale=0.3]{sumss_MPA1405-5507.eps}\label{sumss_MPA1405-5507}
\includegraphics[scale=0.3]{sumss_PHR1420-5933.eps}\label{sumss_PHR1420-5933}
\includegraphics[scale=0.3]{sumss_PHR1438-6140.eps}\label{sumss_PHR1438-6140}
\includegraphics[scale=0.3]{sumss_PHR1504-6113.eps}\label{sumss_PHR1504-6113}
\includegraphics[scale=0.3]{sumss_PHR1514-5436.eps}\label{sumss_PHR1514-5436}
\end{center}
\caption{Radio-continuum contour plots of possibly radio detected MASH PNe in the MGPS-2 survey. None of these objects is catalogued in the MGPS-2 catalogue. Contours are at -2, 2, 3, 5, 8, 12, 17, 23, 30, 60 and 120 $\times\sigma_{rms}$ (where $\sigma_{rms}$ is a local rms noise). Negative contour (at -2$\times\sigma_{rms}$) is presented with a dashed line. Radio-continuum contour images are overlaid with a cross centred in the MASH PN optical position. MASH PN designation is in the upper left corner.}\label{pysumss1}\end{figure}

\begin{figure}
\begin{center}
\includegraphics[scale=0.3]{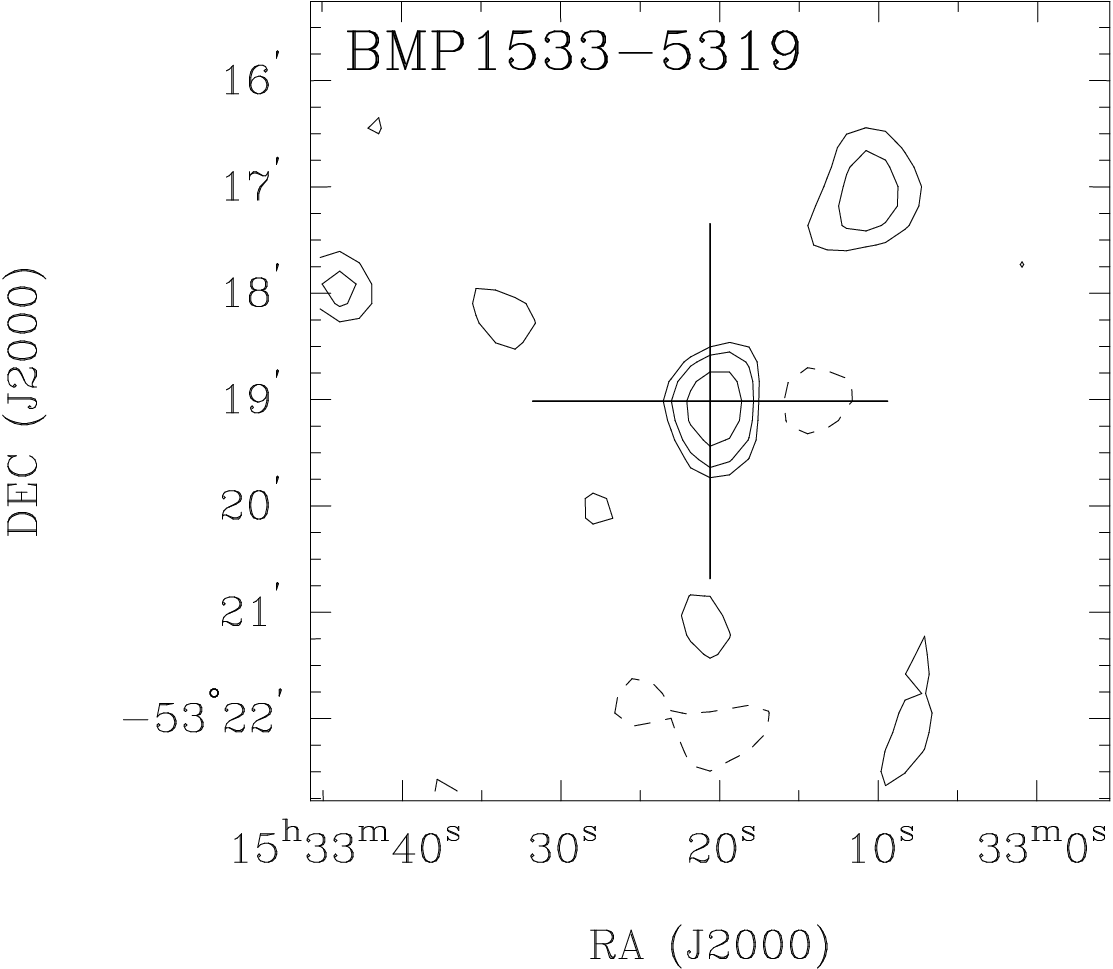}\label{sumss_BMP1533-5319}
\includegraphics[scale=0.3]{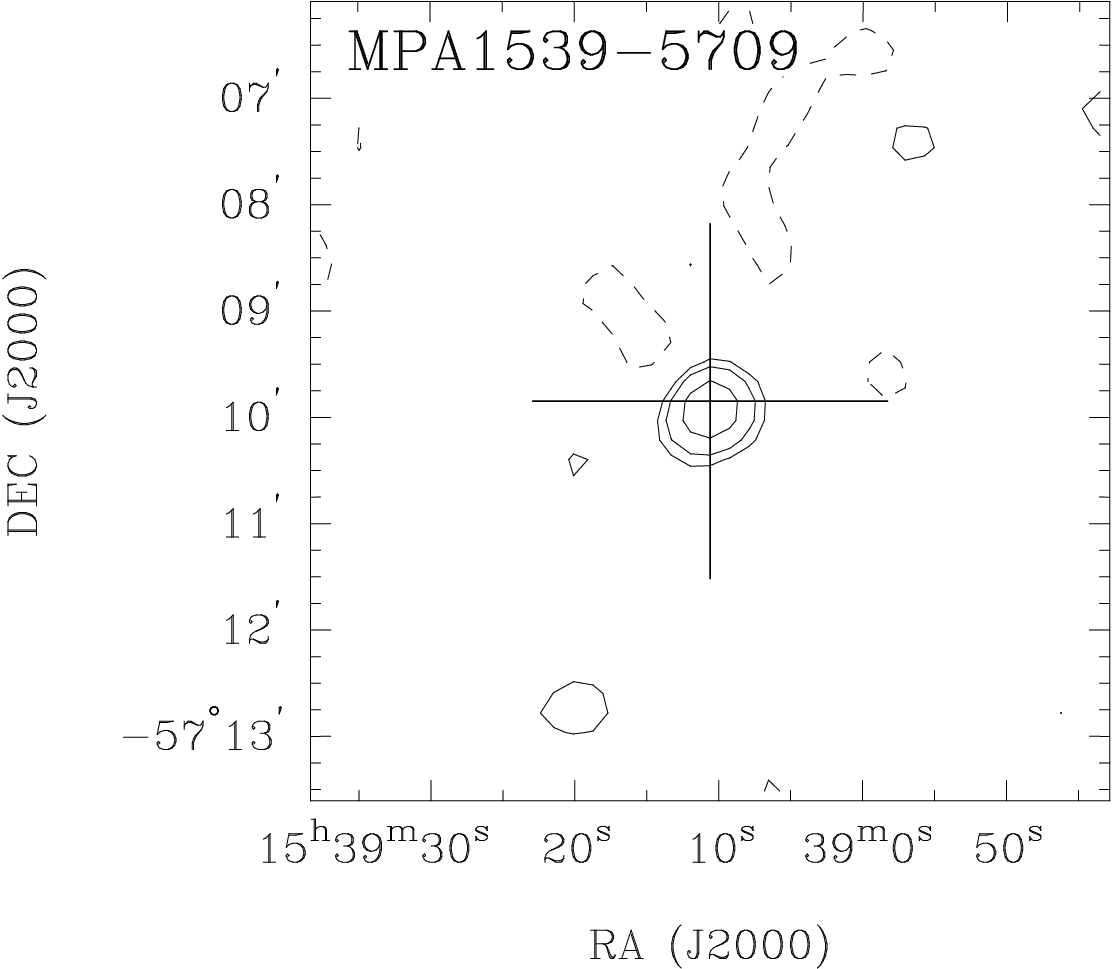}\label{sumss_MPA1539-5709}
\includegraphics[scale=0.3]{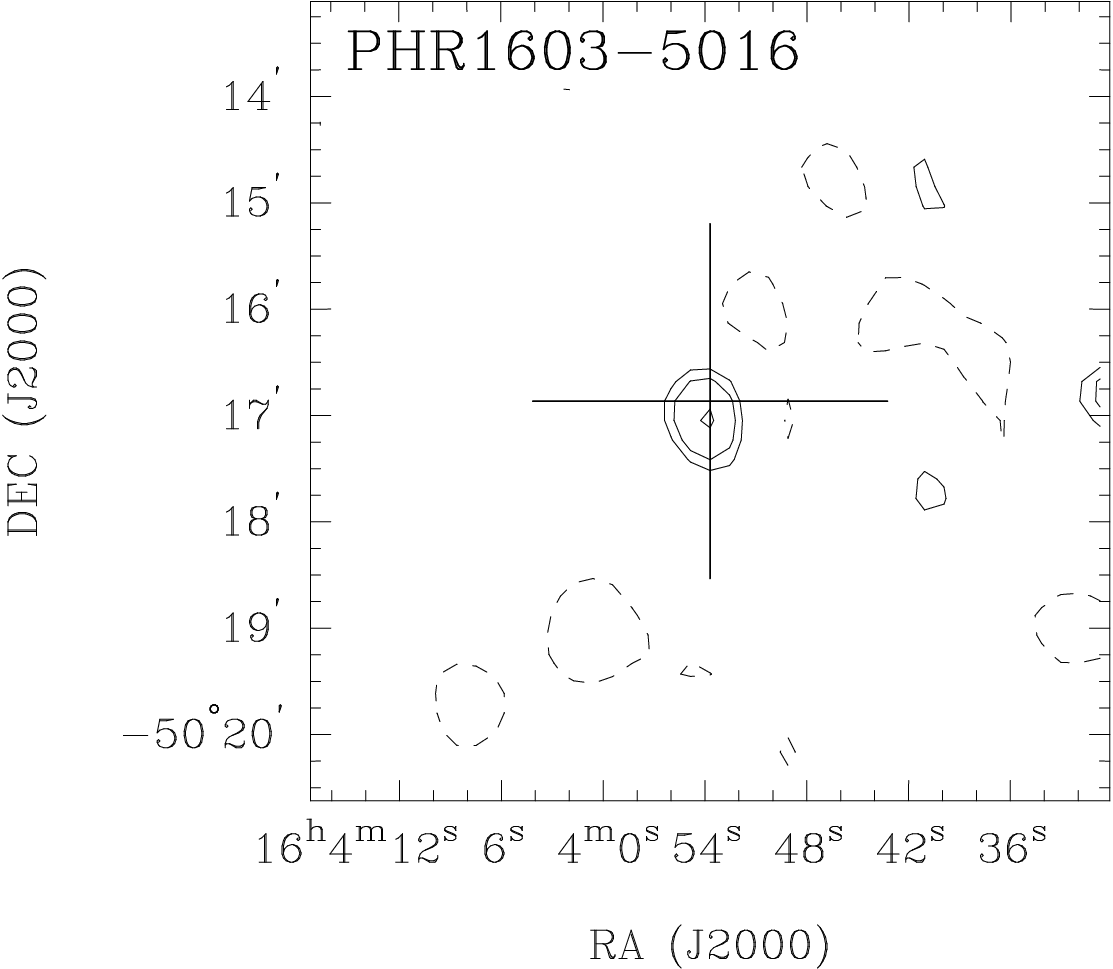}\label{sumss_PHR1603-5016}
\includegraphics[scale=0.3]{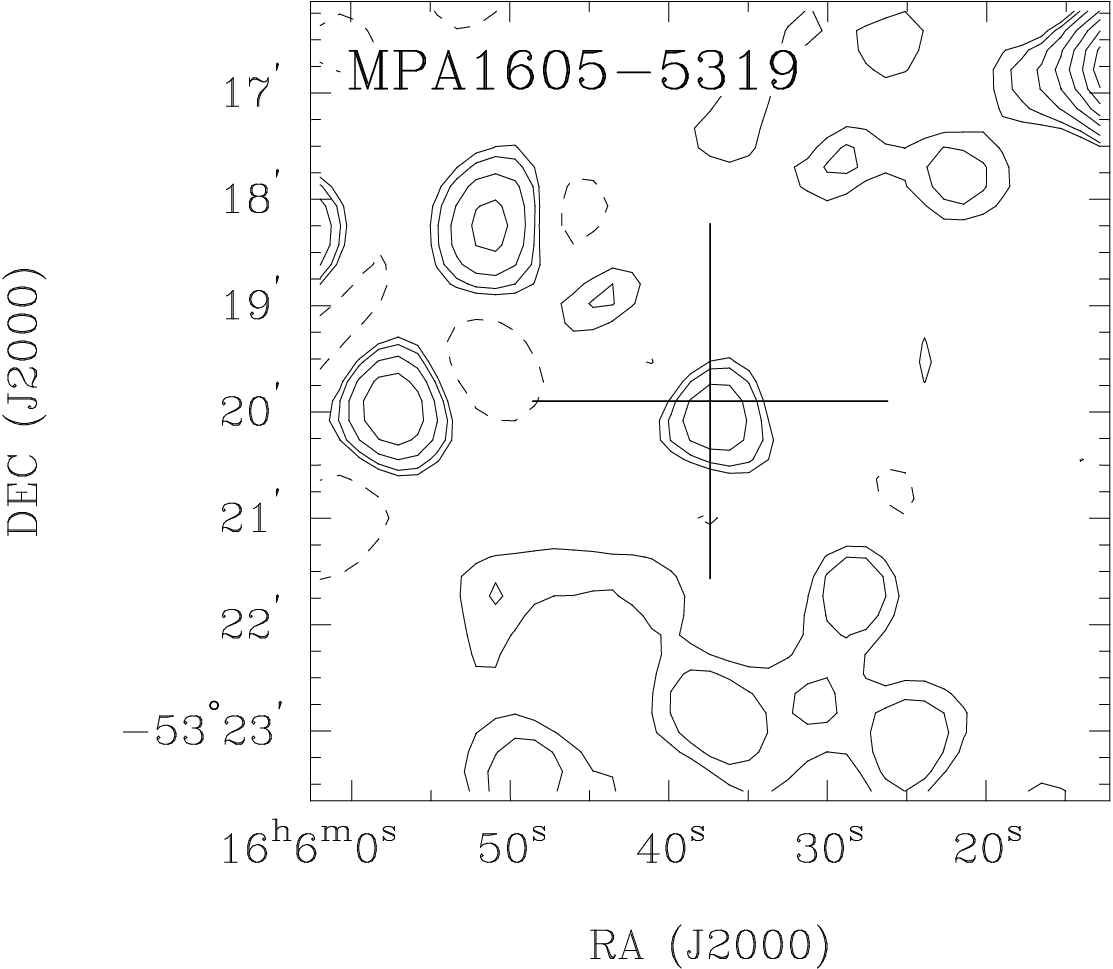}\label{sumss_MPA1605-5319}
\includegraphics[scale=0.3]{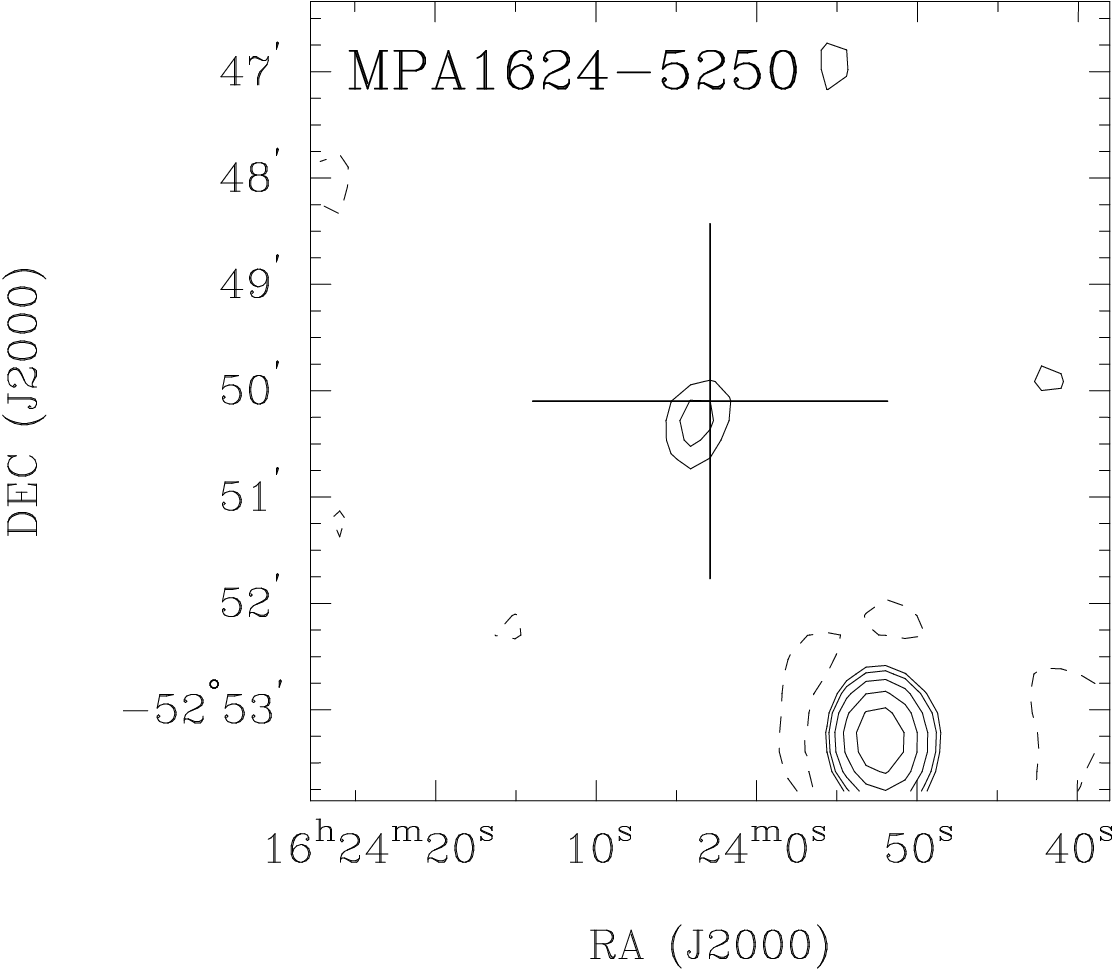}\label{sumss_MPA1624-5250}
\includegraphics[scale=0.3]{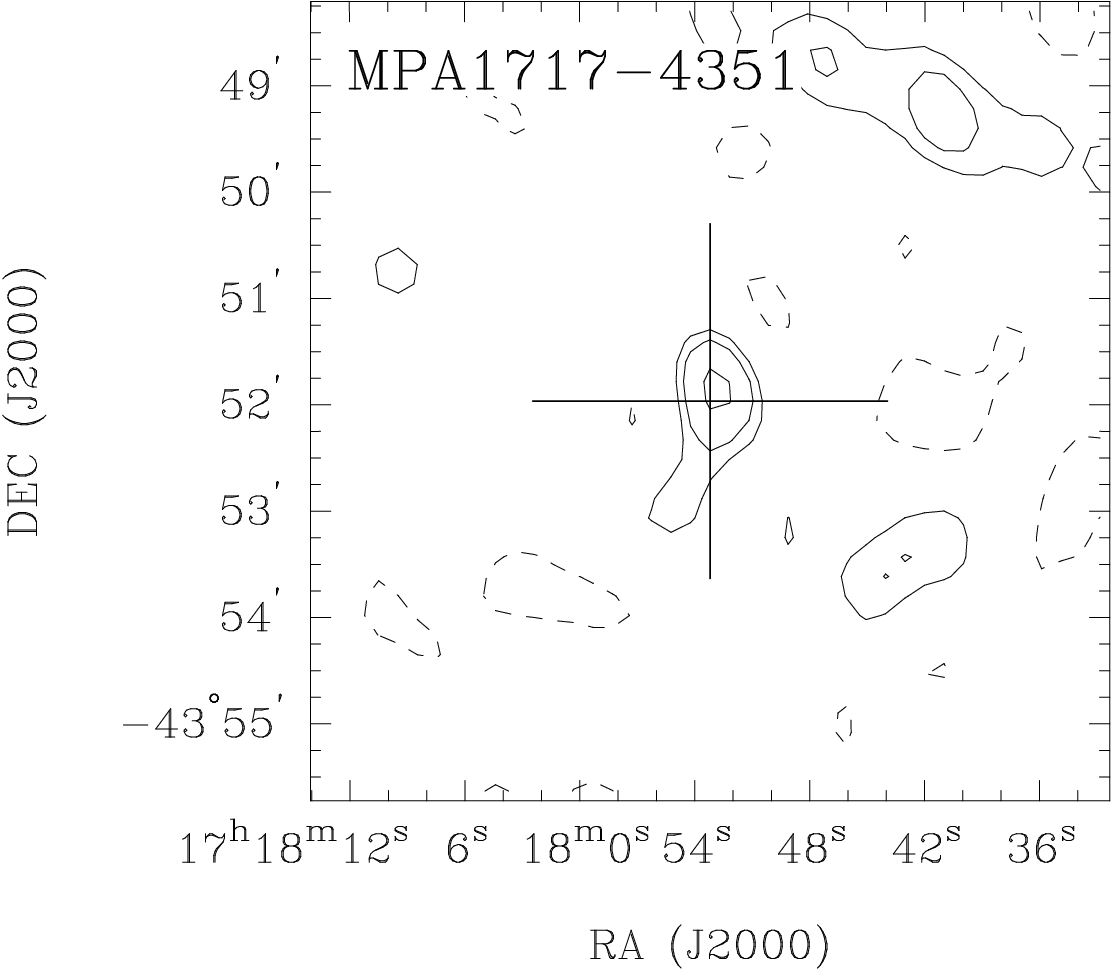}\label{sumss_MPA1717-4351}
\includegraphics[scale=0.3]{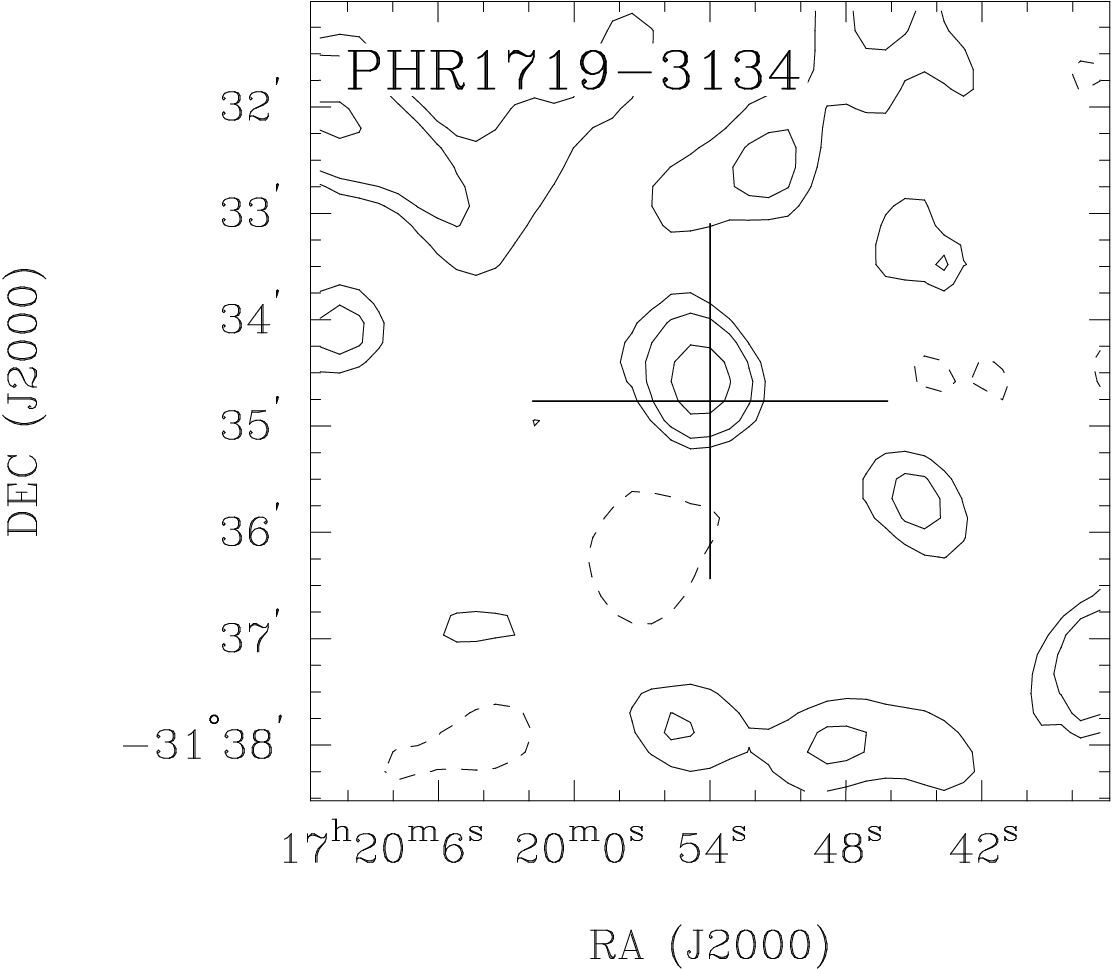}\label{sumss_PHR1719-3134}
\includegraphics[scale=0.3]{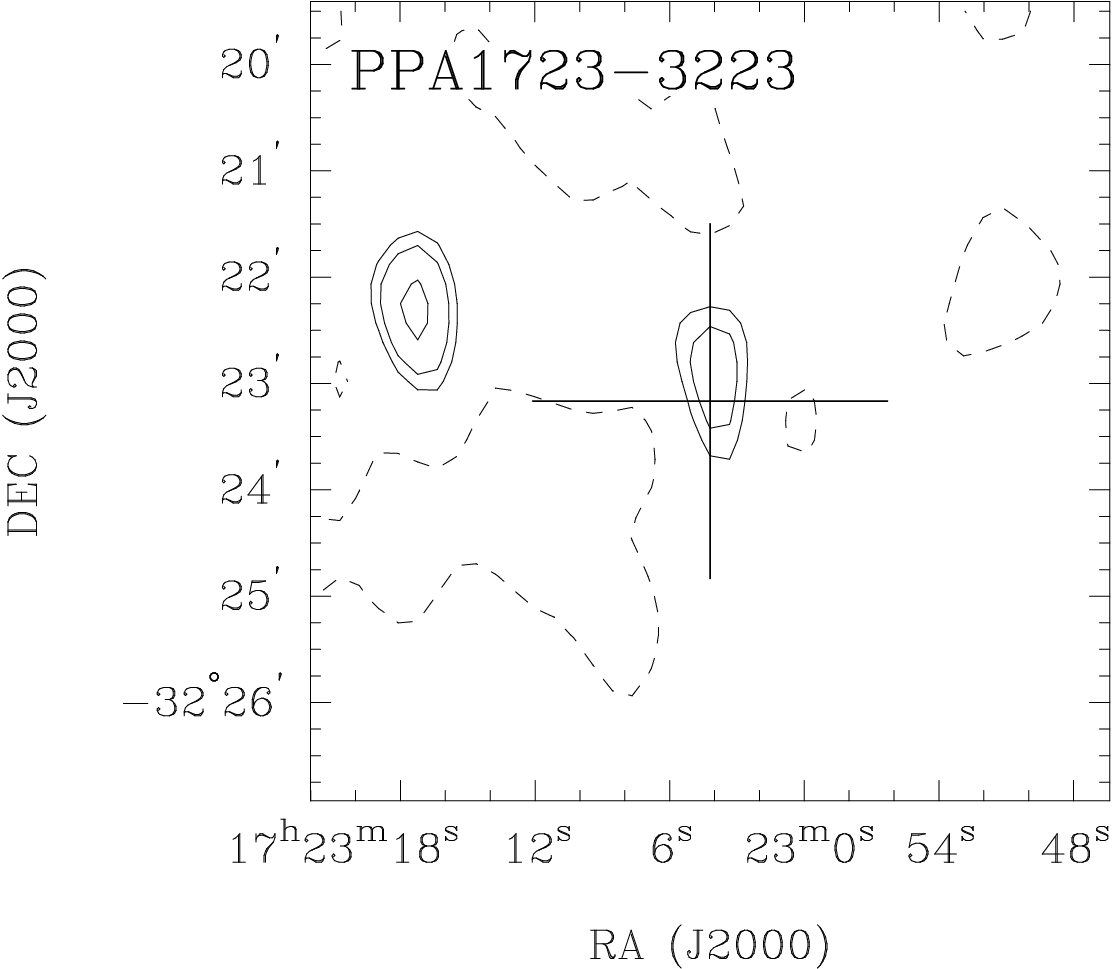}\label{sumss_PPA1723-3223}
\includegraphics[scale=0.3]{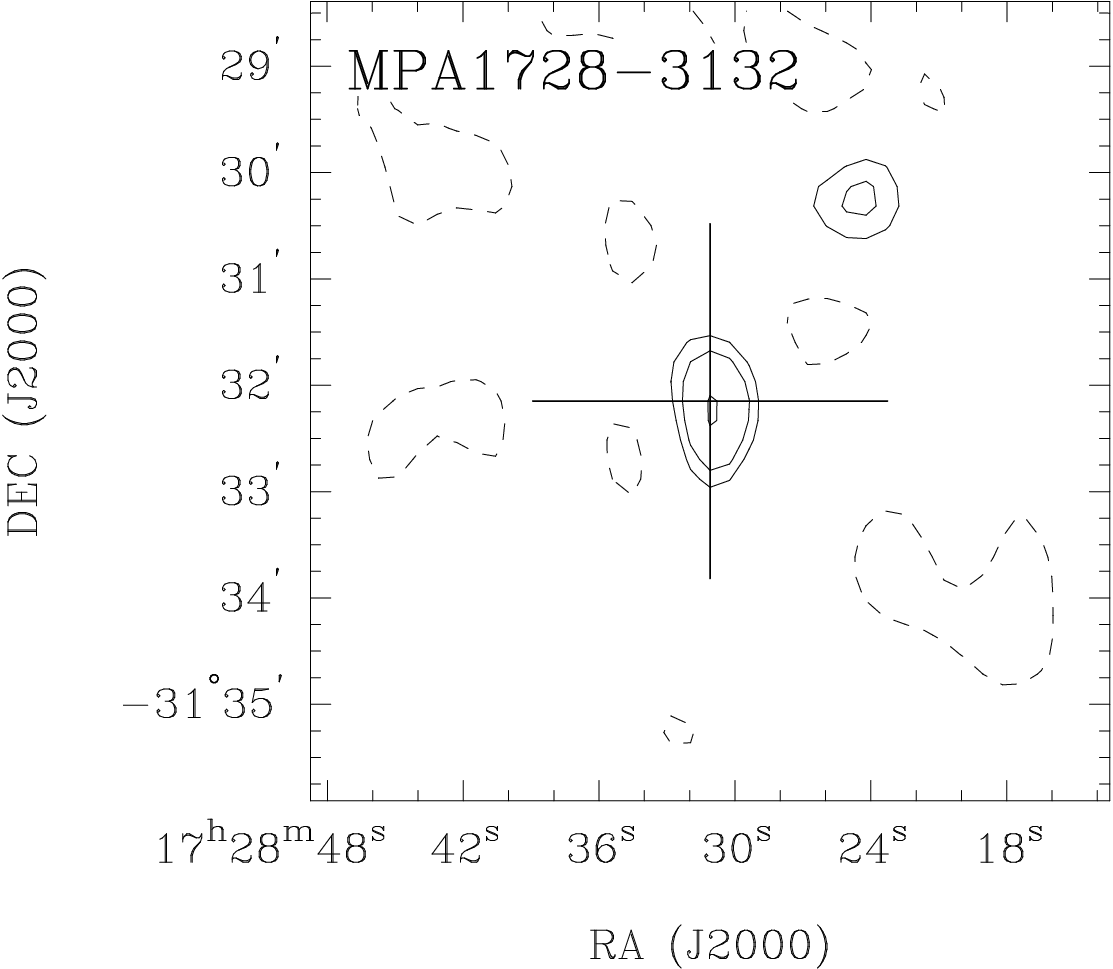}\label{sumss_MPA1728-3132}
\includegraphics[scale=0.3]{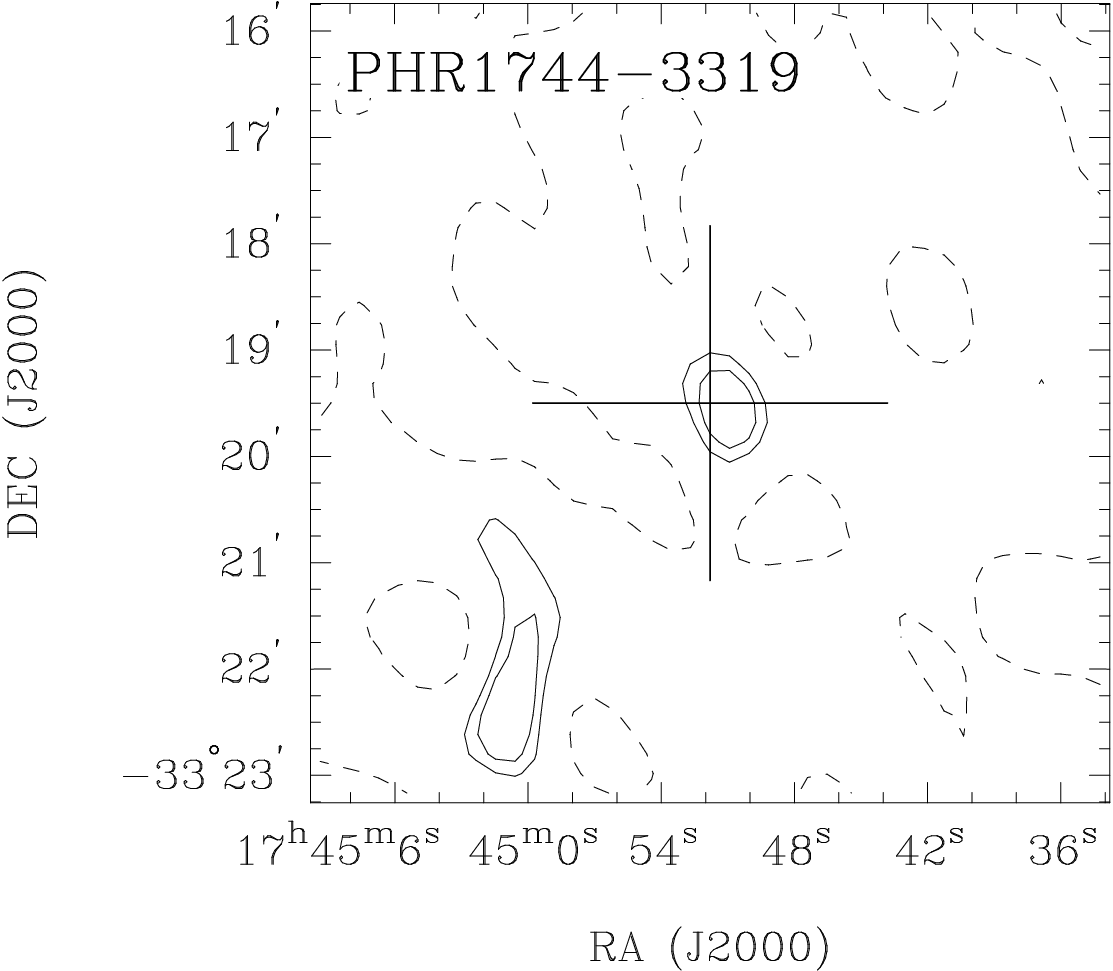}\label{sumss_PHR1744-3319}
\includegraphics[scale=0.3]{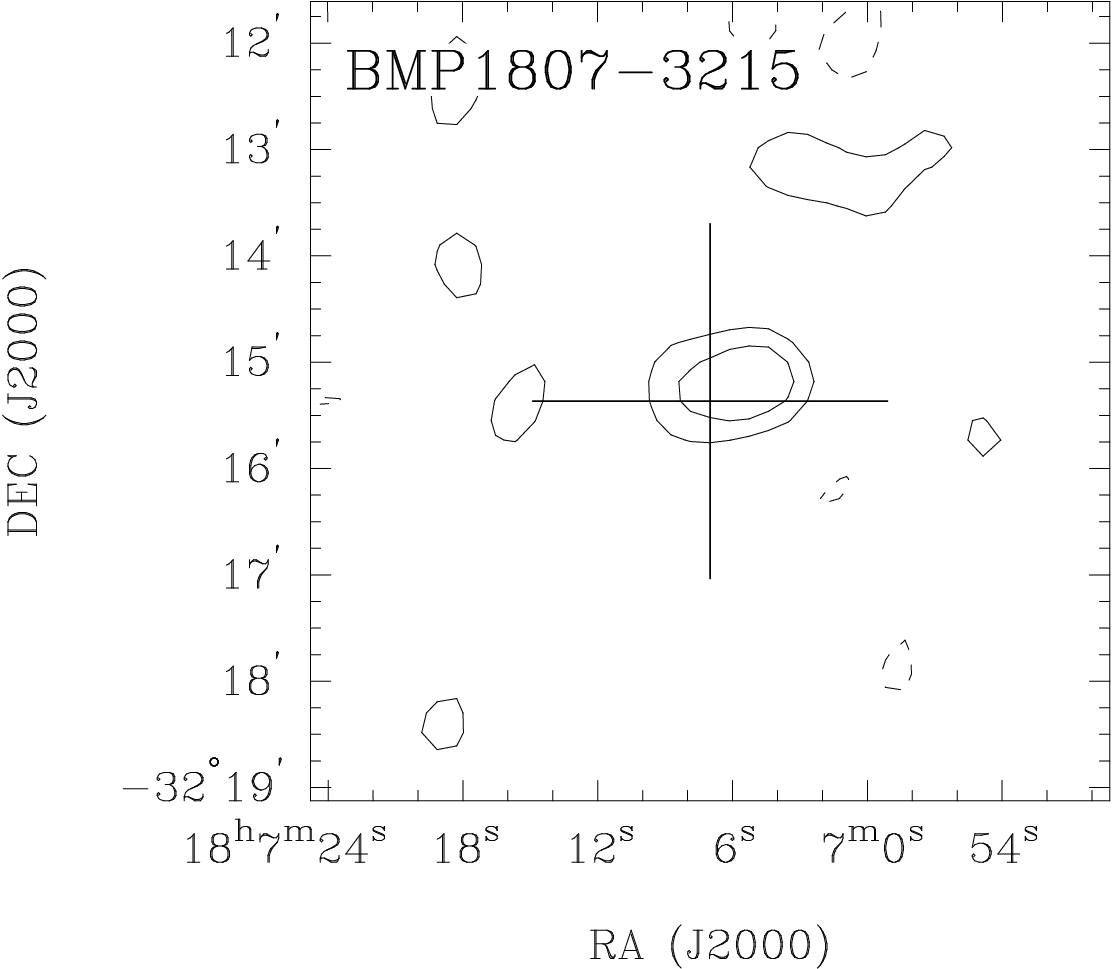}\label{sumss_BMP1807-3215}
\end{center}
\caption{Same as in Fig.~\ref{pysumss1}}\label{pysumss2}\end{figure}

In this appendix we present the MASH PNe finding charts used in this study (radio contour images and optical SHS {\it quotient} images overlaid with radio contours). The SHS {\it quotient} images were constructed as SHS \HA\ divided by SHS short red image. This technique allows better examination of spatial properties of extended and low surface brightness PNe \citep[for more details see][]{2008MNRAS.384..525M}. In Fig.~\ref{pmn1} we present nine MASH PNe detected in the PMN survey. Furthermore, we give more detailed descriptions of some of the positively or possibly identified radio counterparts of MASH PNe in the NVSS and MGPS-2. We discuss MASH PNe radio detections with nearby and bright radio objects which could be the source of a confusion and/or radio detections for which we found some discrepancy in the association with the optical emission.  Detailed descriptions of individual identifications for PNe with comment key {\bf 1}  were given in LCY05. Corresponding radio-continuum contour images, if available, are presented in Figures~\ref{fig:exam1} and \ref{fig:exam2}. In Figures~\ref{fig:detextend} and \ref{fig:detsuspect1} we present SHS {\it quotient} images overlaid with radio contours of several positive and suspect radio detected MASH PNe.  All objects from this set have optical angular diameters comparable or larger than the resolution of a corresponding radio image. Finally, the radio contour plots of 25 possibly detected MASH PNe radio counterparts from the MGPS-2 survey, not catalogued in the MGPS-2 catalogue, are presented in Fig.~\ref{pysumss1} and Fig.~\ref{pysumss2}.

{\bf (2)} G000.0$-$01.3 (PPA1751$-$2933; Fig.~\ref{fig:exam1}): Faint, oval nebula, designated as ``likely'' PN. The optical angular diameter is $\sim$12~arcsec\ while the faint radio source appears to be more extended.

{\bf (3)} G004.8$-$01.1 (PHR1801$-$2522; Fig.~\ref{fig:exam1}): Confirmed, compact PN, very bright at 1.4~GHz ($\sim67$~mJy). Faint wings are very likely image artefacts and not associated with the nebula.

{\bf (4)} G010.0$-$01.5 (PHR1813$-$2057; Fig.~\ref{fig:exam1}): Confirmed, bright, oval and compact PN with possible ansae and MSX detection in all 4 bands. The faint radio extension to the south ($\sim$4~mJy) is very likely not associated with the nebula.

{\bf (5)} G025.9+03.4 (PHR1826$-$0435; Fig.~\ref{fig:exam1}): Faint, slightly oval, true PN with enhanced opposing lobes and bipolar core. Optical angular diameter ($\sim20$~arcsec) is a factor of two smaller than FWHM of the NVSS restoring beam. Thus, the radio ``wing'' is probably a background source.

{\bf (6)} G026.4+02.7 (PHR1829$-$0431; Fig.~\ref{fig:exam1}): Small ($\sim20$~arcsec), faint, semi-circular nebula with bipolar core designated as a true PN. The radio ``wing'', extending to the south of the nebulosity, is not visible in available optical bands and it is very likely a faint background source.

{\bf (7)} G034.1$-$01.6 (MPA1859+0017; Fig.~\ref{fig:exam1}): Faint and compact ($\sim10$~arcsec) confirmed PN. Radio source, located $\sim$90~arcsec\ from the radio detection, appears slightly extended but not correlated to this PN.

{\bf (8)} G322.4$-$00.1a (MPA1523$-$5710; Fig.~\ref{fig:exam1}): Confirmed bipolar PN from the MASH-II supplement super-imposed over the shell of the nearby SNR~G322.5-00.1 discovered in the MOST survey of the southern Galactic plane \citep{1992MNRAS.256..121W}. The southern radio extension is a part of the SNR shell and it is not correlated to this PN. However, the fitted flux could be overestimated as a result of the underlying large scale structure.

{\bf (9)} G332.3$-$00.9 (PHR1619$-$5131; Fig.~\ref{fig:exam1}): While the coincidence of the optical position from the MASH with the radio excess in the SUMSS/MGPS-2 radio image looks quite convincing we didn't find a radio counterpart in the MGPS-2 catalogue.  Thus, this PN is designated with suspect radio identification. The measured flux density of 13.7~mJy is well above the usual {\it rms} noise of 1.0~mJy~Beam$^{-1}$. However, the local {\it rms} noise (or the large scale structures) in the vicinity of this PN is significantly above the average ($\sim4$~mJy~Beam$^{-1}$).

{\bf (10)} G352.6+02.2~(PPA1717$-$3349; Fig.~\ref{fig:exam2}): A compact optical nebula ($\theta_{opt}<10$~arcsec), designated as a ``likely'' PN. The nearby extended radio source has a positional association with the IRAS source 17146-3344. The ratio between IRAS fluxes at 12~$\mu$m and 25~$\mu$m is 1.3, which implies association with an OH/IR star. The extended source is relatively faint ($S_{1.4\textrm{GHz}}<$15~mJy) so the confusion of this PN is probably only a mild one.

{\bf (11)} G354.6$-$01.4 (PPA1737$-$3414; Fig.~\ref{fig:exam2}): The radio flux ($\sim$7~mJy) from the compact ($\theta_{opt}\approx$6~arcsec) true PN, is probably highly confused from the much brighter, and very likely unrelated, nearby object ($S_{1.4\textrm{GHz}}=177$~mJy).

{\bf (12)} G355.8+01.7~(MPA1728$-$3132; Fig.~\ref{fig:exam2}): Confirmed planetary nebula from the MASH-II supplement. Optical size of the nebula is 4~arcsec. The position of the correlated NVSS radio peak is $\sim$6~arcsec\ away from the estimated optical position. Also, the radio source appears to be relatively extended. This implies that the angular size of this PN could be much larger than seen in \HA.

{\bf (13)} G356.5$-$01.8 (PPA1744$-$3252; Fig.~\ref{fig:exam2}): Possible PN. The radio-continuum flux ($\sim6$~mJy) from the compact ($\theta_{opt}\approx$6~arcsec) nebula, is probably only mildly confused from the nearby, stronger, radio source (174423-325116; $S_{1.4\textrm{GHz}}=31$~mJy).

{\bf (14)} G000.3-01.6 (PHR1752$-$2930; Fig.~\ref{fig:exam2}): Compact ($\theta_{opt}\approx$8~arcsec), true PN close to star. Radio detected by \cite{2001A&A...373..536V} with measured flux densities at 6~cm and 3~cm of 8.5~mJy and 2.5~mJy. respectively. Also cross-correlated with NVSS source 175252$-$293000 with flux density of 4.1~mJy. The confusion with a nearby, radio brighter ($\sim$40~mJy at 1.4~GHz) NVSS radio source 175256$-$293044 is possible.

{\bf (15)} G302.3$-$00.5 (PHR1246$-$6324; Fig.~\ref{fig:exam2}): Small bipolar PN, clearly visible in the radio. The radio source is centred on the PN and shows possible  extended structure in a direction opposite to possible bipolar outflows. It is not clear if this structure is related to the PN or is it a faint background source.

{\bf (16)} G291.6$-$00.2 (PHR1115$-$6059; Fig.~\ref{fig:detextend}): Bright, large ($\sim100$~arcsec), circular nebula designated as ``likely'' PN. Radio detection at 0.843~GHz appears to be associated with the \HA\ bright NW edge. Estimated flux density is flagged as the low limit due to the marginal detection.

{\bf (17)} G337.4+02.6 (PHR1625$-$4522; Fig.~\ref{fig:detextend}): Very large, diffuse nebula designated as a ``likely'' PN in MASH. The radio counterpart (not catalogued in the MGPS-2) is a faint ``patch'' of extended emission placed over the brighter nebular region. Estimated flux density is flagged as the low limit.

{\bf (18)} G356.6$-$01.9 (PHR1745$-$3246; Fig.~\ref{fig:detextend}): Confirmed, slightly extended PN with central concentration with a radio counterpart catalogued in the MGPS-2. From Fig.~\ref{fig:detextend} can be seen that radio peak is slightly offset from the brightest part of nebulosity. Apparent extension in the SN direction is an effect of the beam elongation due to the low declination.

{\bf (19)} G013.3+01.1 (PHR1810$-$1647; Fig.~\ref{fig:detextend}): Slightly oval, confirmed PN, with prominent internal structure. The NVSS detection peaks over the brightest part of the possible shell. Estimated flux density is flagged as the low limit due to the marginal detection.

{\bf (20)} G222.5+07.6 (BMP0736$-$0500; Fig.~\ref{fig:detsuspect1}): Relatively large ($\sim80$~arcsec), extremely faint elliptical PN. The suspect radio detection from the NVSS is placed on the brighter part of the nebula.

{\bf (21)} G247.5$-$04.7 (PHR0742$-$3247; Fig.~\ref{fig:detsuspect1}): Large, elliptical, diffuse PN. The NVSS detection is suspected (barely above $1\sigma_{rms}$ local noise level) and coincides with the brightest parts of the shell. 

{\bf (22)} G254.5$-$02.7 (PHR0808$-$3745; Fig.~\ref{fig:detsuspect1}): Large, diffuse nebula designated as ``likely'' PN. The NVSS suspect radio detection is placed on the brighter part of the nebula.

{\bf (23)} G297.0$-$04.9 (PHR1150$-$6704; Fig.~\ref{fig:detsuspect1}): S-bar shaped, confirmed PN, with internal knots. Radio detection (MGPS-2) is suspect due to the relatively large offset from the \HA\ bright region. However, it is important to emphasise the similarity with position and extent of radio-peak offset seen in PHR1739$-$3829.

{\bf (24)} G309.5+00.8 (PHR1346$-$6116; Fig.~\ref{fig:detsuspect1}): Partial arcuate nebula with sharp western edge, designated as ``likely'' PN. A faint radio source found in MGPS-2 (not catalogued) is placed over the fainter region and it is considered as suspect.

{\bf (25)} G324.3+01.1 (PHR1529$-$5458; Fig.~\ref{fig:detsuspect1}): Possible PN, with strongly elongated, irregular emission and with possible superposed arcuate nebula. The associated radio source from MGPS-2 is extended and approximately follows the brightness distribution of \HA\ emission. However, as can be seen from the presented histogram equalised quotient image, the radio peak is placed over the low \HA\ emission region. The nebula is approximately twice the size of the MOST synthesised beam FWHM.

{\bf (26)} G332.3+07.0 (PHR1547$-$4533; Fig.~\ref{fig:detsuspect1}): Very faint, circular nebula designated as a ``likely'' PN in MASH. The radio detection at 0.843~GHz appears extended but covers only the inner (fainter) part of the nebula.

{\bf (27)} G351.1$-$03.9 (PHR1739$-$3829; Figure~\ref{fig:detsuspect1}): Bright, bipolar, confirmed PN. As in the case of PHR1748$-$3538, the radio emission at 1.4~GHz appear to be generally correlated with the \HA\ but with a radio peak $\sim20$~arcsec from the brightest parts of the nebula. In contrast to of the PHR1748$-$3538, the radio ``wings'' are, in this case, in the direction of possible bipolar outflows. It is flagged as a suspect radio detection.

{\bf (28)} G354.5$-$03.9~(PHR1748$-$3538; Fig.~\ref{fig:detsuspect1}): Oval ring PN with faint outer extensions, designated as a true PN with optical dimensions 56$\times$39~arcsec. As stated in LCY05, the bright radio peak is about 60~arcsec away from the optical centroid. A bright offset radio source is catalogued in NVSS as a possible complex object 174818-35375 with a flux density estimate of $S_{1.4\textrm{GHz}}=44.4\pm1.8$~mJy. The flux density of the nebula itself is estimated to be $S_{1.4\textrm{GHz}}=3.2\pm0.6$~mJy. Fig.~\ref{fig:detsuspect1} (left hand side panel in the bottom row) shows the histogram equalised quotient image overlaid with contours from the 1.4~GHz NVSS image. A bright, ring like structure is visible, with possible bipolar outflows. While the radio emission is extended in the direction of the bright region of nebulosity, the radio peak does not appear to be correlated with \HA\ emission. We also found a nearby bright radio source in cross-correlation with the MGPS-2 catalogue. It peaks at a similar position to the NVSS bright source with a flux density at 0.843~GHz of 70.8$\pm$2.6~mJy. The spectral index ($\alpha\approx-0.8$) of this object clearly points out to the non-thermal origin. Thus, we can definitely conclude non-association with the PN. Unfortunately, the mosaic containing this region (J1742M36) was not available and we have not been able to examine if similar extended emission, seen at 1.4~GHz, exists at 0.843~GHz. A faint radio source, to the left of the nebulosity, also does not appear to be associated with this PN.

{\bf (29)} G015.5+02.8 (BMP1808$-$1406; Fig.~\ref{fig:detsuspect1}): Very large ($\sim470$~arcsec), confirmed elliptical PN with enhanced opposing edges. The detection of the NVSS source placed over the bright SW \HA\ emission region is flagged as a suspect and, regarding very low surface brightness of this PN, it is very likely that it is due to chance coincidence.  

\clearpage

\clearpage

\section{MASH PNe with steep negative radio spectra}\label{app:negspectra}

Several, apparently ``True'' MASH PNe,  show a steep negative radio-continuum spectra. The negative spectral index implies nonthermal or strongly variable radio-continuum emission, both uncommon for PNe. We examined this set of objects in more detail.

{\bf PPA1722$-$3317} (PNG353.6+01.7; $\alpha^*_{0.843/1.4}=-0.5\pm0.3$) is a compact (4~arcsec) true PN. It is detected in NVSS with flux density $S_{1.4\textrm{GHz}}=14.6\pm0.7$~mJy and it has a catalogued counterpart in MGPS-2 with a flux density $S_{0.843\textrm{GHz}}=18.6\pm2.9$~mJy ($F_{0.843\textrm{GHz}}=17.9\pm2.7$~mJy Beam$^{-1}$). The peak position of the NVSS source is within 1~arcsec from its optical counterpart. However, the MGPS-2 detection appears to be offset with distance from the optical centroid by more than 10~arcsec. The radio object has a high brightness temperature of more than 10$^3$~K at frequencies below 1~GHz. Thus, if this object is a true PN, as seems likely, then, according to its high brightness temperature, it is very likely a young and compact nebula. Thus, we would expect to see a rapid rise in flux densities toward higher frequencies. In order to resolve this discrepancy this object is scheduled to be observed at 3~cm and 6~cm (Boji\v{c}i\'c et al. in prep.).

{\bf PPA1725$-$3216} (PNG354.8+01.8; $\alpha^*_{0.843/1.4}=-1.1\pm0.4$) is a confirmed, compact ($\theta_{opt}\approx8\times6$~arcsec) PN with strong emission lines. Radio-continuum detections at 1.4~GHz ($S_{1.4\textrm{GHz}}=10.0\pm0.6$~mJy) and 0.843~GHz ($S_{0.843\textrm{GHz}}=17.6\pm3.4$~mJy) are well aligned with the optical position with small angular offsets of $\sim$2.5~arcsec and $\sim$3.5~arcsec. The steep negative radio spectra points to non-thermal emission as the main emission mechanism at cm wavelengths. Additional observations are needed to resolve the true nature of this object.

{\bf PPA1729$-$3152} (PNG355.6+01.4; $\alpha^*_{0.843/1.4}=-0.7\pm0.4$) is a compact ($6\times5$~arcsec), probable very low excitation (VLE) PN with very strong lines in the red including [ArIII]. This object is designated only as ``likely'' PN. The radio counterpart seen at 1.4~GHz ($S_{1.4\textrm{GHz}}=8.2\pm0.6$~mJy) and at 0.843~GHz ($S_{0.843\textrm{GHz}}=11.6\pm2.1$~mJy and peak flux $F_{0.843\textrm{GHz}}=11.2\pm2.1$~mJy Beam$^{-1}$) is within 3~arcsec radius from the optical centroid. The MGPS-2 detection is close to the catalogue detection threshold and, additionally, the object appears to be placed in a noisy region. In order to resolve the true nature of the radio-continuum emission mechanism from this object additional observations are needed.

{\bf PHR1753$-$3443} (PNG355.9$-$04.4; $\alpha^*_{0.843/1.4}=-0.6\pm0.2$) is a confirmed, bipolar PN with newly identified WR spectral features from the central star \citep{DEPEW2010} and an angular size of 27$\times$15~arcsec. The found flux densities are  $S_{1.4\textrm{GHz}}=15.0\pm0.7$~mJy and $S_{0.843\textrm{GHz}}=20.9\pm1.8$~mJy. Both NVSS and MPGS-2 peak positions are slightly offset from the optical centroid ($\sim$6~arsec and $\sim$7~arcsec, respectively) but still well within the visible nebulosity. Even though a possibility of chance coincidence detection exists it is very likely that the radio-continuum detections are genuine. A small offset indicates the possibility that the detected radio-continuum emission is coming from a region much smaller than the visible nebular extent.

{\bf PHR1755$-$2904} (PNG001.0$-$01.9; $\alpha^*_{0.843/1.4}=-2.3\pm0.4$) is a compact, bright, slightly oval true PN with faint outer halo and angular diameter of 14.5$\times$12.5 arcsec. It is detected in NVSS with measured flux density $S_{1.4\textrm{GHz}}=4.8\pm0.6$~mJy and it has a catalogued counterpart in MGPS-2 with a flux density $S_{0.843\textrm{GHz}}=15.5\pm2.3$~mJy. Radio peak offsets, for both radio-continuum detections, are smaller than 3~arcsec. Unfortunately, the SUMSS/MGPS-2 mosaic cut-out is not available from the postage stamp server so we could not examine the possibility of some confusing source at 0.843~GHz. In order to resolve this discrepancy this object was observed at 3~cm and 6~cm (Boji\v{c}i\'c et al. in prep.).

{\bf PHR1758$-$1841} (PNG010.2+02.7; $\alpha^*_{1.4/5}=-0.9\pm0.2$) is a spectroscopically confirmed, compact, circular PN with angular diameter of 8~arcsec. The NVSS detection, with flux density $S_{1.4\textrm{GHz}}=18.3\pm0.7$~mJy, and detection at 5~GHz, with flux density $S_{5\textrm{GHz}}=6.1\pm1.5$~mJy \citep{1991A&AS...91..481R} are placed over the central part of this nebula with offsets between the radio peak and optical centroid of 1 and 3 seconds of arc, respectively. The WSRT observation (5~GHz) was performed with the synthesised beam FWHM of about 3-6 arcsec in $\alpha$ and 15-35 arcsec in $\delta$ \citep{1991A&AS...91..481R}. Thus, due to the small angular diameter of this PN, we believe that the large negative spectral index cannot be accounted to the missing flux problem. In order to resolve the true nature of this object additional observations are needed.

\newpage
\onecolumn

\end{document}